\documentclass[aps,prb,onecolumn,showpacs,preprintnumbers,amsmath,amssymb,groupedaddress]{revtex4}

\pdfoutput=1

\usepackage{amssymb}
\usepackage{amsthm}
\usepackage{amscd,color,bm}
\usepackage{colordvi}
\usepackage{graphicx}
\usepackage{enumerate}
\usepackage{psfrag}
\usepackage{amsmath}
\usepackage{multirow}
\usepackage[normalem]{ulem}

\makeatletter
\newcommand{\rmnum}[1]{\romannumeral #1}
\newcommand{\Rmnum}[1]{\expandafter\@slowromancap\romannumeral #1@}
\makeatother

\newcommand{\eg}{{\it e.g.}}
\newcommand{\etal}{{\it et al.}}
\newcommand{\vect}[1]{\mathbf{#1}}
\newcommand{\hvect}[1]{\hat{\mathbf{#1}}}
\newcommand{\Fourier}[1]{\widetilde{#1}}

\newcommand{\BetaP}{\beta^{\rm p}}
\newcommand{\BetaPo}{\beta^{\rm p0}}
\newcommand{\BetaPI}{\beta^{\rm p,I}}
\newcommand{\BetaPH}{\beta^{\rm p,H}}
\newcommand{\KBetaP}{\Fourier{\beta}^{\rm p}}
\newcommand{\KBetaPI}{\Fourier{\beta}^{\rm p,I}}
\newcommand{\KBetaPH}{\Fourier{\beta}^{\rm p,H}}
\newcommand{\BetaE}{\beta^{\rm e}}
\newcommand{\KBetaE}{\Fourier{\beta}^{\rm e}}
\newcommand{\Lam}{\vect{\Lambda}}

\newcommand{\Krho}{\Fourier{\rho}}
\newcommand{\vrho}{\varrho}

\newcommand{\vu}{\bm u}
\newcommand{\vf}{\bm f}
\newcommand{\vv}{\bm v}

\newcommand{\E}{\mathcal{E}}
\newcommand{\F}{\mathcal{F}}

\newcommand{\MC}{\mathcal{C}}
\newcommand{\bx}{\mathbf{x}}
\newcommand{\by}{\mathbf{y}}
\newcommand{\bk}{\mathbf{k}}
\newcommand{\bz}{\mathbf{z}}
\newcommand{\bs}{\mathbf{s}}
\newcommand{\bD}{\mathbf{\Delta}}

\newcommand{\rS}{\ensuremath{\rho^\Sigma}}

\newcommand{\sfE}{\overset{sf}{=}}

\begin{document}

\title{Scaling theory of continuum dislocation dynamics in three dimensions: Self-organized fractal pattern formation}

\author{Yong S. Chen}
\email{yc355@cornell.edu}
\author{Woosong Choi}
\affiliation{Laboratory of Atomic and Solid State Physics (LASSP),
Clark Hall, Cornell University, Ithaca, New York 14853-2501, USA}
\author{Stefanos Papanikolaou}
\affiliation{Department of Mechanical Engineering and Materials Science \& Department of Physics,
Yale University, New Haven, Connecticut, 06520-8286, USA}
\author{Matthew Bierbaum}
\author{James P. Sethna}
\email{sethna@lassp.cornell.edu}
\affiliation{Laboratory of Atomic and Solid State Physics (LASSP),
Clark Hall, Cornell University, Ithaca, New York 14853-2501, USA}

\date{\today}

\begin{abstract}
We focus on mesoscopic dislocation patterning via a continuum dislocation dynamics theory 
(CDD) in three dimensions (3D).
We study three distinct physically motivated dynamics which consistently lead to 
fractal formation in 3D with rather similar morphologies, and therefore we suggest that this is a general feature of 
the 3D collective behavior of geometrically necessary dislocation (GND) ensembles. 
The striking self-similar features are measured 
in terms of correlation functions of physical observables, such as the GND
density, the plastic distortion, and the crystalline orientation.
Remarkably, all these correlation functions exhibit spatial power-law behaviors, sharing a single underlying universal critical exponent
for each type of dynamics. 
\end{abstract}

\maketitle

\section{\label{sec:intro}Introduction}
Dislocations in plastically deformed crystals, driven by their long-range interactions, 
collectively evolve into complex heterogeneous structures where 
dislocation-rich cell walls or boundaries surround dislocation-depleted cell interiors.
These have been observed both in single crystals~\cite{YawasakiSM80,MughrabiPMA86,SchwinkSM92} and polycrystals~\cite{UngarAM86}
using transmission electron microscopy (TEM). 
The mesoscopic cellular structures have been recognized as scale-free patterns through fractal
analysis of TEM micrographs~\cite{SevillanoSM91,SevillanoRS93,ZaiserPRL98,ZaiserAM99}.
The complex collective behavior of dislocations has been a challenge for understanding the
underlying physical mechanisms responsible for the development of emergent dislocation morphologies.

Complex dislocation microstructures, as an emergent mesoscale phenomenon, 
have been previously modeled using various theoretical 
and numerical approaches~\cite{AnanthakrishnaPR07}. Discrete dislocation dynamics (DDD) models have provided insights into the 
dislocation pattern formations: parallel edge dislocations in a two-dimensional system
evolve into `matrix structures' during single slip~\cite{GromaPRB99}, and `fractal and cell structures' during
multiple slip~\cite{GromaPRL07,BakoPRB07}; random dislocations in a three-dimensional system self-organize themselves
into microstructures through junction formation, cross-slip, and short-range interactions~\cite{DevincreSM02,DevincrePRL06}. 
However, DDD simulations are limited by the computational challenges on the relevant scales of length and strain.
Beyond these micro-scale descriptions, CDD has also been used
to study complex dislocation structures. Simplified reaction-diffusion models have described persistent
slip bands~\cite{WalgraefIJES85}, dislocation cellular structures during multiple slip~\cite{HahnerAPA96}, and dislocation vein
structures~\cite{SaxlovaMSEA97}. Stochasticity in CDD models~\cite{ZaiserPRL98,GromaPRB99,GromaPRL00} 
or in the splittings and rotations of the macroscopic cells~\cite{Pantleon96,Pantleon98,JSPRB03} have been suggested as an explanation for 
the formation of organized dislocation structures. The source of the noise in these stochastic theories is derived from 
either extrinsic disorder or short-length-scale fluctuations.
\begin{figure}[!t]
\centering
\includegraphics[width=0.85\columnwidth]{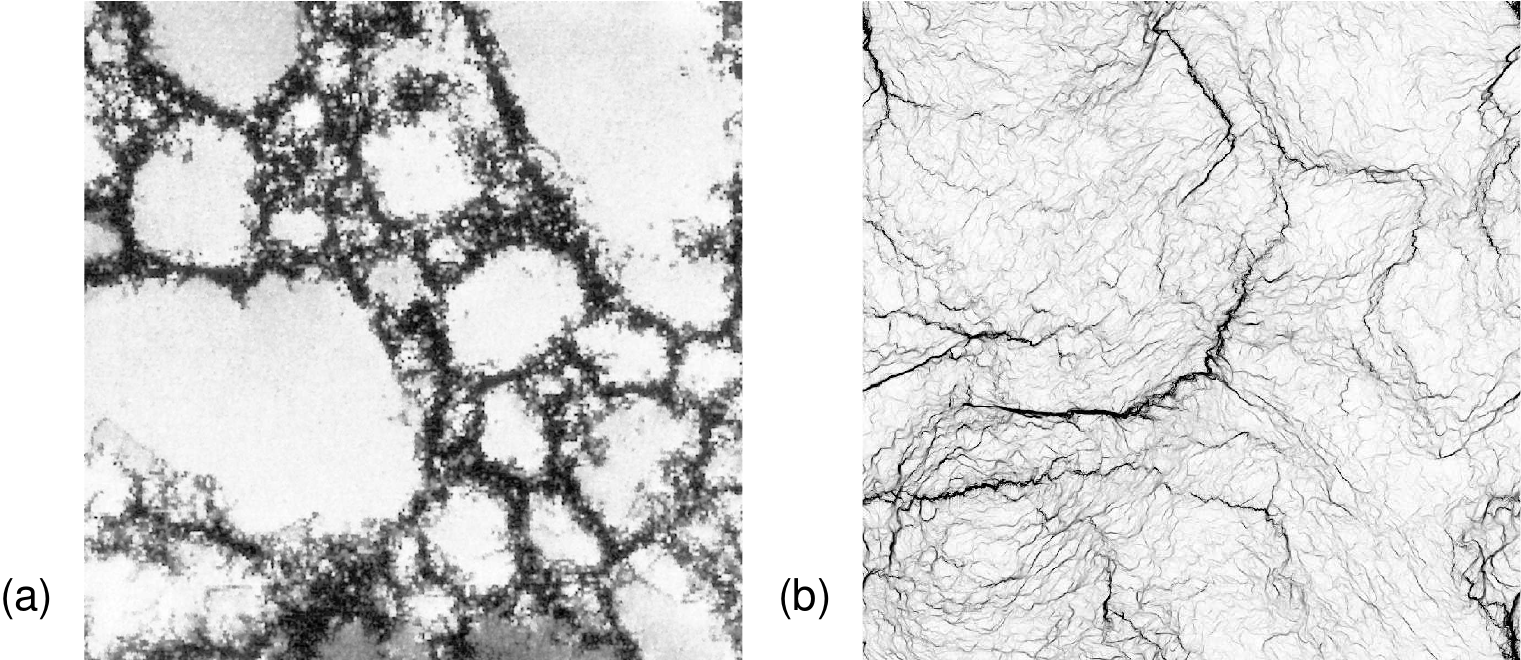}
\caption{{\bf Experimental and simulated dislocation cellular structures.} 
In (a), a typical TEM micrograph at a micron scale is taken from a Cu single
crystal after $[100]$ tensile deformation to a stress of $76.5$ MPa~\cite{ZaiserPRL98};
in (b), a simulated GND density plot is shown. 
Note the striking morphological similarity
between theory and experiment.}
\label{fig:comparison}
\end{figure}

In a recent manuscript~\cite{YSCPRL10}, we analyzed the behavior of a
grossly simplified continuum dislocation model for
plasticity~\cite{AcharyaJMPS01,AcharyaJMPS05,AcharyaJMPS06I,JSPRL06,YSCPRL10} -- a 
physicist's `spherical cow' approximation designed to explore the minimal
ingredients necessary to explain key features of the dynamics of deformation.
Our simplified model ignores many features known to be important for 
cell boundary morphology and evolution, including slip systems and
crystalline anisotropy, dislocation nucleation, lock formation and entanglement,
line tension, geometrically unnecessary forest dislocations, etc. However,
our model does encompass a realistic order parameter field (the Nye-Kr\"oner dislocation
density tensor~\cite{Nye53,Kroner58} embodying the GNDs), which
allows detailed comparisons of local rotations and deformations, stress, and 
strain.  It is not a realistic model of a real material,
but it is a model material with a physically sensible evolution law. 
Given these simplifications, our model exhibited a surprisingly realistic
evolution of cellular structures. We analyzed these structures 
in two-dimensional simulations (full three-dimensional rotations and
deformations, but uniform along the $z$-axis) using both the fractal
box counting method~\cite{SevillanoSM91,SevillanoRS93,ZaiserPRL98,ZaiserAM99} and the single-length-scale scaling
methods~\cite{HughesAM97,HughesPRL98,DawsonAM99,HughesPRL01} used in previous theoretical analyses of 
experimental data. Our model qualitatively reproduced the self-similar,
fractal patterns found in the former, and the scaling behavior of the cell
sizes and misorientations under strain found in the latter (power-law
refinement of the cell sizes, power-law increases in misorientations, and
scaling collapses of the distributions).

There are many features of real materials which are not explained by our
model. We do not observe distinctions between `geometrically necessary'
and `incidental' boundaries, which appear experimentally to scale in
different ways. The fractal scaling observed in our model may well be
cut off or modified by entanglement, slip-system physics, quantization
of Burgers vector~\cite{KuhlmannMMTA85} or anisotropy
-- we cannot predict that real materials should have fractal cellular
structures; we only observe that our model material does so naturally.
Our spherically symmetric model obviously cannot reproduce 
the dependence of morphological evolution on the axis of applied
strain (and hence the number of activated slip systems); indeed, the fractal
patterns observed in some experiments~\cite{ZaiserPRL98,ZaiserAM99} could be associated
with the high-symmetry geometry they studied~\cite{PoulsenMT07,HansenMMTA2011}.
While many realistic features of materials that we ignore may be important
for cell-structure formation and evolution, our model
gives clear evidence that these features are not essential to the formation
of cellular structures when crystals undergo plastic deformation.

In this longer manuscript, we provide an in-depth analysis of three
plasticity models. We show how they (and more traditional models) can
be derived from the structures of the broken symmetries and order parameters.
We extend our simulations to 3D, where the behavior is
qualitatively similar with a few important changes. Here we focus our 
attention on relaxation (rather than strain), and on correlation functions
(rather than fractal box counting or cell sizes and misorientations).

Studying simplified `spherical cow' models such as ours is justified
if they capture some key phenomenon, providing a perspective or explanation
for the emergent behavior. Under some circumstances, these simplified 
models can capture the long-wavelength behavior precisely -- the model
is said to be in the same universality class as the observed
behavior~\citep[Chapter~12]{Sethna06}.  The Ising model for magnetism, 
two-fluid criticality, and order-disorder transitions;
self-organized critical models for magnetic Barkhausen
noise~\cite{SethnaNature01,Durin06} and dislocation avalanches~\cite{ZaiserAIP06}
all exhibit the same type of emergent scale-invariant behavior as 
observed in some experimental cellular structures~\cite{ZaiserPRL98}.
For all of these systems, `spherical cow' models provide quantitative
experimental predictions of all phenomena on long length and time scales,
up to overall amplitudes, relevant perturbations, and corrections to 
scaling. Other experimental cellular structures~\cite{HughesPRL98} have been
interpreted in terms of scaling functions with a characteristic scale,
analogous to those seen in crystalline grain growth. Crystalline
grain growth also has a `universal' description, albeit one which depends
upon the entire anisotropic interfacial energy and mobility~\cite{RutenbergPRL99}
(and not just temperature and field).%
  \footnote{See, however, Ref.~\onlinecite{KacherMSEA11} for experimental observations
  of bursty grain growth is incompatible with these theories.}
We are cautiously optimistic that a model like ours (but with metastability
and crystalline) could indeed describe the emergent complex dislocation
structures and dynamics in real materials. Indeed, recent
work on dislocation avalanches suggests that even the yield stress
may be a universal critical point~\cite{FriedmanPRL12}. 

Despite universality, we must justify and explain the form of the CDD
model we study. In Sec.~\ref{sec:theory}
we take the continuum, `hydrodynamic' limit approach, traditionally
originating with Landau in the study of systems near thermal equilibrium
(clearly not true of deformed metals!). All degrees of freedom are 
assumed slaves to the order parameter, which is systematically
constructed from conserved quantities and broken
symmetries~\cite{Martin68,Forster75,HohenbergRMP77} -- this is the fundamental tool
used in the physics community to derive
the diffusion equation, the Navier-Stokes equation, 
and continuum equations for superconductors, superfluids, liquid
crystals,~etc. Ref.~\onlinecite{rickman97} have utilized this general approach
to generate CDD theories, and in
Sec.~\ref{sec:theory} we explain how our approach differs from theirs.

In Sec.~\ref{sec:CoarseGraining} we explore the validity of several
approximations in our model, starting in the engineering language of
state variables. Here local equilibration is not presumed; the state 
of the system depends in some arbitrarily complex way on the history. 
Conserved quantities and broken symmetries can be supplemented by
internal state variables -- statistically stored dislocations (SSDs), yield surfaces, void fractions, etc.,
whose evolution laws are judged according to their success in matching
experimental observations. (Eddy viscosity theories of turbulence are
particular successful examples of this framework.) 
The `single-velocity' models we use were originally developed by 
Acharya \etal~\cite{AcharyaJMPS01,AcharyaJMPS05}, and we discuss their
microscopic derivation~\cite{AcharyaJMPS01} and the
correction term $L^p$ resulting from coarse-graining and multiple
microscopic velocities~\cite{AcharyaJMPS06I}. This term is
usually modeled by the effects of SSDs using crystal plasticity 
models.  We analyze experiments to suggest that ignoring SSDs may
be justified on the length-scales needed in our modeling. However, we 
acknowledge the near certainty that Acharya's $L^p$ will be important --
the true coarse-grained evolution laws will incorporate multiple
velocities. Our model should be viewed as a physically sensible model material,
not a rigorous continuum limit of a real material.

In this manuscript, we study fractal cell structures that form
upon relaxation from randomly deformed initial conditions
(Sec.~\ref{sec:corrfuncVI}). One might be concerned that relaxation
of a randomly imposed high-stress dislocation structure (an
instantaneous hammer blow) could yield qualitatively different behavior
from realistic deformations, where the dislocation structures evolve
continuously as the deformation is imposed.
In Sec.~\ref{sec:corrfuncVI} we note that this alternative `slow hammering'
gives qualitatively the same fractal dislocation patterns. 
Also, the resulting cellular structures
are qualitatively very similar to those we observe under subsequent
uniform external strain~\cite{YSCPRL10,tobepublished2}, except that the
relaxed structures are statistically isotropic.
We also
find that cellular structures form immediately at small deformations.
Cellular structures in real materials emerge only after significant deformation;
presumably this feature is missing in our model because
our model has no impediment to cross-slip or multiple slip, and
no entanglement of dislocations.
This initial relaxation should not be viewed as annealing or dislocation creep.
A proper description of annealing must include dislocation line tension 
effects, since the driving force for annealing is the reduction in total
dislocation density -- our dislocations annihilate when their
Nye Burgers vector
density cancels under evolution, not because of the dislocation core energies.
Creep involves dislocation climb, which (for two of our three models) is
forbidden.

We focus here on correlation functions, rather than the methods used
in previous analyses of experiments. Correlation functions have a long,
dignified history in the study of systems exhibiting emergent scale
invariance -- materials at continuous thermodynamic phase
transitions~\cite{Chaikin1995}, fully developed
turbulence~\cite{LvovPR91,ChoiCiSE11,SalmanIJES12}, and crackling noise
and self-organized criticality~\cite{SethnaNature01}. We study not
only numerical simulations of these correlations, but provide also
extensive analysis of the relations between the correlation
functions for different physical quantities and their (possibly universal)
power-law exponents. The decomposition
of the system into cells (needed for the cell-size and misorientation
distribution analyses~\cite{HughesAM97,HughesPRL98,DawsonAM99,HughesPRL01}) demands the introduction of
an artificial cutoff misorientation angle, and demands either laborious
human work or rather sophisticated numerical algorithms~\cite{MattWebSite}. 
These sections of the current manuscript may be viewed both as a full
characterization of the behavior of our simple model, and as an illustration
of how one can use correlation functions to analyze the complex morphologies
in more realistic models and in experiments providing 2D or 3D real-space data.
We believe that analyses that explicitly decompose structures into cells
remain important for systems with single changing length-scale: grain boundary
coarsening should be studied both with correlation functions and with 
explicit studies of grain shape and geometry evolution, and the same should
apply to cell-structure models and experiments that are not fractal. But
our model, without such an intermediate length-scale, is best analyzed
using correlation functions.

Our earlier work~\cite{YSCPRL10} focused on 2D. 
How different are our predictions in 3D? 
In this paper, we explore three different CDDs that display similar dislocation fractal formation in 3D 
and confirm analytically that correlation functions of the GND density, the plastic distortion, and the 
crystalline orientation, all share a single underlying critical exponent, up to exponent relations, 
dependent only on the type of dynamics. Unlike our 2D simulations, where forbidding climb led to rather 
distinct critical exponents, all three dynamics in 3D share quite similar scaling behaviors. 

We begin our discussion in~Sec.~\ref{sec:orderP} by defining the various dislocation, distortion, and orientation fields. 
In~Sec.~\ref{sec:engineermodels}, we derive standard local dynamical evolution laws using
traditional condensed matter approaches, starting from both the non-conserved plastic distortion and the conserved 
GND densities as order parameters. Here, we also explain why these resulting dynamical laws are inappropriate at 
the mesoscale. In~Sec.~\ref{sec:cdd}, we show how to extend this approach by defining appropriate 
constitutive laws for the dislocation flow velocity to build novel dynamics~\cite{Landau75}. 
There are three different dynamics we study: \rmnum{1}) isotropic 
climb-and-glide dynamics~(CGD)~\cite{AcharyaJMPS01,AcharyaPRSA03,AcharyaJMPS04,AcharyaJMPS05,JSPRL06}, 
\rmnum{2}) isotropic glide-only dynamics, where we define the part of the local dislocation density that 
participates in the local {\it mobile dislocation population}, keeping the local volume 
conserved at all times~(GOD-MDP)~\cite{YSCPRL10}, \rmnum{3}) isotropic glide-only dynamics, 
where glide is enforced by a {\it local vacancy pressure} due to a co-existing background of 
vacancies that have an infinite energy cost~(GOD-LVP)~\cite{AcharyaJMPS06I}. 
All three types of dynamics present physically valid alternative approaches for 
deriving a coarse-grained continuum model for GNDs. 
In~Sec.~\ref{sec:CoarseGraining}, we explore the effects of coarse-graining,
explain our rationale for ignoring SSDs at
the mesoscale, and discuss the single-velocity approximation we use.
In~Sec.~\ref{sec:results}, we discuss the details of numerical simulations in both two and three dimensions, and characterize the self-organized critical
complex patterns in terms of correlation functions of the order parameter fields.
In~Sec.~\ref{sec:staticscaling},
we provide a scaling theory, derive relations among the
critical exponents of these related correlation functions, study
the correlation function as a scaling function of coarse-graining length scale,
and conclude in 
Sec.~\ref{sec:conclusion}. 

In addition, we provide extensive details of our study in Appendices. In \ref{sec:physParameters}, 
we collect useful formulas from the literature relating different physical quantities within 
traditional plasticity, while in \ref{sec:varmethod} we show how functional derivatives 
and the dissipation rate can be calculated using this formalism, 
leading to our proof that our CDDs are strictly dissipative (lowering the appropriate free energy with time). 
In \ref{sec:modelex}, we show the flexibility of our CDDs by extending our dynamics: 
In particular, we show how to add vacancy diffusion in the structure of CDD, 
and also, how external disorder can be in principle incorporated (to be explored in future work). In \ref{sec:detail}, 
we elaborate on numerical details -- we demonstrate the statistical convergence 
of our simulation method and also we explain how we construct the Gaussian random initial conditions.
Finally, in \ref{sec:othercorrfunc}, we discuss the scaling properties of several 
correlation functions in real and Fourier spaces, including the strain-history-dependent plastic deformation and distortion fields,
the stress-stress correlation functions, the elastic energy
density spectrum, and the stressful part of GND density.

\section{\label{sec:theory}Continuum models}
\subsection{\label{sec:orderP}Order parameter fields}
\subsubsection{\label{sec:rho}Conserved order parameter field}
\begin{figure}[!t]
\centering
\includegraphics[width=0.5\columnwidth]{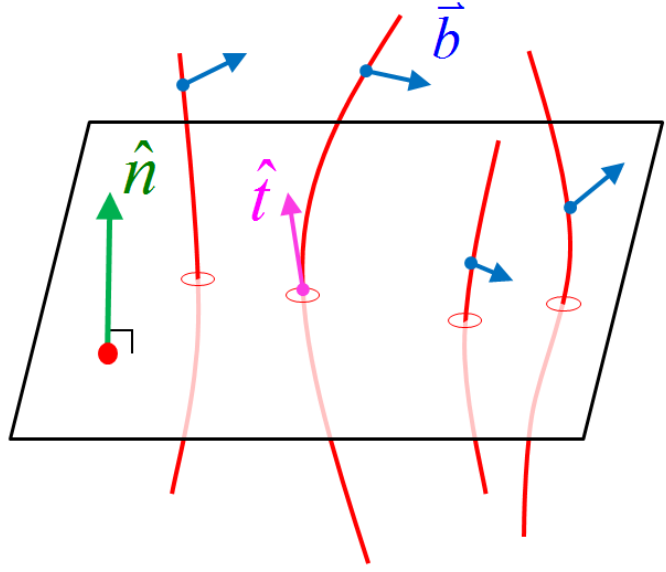}
\caption{(Color online)~{\bf Representation of the crystalline line defect --- dislocation.} 
Each curved line represents a dislocation line with the tangent direction $\hvect{t}$, 
and the Burgers vector $\vect{b}$ which characterizes the magnitude and direction of the distortion to the lattice.
The two-index GND density $\rho_{km}$~\cite{Nye53,Kroner58} (Eqs.~\ref{eq:rho_Nye0} and ~\ref{eq:rho_Nye}) is the net flux of the Burgers
vector density $\vect{b}$ along $\hvect{e}^{(m)}$ through an infinitesimal piece of a plane with normal direction $\hvect{n}$ along
$\hvect{e}^{(k)}$. The three-index version $\vrho_{ijm}$ (Eqs.~\ref{eq:varrho_Nye0} and ~\ref{eq:varrho_Nye1}) is the flux density
through the plane along the axes $\hvect{e}^{(i)}$ and $\hvect{e}^{(j)}$, with the unit bivector $\hat{E}=\hvect{e}^{(i)}\wedge\hvect{e}^{(j)}$.
}
\label{fig:rhotensor}
\end{figure}

A dislocation is the topological defect of a crystal lattice. 
In a continuum theory, it can be described by a coarse-grained variable,
the GND density,
   \footnote{ Dislocations which cancel at the macroscale
   may be geometrically necessary at the mesoscale. See 
   Sec.~\ref{sec:CoarseGraining} for our rationale for not including
   the effects of SSDs (whose Burgers vectors
   cancel in the coarse-graining process).}
(also called the net dislocation density or the Nye-Kr\"oner dislocation
density), 
which can be defined by the GND density tensor
\begin{equation}
\rho(\bx)=\sum_\alpha (\hvect{t}^\alpha\!\!\cdot\hvect{n})\hvect{n}\otimes\vect{b}^\alpha\delta(\bx-{\bm\xi}^\alpha),
\label{eq:rho_Nye0}
\end{equation}
so
\begin{equation}
\rho_{km}(\bx)=\sum_\alpha \hat{t}_k^\alpha b_m^\alpha\delta(\bx-{\bm\xi}^\alpha),
\label{eq:rho_Nye}
\end{equation}
measuring the sum of the net flux of dislocations $\alpha$ located at $\bm \xi$, tangent to $\hvect{t}$, with Burgers
vector $\vect{b}$, in the neighborhood of $\bx$, through an infinitesimal plane with the normal direction along $\hvect{n}$, 
seen in Fig.~\ref{fig:rhotensor}. In the continuum, the discrete sum of line singularities in Eqs.~(\ref{eq:rho_Nye0}) and~(\ref{eq:rho_Nye})
is smeared into a continuous (nine-component) field, just as the continuum density of a liquid is at root a sum of point 
contributions from atomic nuclei.

Since the normal unit pseudo-vector $\hvect{n}$ is equivalent to an antisymmetric unit bivector $\hat{E}$,
$\hat{E}_{ij}=\varepsilon_{ijk}\hat{n}_k$, we can reformulate the GND density as a three-index tensor
\begin{equation}
\vrho(\bx)=\sum_\alpha (\hvect{t}^\alpha\!\!\cdot\hvect{n})\hat{E}\otimes\vect{b}^\alpha\delta(\bx-{\bm\xi}^\alpha),
\label{eq:varrho_Nye0}
\end{equation}
so
\begin{equation}
\vrho_{ijm}(\bx)=\sum_\alpha (\hvect{t}^\alpha\!\!\cdot\hvect{n})\hat{E}_{ij}b_m^\alpha\delta(\bx-{\bm\xi}^\alpha),
\label{eq:varrho_Nye1}
\end{equation}
measuring the same sum of the net flux of dislocations in the neighborhood of $\bx$, through the infinitesimal plane indicated
by the unit bivector $\hat{E}$. This three-index variant will be useful in Sec.~\ref{sec:GOD}, where we adapt the equations of 
Refs.~\onlinecite{AcharyaJMPS05}~and~\onlinecite{JSPRL06} to forbid dislocation climb (GOD-MDP).

According to the definition of $\hat{E}$, we can find the relation between $\rho$ and $\vrho$
\begin{equation}
\vrho_{ijm}(\bx)=\sum_\alpha (\hat{t}_l^\alpha\hat{n}_l)\varepsilon_{ijk}\hat{n}_kb_m^\alpha\delta(\bx-{\bm\xi}^\alpha)=\varepsilon_{ijk}\rho_{km}(\bx).
\label{eq:varrho_Nye2}
\end{equation}

It should be noted here that dislocations cannot terminate within the
crystal, implying that 
\begin{equation}
\partial_{i}\rho_{ij}(\bx)=0,
\label{eq:cons_rho}
\end{equation}
or
\begin{equation}
\varepsilon_{ijk}\partial_{k}\vrho_{ijl}(\bx)=0.
\end{equation}

Within plasticity theories, the gradient of the total displacement field $\vect{u}$ 
represents the compatible total distortion field~\cite{Kroner58,Kroner80} $\beta_{ij}=\partial_i u_j$, which is 
the sum of the elastic and the plastic distortion fields~\cite{Kroner58,Kroner80}, $\beta=\BetaP+\BetaE$.
Due to the presence of dislocation lines, both $\BetaP$ and $\BetaE$ are incompatible, characterized
by the GND density $\rho$
\begin{eqnarray}
\rho_{ij}&=&\epsilon_{ilm}\partial_l\BetaE_{mj},\label{eq:betaE_rho}\\
&=&-\epsilon_{ilm}\partial_l\BetaP_{mj}.\label{eq:betaP_rho}
\end{eqnarray} 

The elastic distortion field $\BetaE$ is the sum of its symmetric strain and antisymmetric
rotation fields, 
\begin{equation}
\BetaE=\epsilon^{\rm e}+\omega^{\rm e},
\label{eq:betaEDecom}
\end{equation}
where we assume linear elasticity, ignoring the `geometric nonlinearity' in these tensors. 
Substituting the sum of two tensor fields into the incompatibility relation Eq.~(\ref{eq:betaE_rho}) gives
\begin{equation}
\rho_{ij}=\varepsilon_{ikl}\partial_k\omega_{lj}^{\rm e}+\varepsilon_{ikl}\partial_k\epsilon_{lj}^{\rm e}.
\label{eq:rholambda}
\end{equation}

The elastic rotation tensor $\omega^{\rm e}$ can be rewritten as an axial vector, 
the crystalline orientation vector $\Lam$
\begin{equation}
\Lambda_k=\frac{1}{2}\varepsilon_{ijk}\omega^{\rm e}_{ij},
\label{eq:LambdaDef}
\end{equation}
or 
\begin{equation}
\omega^{\rm e}_{ij}=\varepsilon_{ijk}\Lambda_k.
\label{eq:strain_lambda}
\end{equation}
Thus we can substitute Eq.~(\ref{eq:strain_lambda}) into Eq.~(\ref{eq:rholambda})
\begin{equation}
\rho_{ij}=(\delta_{ij}\partial_k\Lambda_k-\partial_j\Lambda_i)+\varepsilon_{ikl}\partial_k\epsilon_{lj}^{\rm e}.
\label{eq:rholambda1}
\end{equation}
For a system without residual elastic stress, the GND density thus depends only on the varying crystalline orientation~\cite{JSPRB07}.

Dynamically, the time evolution law of the GND density emerges from the conservation of the Burgers vector~\cite{Kosevich79,Lazar11}
\begin{equation}
\frac{\partial}{\partial t}\rho_{ik} = -\varepsilon_{ijq}\partial_{j}J_{qk},
\label{eq:vrhoPDE}
\end{equation}
or 
\begin{equation}
\frac{\partial}{\partial t}\vrho_{ijk} = -\varepsilon_{ijm}\varepsilon_{mpq}\partial_{p}J_{qk}=-g_{ijpq}\partial_{p}J_{qk},
\label{eq:rhoPDE}
\end{equation}
where $J$ represents the Burgers vector flux, and the symbol $g_{ijpq}$ indicates 
$\varepsilon_{ijm}\varepsilon_{mpq}=\delta_{ip}\delta_{jq}-\delta_{iq}\delta_{jp}$.

\subsubsection{\label{sec:betaP}Non-conserved order parameter field}
The natural physicist's order parameter field $\vrho$,
characterizing the incompatibility, can be written in terms of the plastic distortion field $\BetaP$
\begin{equation}
\vrho_{ijk}=\varepsilon_{ijm}\rho_{mk}=-g_{ijls}\partial_l\BetaP_{sk}.
\label{eq:vrho_betaP}
\end{equation}
In the linear approximation, the alternative order parameter field $\BetaP$ fully specifies the local deformation $\vu$ of the material, 
the elastic distortion $\BetaE$, the internal long-range stress field $\sigma^{\rm int}$ and the 
crystalline orientation (the Rodrigues vector $\Lam$ giving the axis and angle of rotation), 
as summarized in \ref{sec:physParameters}.

It is natural, given Eq.~(\ref{eq:betaP_rho}) and Eq.~(\ref{eq:vrhoPDE}), to
use the flux $J$ of the Burgers vector density to define the dynamics of the
plastic distortion tensor $\BetaP$~\cite{Kosevich79,JSPRL06,Lazar11}:
\begin{equation}
\frac{\partial \BetaP_{ij}}{\partial t} = J_{ij}.
\label{eq:Jeq}
\end{equation}
As noted by Ref.~\onlinecite{AcharyaJMPS04}, Eq.~(\ref{eq:betaP_rho}) and
Eq.~(\ref{eq:vrhoPDE}) equate a curl of $\BetaP$ to a curl of $J$, so an arbitrary
divergence may be added to Eq.~(\ref{eq:Jeq}): the evolution of the plastic
distortion $\BetaP$ is not determined by the evolution of the GND 
density. Ref.~\onlinecite{AcharyaJMPS04} resolves this ambiguity using a Stokes-Helmholtz 
decomposition
of $\BetaP$. In our notation, $\BetaP = \BetaPI + \BetaPH$. The
`intrinsic' plastic distortion $\BetaPI$ is divergence-free 
($\partial_i\BetaPI_{ij}=0$, i.e., $k_i\KBetaPI_{ij}=0$), and determined
by the GND density $\rho$. The `history-dependent'
   \footnote{Changing the initial reference state through a curl-free
   plastic distortion (leaving behind no dislocations) will change
   $\BetaPH$ but not $\BetaPI$; the former depends on the history
   of the material and not just the current state, motivating our nomenclature.}
$\BetaPH$ is
curl-free ($\epsilon_{\ell ij} \partial_\ell\BetaPI_{ij}=0$, 
$\epsilon_{\ell ij} k_\ell\KBetaPI_{ij}=0$). In Fourier space, we can do this
decomposition explicitly, as
\begin{eqnarray}
\KBetaP_{ij}(\bk)&=&-i\varepsilon_{ilm}\frac{k_l}{k^2}\widetilde{\rho}_{mj}(\bk)+ik_i\widetilde{\psi}_j(\bk)\nonumber\\
&\equiv&\KBetaPI_{ij}(\bk)+\KBetaPH_{ij}(\bk).
\label{eq:betaPinrho}
\end{eqnarray}
This decomposition will become important to us in Sec.~\ref{sec:corrfuncBetaP},
where the correlation functions of $\BetaPI$ and $\BetaPH$ will
scale differently with distance.

Ref.~\onlinecite{AcharyaJMPS04} treats the evolution of the two components $\BetaPI$ and $\BetaPH$
separately. Because our simulations have periodic boundary conditions, 
the evolution of $\BetaPH$ does not affect the evolution of $\rho$. As
noted by Ref.~\onlinecite{AcharyaJMPS04}, in more general situations $\BetaPH$ will alter the 
shape of the body, and hence interact with the boundary conditions
   \footnote{For our simulations with external shear~\cite{YSCPRL10},
   the $\bk=0$ of $\BetaPH$ couples to the boundary condition. We determine
   the plastic evolution of the $\bk=0$ mode explicitly in that case.
   For correlation functions presented here, the $\bk=0$ mode is unimportant
   because we subtract $\BetaP$ fields at different sites before correlating.}.
Hence in the simulations presented here, we use Eq.~(\ref{eq:Jeq}), with
the warning that the plastic deformation fields shown in the figures
are arbitrary up to an overall divergence. The correlation functions we study
of the intrinsic plastic distortion $\BetaPI$ are independent of this ambiguity,
but the correlation functions of $\BetaPH$ we discuss in the 
Appendix \ref{sec:betaPH} will depend on this choice.
 
In the presence of external loading, we can express the appropriate free energy $\F$ as
the sum of two terms: the elastic interaction energy of GNDs, and the energy of interaction with the applied stress field.
The free energy functional is
\begin{equation}
\F = \int d^3\bx\biggl(\frac{1}{2}\sigma_{ij}^{\rm int}\epsilon_{ij}^{\rm e}-\sigma_{ij}^{\rm ext}\epsilon_{ij}^{\rm p}\biggr).
\label{eq:HFR}
\end{equation}
Alternatively, it can be rewritten in Fourier space
\begin{equation}
\F =-\int\frac{d^3\bk}{(2\pi)^3}\biggl(\frac{1}{2}M_{ijmn}(\bk)\KBetaP_{ij}(\bk)\KBetaP_{mn}(-\bk)+\Fourier{\sigma}_{ij}^{\rm ext}(\bk)\KBetaP_{ij}(-\bk)\biggr),
\label{eq:HFE}
\end{equation}
as discussed in \ref{sec:FourierF}.

\subsection{\label{sec:engineermodels}Traditional dissipative continuum dynamics}
There are well known approaches for deriving continuum
equations of motion for dissipative systems, which in this case produce a traditional von Mises-style theory~\cite{rickman97},
useful at longer scales. We begin by reproducing these standard equations.

For the sake of simplicity, we ignore external stress ($\sigma_{ij}$ simplified to $\sigma^{\rm int}_{ij}$) 
in the following three subsections. We start by using 
the standard methods applied to the non-conserved order parameter $\BetaP$, and then turn to the conserved order parameter
$\vrho$.

\subsubsection{Dissipative dynamics built from the non-conserved order parameter field $\BetaP$}
The plastic distortion $\BetaP$ is a non-conserved order parameter field, which is utilized by the engineering
community to study texture evolution and plasticity of mechanically deformed 
structural materials. The simplest dissipative dynamics in terms of $\BetaP$ 
minimizes the free energy by steepest descents
\begin{equation}
\frac{\partial}{\partial t}\BetaP_{ij} = -\Gamma\frac{\delta\F}{\delta\BetaP_{ij}},
\end{equation}
where $\Gamma$ is a positive material-dependent constant.
We may rewrite it in Fourier space, giving
\begin{equation}
\frac{\partial}{\partial t}\KBetaP_{ij}(\bk) = -\Gamma\frac{\delta\F}{\delta\KBetaP_{ij}(-\bk)}.
\end{equation}
The functional derivative $\delta\F/\delta\KBetaP_{ij}(-\bk)$ is the negative of the long-range stress 
\begin{equation}
\frac{\delta\F}{\delta\KBetaP_{ij}(-\bk)}= - M_{ijmn}(\bk)\KBetaP_{mn}(\bk)\equiv-\Fourier{\sigma}_{ij}(\bk).
\label{eq:EFBetaP}
\end{equation}
This dynamics implies a simplified version of von Mises plasticity
\begin{equation}
\frac{\partial}{\partial t}\KBetaP_{ij}(\bk) = \Gamma\Fourier{\sigma}_{ij}(\bk).
\label{eq:vonMises}
\end{equation}

\subsubsection{Dissipative dynamics built from the conserved order parameter field $\vrho$}
We can also derive an equation of motion starting from the GND density $\vrho$, as was
done by Ref.~\onlinecite{rickman97}.
For this dissipative dynamics Eq.~(\ref{eq:rhoPDE}), the simplest expression for $J$ is
\begin{equation}
J_{qk} = -\Gamma'_{ablq}\partial_{l}\frac{\delta\F}{\delta\vrho_{abk}},
\label{eq:jrho}
\end{equation}
where the material-dependent constant tensor $\Gamma'$ must be chosen to guarantee a decrease of 
the free energy with time. 

The infinitesimal change of $\F$ with respect to the GND density $\vrho$ is
\begin{equation}
\delta\F[\vrho] = \int\!d^3\bx\,\,\frac{\delta\F}{\delta\vrho_{ijk}}\delta\vrho_{ijk}.
\label{eq:EF0}
\end{equation}
The free energy dissipation rate is thus $\delta\F/\delta t$ for $\delta\vrho=\frac{\partial \vrho}{\partial t}\delta t$, hence
\begin{equation}
\frac{\partial}{\partial t}\F[\vrho] = \int\!d^3\bx\,\,\frac{\delta\F}{\delta\vrho_{ijk}}\frac{\partial\vrho_{ijk}}{\partial t}.
\label{eq:EF1}
\end{equation}
Substituting Eq.~(\ref{eq:rhoPDE}) into Eq.~(\ref{eq:EF1}) and integrating by parts gives
\begin{equation}
\frac{\partial}{\partial t}\F[\vrho] = \int\!d^3\bx\,\,\biggl(g_{ijpq}\partial_p\frac{\delta\F}{\delta\vrho_{ijk}}\biggr)J_{qk}.
\label{eq:EF2}
\end{equation}
Substituting Eq.~(\ref{eq:jrho}) into Eq.~(\ref{eq:EF2}) gives
\begin{equation}
\frac{\partial}{\partial t}\F[\vrho] = -\int\!d^3\bx\,\,\biggl(g_{ijpq}\partial_p\frac{\delta\F}{\delta\vrho_{ijk}}\biggr)
\biggl(\Gamma'_{ablq}\partial_{l}\frac{\delta\F}{\delta\vrho_{abk}}\biggr).
\label{eq:EF3}
\end{equation}
Now, to guarantee that energy never increases, we choose 
$\Gamma'_{ablq}=\Gamma g_{ablq}$, ($\Gamma$ is a positive material-dependent constant), which yields the rate
of change of energy as a negative of a perfect square
\begin{equation}
\frac{\partial}{\partial t}\F[\vrho] = -\int\!d^3\bx\,\,\Gamma\sum_{q,k}
\biggl(g_{ablq}\partial_{l}\frac{\delta\F}{\delta\vrho_{abk}}\biggr)^2.
\end{equation}
Using Eqs.~(\ref{eq:rhoPDE}) and (\ref{eq:jrho}), we can write the dynamics in terms of $\vrho$
\begin{equation}
\frac{\partial}{\partial t}\vrho_{ijk} = \Gamma g_{ijpq}g_{ablq}\partial_p\partial_l\frac{\delta\F}{\delta\vrho_{abk}}.
\label{eq:vrhodynamics}
\end{equation}
Substituting the functional derivative $\delta\F/\delta\vrho_{abk}$, Eq.~(\ref{eq:sigmarho}), derived in \ref{sec:calDFDrho}, into
Eq.~(\ref{eq:vrhodynamics}) and comparing to Eq.~(\ref{eq:rhoPDE}) tells us 
\begin{equation}
\frac{\partial}{\partial t}\vrho_{ijk}(\bx) = -\Gamma g_{ijpq}\partial_p\sigma_{qk}(\bx)=-g_{ijpq}\partial_pJ_{qk}(\bx),
\label{eq:vrhodynamics1}
\end{equation}
where
\begin{equation}
J_{qk} = \Gamma\sigma_{qk}
\end{equation}
duplicating the von Mises law (Eq.~\ref{eq:vonMises}) of the previous subsection.
The simplest dissipative dynamics of either non-conserved or conserved order parameter fields thus turns out to be
the traditional linear dynamics, a simplified von Mises law. 

The problem with this law for us is that it allows for plastic deformation in the absence of dislocations, i.e., the Burgers vector
flux can be induced through the elastic loading on the boundaries, even in a defect-free medium.
This is appropriate on engineering length scales above or around a micron, where SSDs 
dominate the plastic deformation. (Methods to incorporate their effects 
into a theory like ours have been provided by Acharya \etal~\cite{AcharyaJMPS06I,AcharyaJMPS06II} and 
Varadhan \etal~\cite{BeaudoinPS2005}) 

By ignoring the SSDs, our theory assumes
that there is an intermediate coarse-grain length scale, large compared
to the distance between dislocations and small compared to the distance
where the cancelling of dislocations with different Burgers vectors
dominates the dynamics, discussed in Sec.~\ref{sec:CoarseGraining}. We believe this latter length scale is given by
the distance between cell walls (as discussed in Sec.~\ref{sec:corrfuncVI}).
The cell wall misorientations are geometrically necessary. On the one 
hand, it is known~\cite{KuhlmannSMM91,HughesMMTA93} that neighboring cell walls often have misorientations
of alternating signs, so that on coarse-grain length scales just above
the cell wall separation one would expect explicit treatment of the
SSDs would be necessary. On the other hand, the
density of dislocations in cell walls is high, so that a coarse-grain
length much smaller than the interesting structures (and hence where we
believe SSDs are unimportant) should be possible~\cite{BenzergaAM11}.
(Our cell structures are fractal, with no characteristic `cell size'; this
coarse-grain length sets the minimum cutoff scale of the fractal, and the
grain size or inhomogeneity length will set the maximum scale.) With this
assumption, to treat the formation of cellular structures, we turn to
theories of the form given in
Eq.~(\ref{eq:vrhoPDE}), defined in terms of dislocation currents $J$ that depend directly on the local GND density.

\subsection{\label{sec:cdd}Our CDD model}
The microscopic motion of a dislocation under external strain depends upon temperature. In general,
it moves quickly along the glide direction, and slowly (or not at all) along the climb direction
where vacancy diffusion must carry away the atoms. The glide speed can be limited by phonon drag at
higher temperatures, or can accelerate to nearly the speed of sound at low temperatures~\cite{HirthBook}. 
It is traditional to assume that the dislocation velocity is over-damped, and proportional to the component of
the force per unit dislocation length in the glide plane.%
  \footnote{In real materials the dislocation dynamics is 
  intermittent, as dislocations bow out or depin from junctions and
  disorder, and engage in complex dislocation avalanches.}

To coarse-grain this microscopics, for reasons described 
in Sec.~\ref{sec:CoarseGraining} , we
choose a CDD model whose dislocation currents vanish when the GND
density vanishes, without considering SSDs. Ref.~\onlinecite{JSPRL06} 
derived a dislocation current $J$ for this case
using a closure approximation of the underlying microscopics. Their work
reproduced (in the case of both glide and climb) an earlier dynamical
model proposed by Acharya \etal~\cite{AcharyaJMPS01,AcharyaJMPS05,AcharyaJMPS06I}, who 
also incorporate the effects of SSDs.
We follow the general approach of Acharya and
collaborators~\cite{AcharyaJMPS01,AcharyaPRSA03,AcharyaJMPS04,AcharyaJMPS05,%
BeaudoinPS2005,AcharyaJMPS06I} in Sec.~\ref{sec:GCD} to derive an
evolution law for dislocations allowed both to glide and climb, and
then modify it to remove climb in Sec.~\ref{sec:GOD}.
We derive a second variant of glide-only dynamics in Sec.~\ref{sec:GODV}
by coupling climb to vacancies and then taking the limit of infinite vacancy
energy, which reproduces a model proposed earlier by Ref.~\onlinecite{AcharyaJMPS06I}.

In our CGD and GOD-LVP dynamics (Sections~\ref{sec:GCD} and~\ref{sec:GODV}
below), all dislocations in the infinitesimal volume at $\bx$ are moving
with a common velocity $\vv(\bx)$. We discuss the validity of this
single-velocity form for the equations of motion at length in
Sec.~\ref{sec:CoarseGraining}, together with a discussion of the 
coarse-graining and the emergence of SSDs. 
We view our simulations as physically sensible `model materials' -- 
perhaps not the correct theory for any particular material, but a sensible
framework to generate theories of plastic deformation and explain
generic features common to many materials.

\subsubsection{\label{sec:GCD}Climb-glide dynamics (CGD)}
We start with a model presuming (perhaps unphysically) that vacancy diffusion is so fast that dislocations climb and glide with
equal mobility. 
The elastic Peach-Koehler force due to the stress $\sigma(\bx)$ on the local GND density is given
by $f^{PK}_u=\sigma_{mk}\vrho_{umk}$. We assume that the velocity $\vv\propto\vf^{PK}$, giving a local constitutive relation
\begin{equation}
v_u\propto\sigma_{mk}\vrho_{umk}.
\label{eq:vGC}
\end{equation}

How should we determine the proportionality constant between velocity and force? In experimental systems, this is
complicated by dislocation entanglement and short-range forces between dislocations. Ignoring these features,
the velocity of each dislocation should 
depend only on the stress induced by the other dislocations, not the local density of dislocations~\cite{ZapperiZaiser}. 
We can incorporate this in an approximate way by making the 
proportionality factor in Eq.~(\ref{eq:vGC}) inversely proportional to
the GND density. We measure the latter by summing the square of all components of $\vrho$, hence $|\vrho|=\sqrt{\vrho_{ijk}\vrho_{ijk}/2}$ 
and $v_u=\frac{D}{|\vrho|}\sigma_{mk}\vrho_{umk}$, 
where $D$ is a positive material-dependent constant. 
This choice has the additional important feature that the evolution of a sharp domain wall whose width is limited by the lattice
cutoff is unchanged when the lattice cutoff is reduced.

The flux $J$ of the Burgers vector is thus~\cite{Kosevich79}
\begin{equation}
J_{ij}=v_u\vrho_{uij}=\frac{D}{|\vrho|}\sigma_{mk}\vrho_{umk}\vrho_{uij}.
\label{eq:GCdynamics}
\end{equation}
Notice that this dynamics satisfies our criterion that $J=0$ when there are no GNDs (i.e., $\vrho=0$). 
Notice also that we do not incorporate the
effects of SSDs 
(Acharya's $L^p$~\cite{AcharyaJMPS06I}); we discuss this further in
Sec.~\ref{sec:CoarseGraining}.

Substituting this flux $J$ (Eq.~\ref{eq:GCdynamics}) into the free energy dissipation rate (Eq.~\ref{eq:dfdt}) gives
\begin{equation}
\frac{\partial\mathcal{F}}{\partial t}=-\int\!d^3\bx\,\,\sigma_{ij}J_{ij}
=-\int\!d^3\bx\,\,\frac{|\vrho|}{D}v^2\leq0. 
\end{equation}
Details are given in \ref{sec:dfdt}.

\subsubsection{\label{sec:GOD}Glide-only dynamics: mobile dislocation population (GOD-MDP)}
When the temperature is low enough, dislocation climb is negligible, i.e., dislocations can only
move in their glide planes. Fundamentally, dislocation glide conserves the total
number of atoms, which leads to an unchanged local volume. Since the local volume change in time is represented
by the trace $J_{ii}$ of the flux of the Burgers vector, conservative motion of GNDs demands $J_{ii}=0$.
Ref.~\onlinecite{JSPRL06} derived the equation of motion for dislocation glide only, by
removing the trace of $J$ from Eq.~(\ref{eq:GCdynamics}). 
However, their dynamics fails to guarantee that the free energy monotonically decreases.
Here we present an alternative approach.

We can remove the trace of $J$ by modifying the first equality in Eq.~(\ref{eq:GCdynamics}),
\begin{equation}
J'_{ij} =v'_u\biggl(\vrho_{uij}-\frac{1}{3}\delta_{ij}\vrho_{ukk}\biggr),
\label{eq:GOdynamics}
\end{equation}
where $\vrho'_{uij}=\vrho_{uij}-\frac{1}{3}\delta_{ij}\vrho_{ukk}$ can be viewed as a subset of `mobile' dislocations 
moving with velocity $\vv'$.

Substituting the current (Eq.~\ref{eq:GOdynamics}) into the free energy dissipation rate (Eq.~\ref{eq:dfdt}) gives
\begin{equation}
\frac{\partial\mathcal{F}}{\partial t}=-\int\!d^3\bx\,\,\sigma_{ij}\bigl(v'_u\vrho'_{uij}
\bigr). 
\end{equation}
If we choose the velocity $v'_u\propto\sigma_{ij}\vrho'_{uij}$,
the appropriate free energy monotonically decreases in time. 
We thus express $v'_u=\frac{D}{|\vrho|}\vrho'_{uij}\sigma_{ij}$, 
where $D$ is a positive material-dependent constant, 
and the prefactor $1/|\vrho|$ is added for the same reasons, as discussed in the second paragraph of
Sec.~\ref{sec:GCD}. 

The current $J'$ of the Burgers vector is thus written~\cite{YSCPRL10}
\begin{equation}
J'_{ij} = v'_u\vrho'_{uij}=\frac{D}{|\vrho|}\sigma_{mn}\biggl(\vrho_{umn}-\frac{1}{3}\delta_{mn}\vrho_{ull}\biggr)
\biggl(\vrho_{uij}-\frac{1}{3}\delta_{ij}\vrho_{ukk}\biggr).
\label{eq:GOdynamics1}
\end{equation}
This natural evolution law becomes much less self-evident when expressed
in terms of the traditional two-index version $\rho$ (Eqs.~\ref{eq:rho_Nye0}\&\ref{eq:rho_Nye})
\begin{eqnarray}
J'_{ij}&=&\frac{D}{|\vrho|}\biggl(\sigma_{in}\rho_{mn}\rho_{mj}-\sigma_{mn}\rho_{in}\rho_{mj}-\frac{1}{3}\sigma_{mm}\rho_{ni}\rho_{nj}+\frac{1}{3}\sigma_{mm}\rho_{in}\rho_{nj}\nonumber\\
&&
-\frac{\delta_{ij}}{3}\Bigl(\sigma_{kn}\rho_{mn}\rho_{mk}-\sigma_{mn}\rho_{kn}\rho_{mk}-\frac{1}{3}\sigma_{mm}\rho_{nk}\rho_{nk}
+\frac{1}{3}\sigma_{mm}\rho_{kn}\rho_{nk}\Bigr)\biggr),\nonumber\\
\label{eq:GOdynamics2}
\end{eqnarray}
(which is why we introduce the three-index variant $\vrho$). 

This current $J'$ makes the free energy dissipation rate the negative of a perfect square in Eq.~(\ref{eq:dfdt2}). 
Details are given in \ref{sec:dfdt}. 

\subsubsection{\label{sec:GODV}Glide-only dynamics: local vacancy-induced pressure (GOD-LVP)}
At high temperature, the fast vacancy diffusion leads to
dislocation climb out of the glide direction. As the temperature decreases, vacancies are frozen out so that dislocations
only slip in the glide planes. In \ref{sec:vacancy}, we present a dynamical model coupling the vacancy diffusion to our CDD model. 
Here we consider the limit of frozen-out vacancies with infinite energy costs, 
which leads to another version of glide-only dynamics.

According to the coupling dynamics Eq.~(\ref{eq:vd_dynamics}), we write down the general form of dislocation current 
\begin{equation}
J''_{ij} = \frac{D}{|\vrho|}\biggl(\sigma_{mn}-\delta_{mn}p\biggr)\vrho_{umn}\vrho_{uij},
\label{eq:nGOdynamics0}
\end{equation}
where $p$ is the local pressure due to vacancies.

The limit of infinitely costly vacancies ($\alpha\to\infty$ in \ref{sec:vacancy}) leads to the traceless current, $J''_{ii}=0$. Solving this
equation gives a critical local pressure $p^c$
\begin{equation}
p^c = \frac{\sigma_{pq}\vrho_{spq}\vrho_{skk}}{\vrho_{uaa}\vrho_{ubb}}.
\end{equation}

\begin{figure}[!t]
\centering
\includegraphics[width=.55\columnwidth]{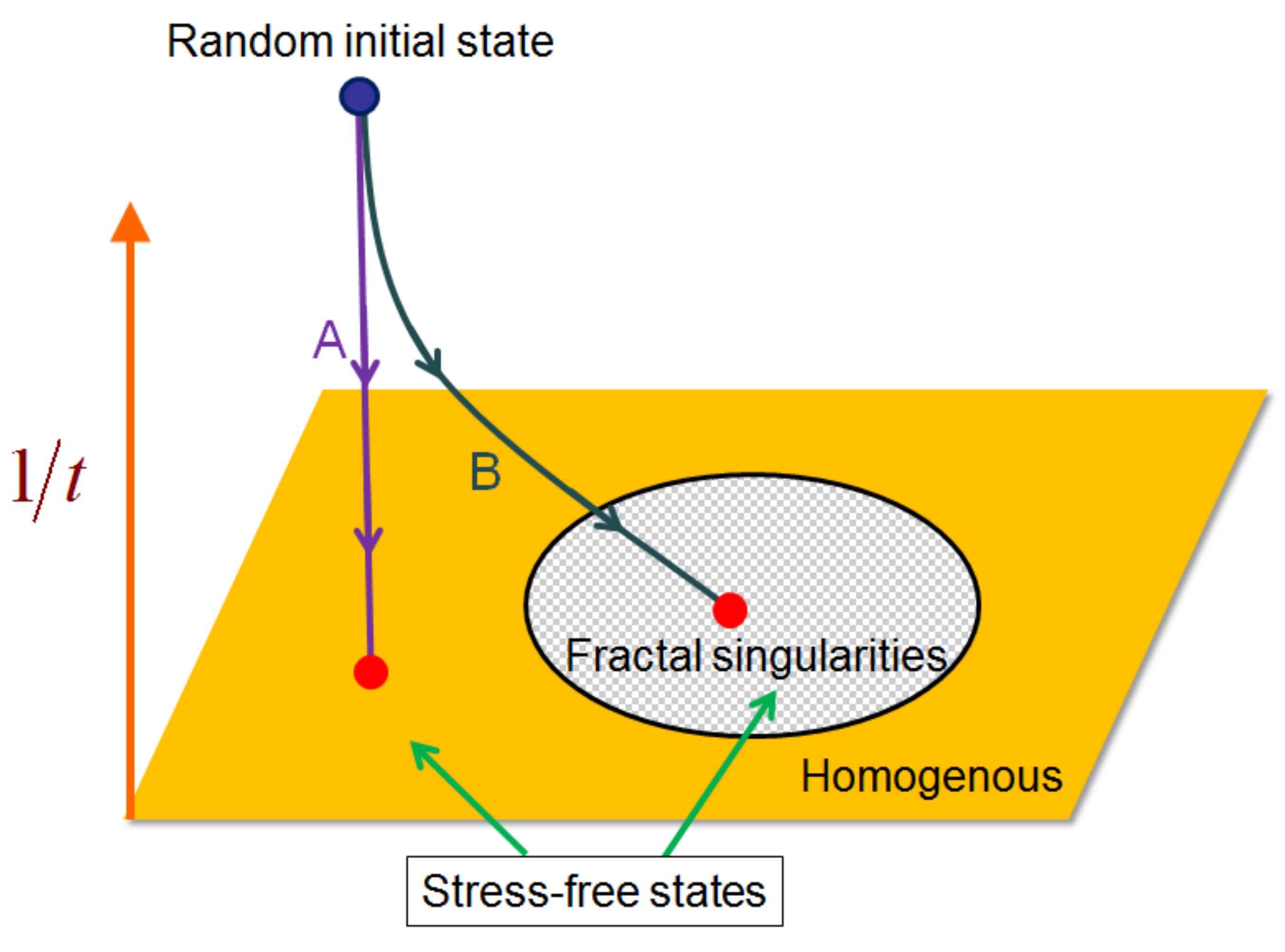}
\caption{(Color online)~{\bf Relaxation of various CDD models.} 
The blue dot represents the initial random plastically-deformed state; the red dots
indicate the equilibrated stress-free states driven by different dynamics. 
Curve A: steepest decent dynamics leads to the trivial homogeneous equilibrated 
state, discussed in Sec.~\ref{sec:engineermodels}. Curve B: our CDD models settle the system 
into non-trivial stress-free states with wall-like singularities of the GND density, discussed in
Sec.~\ref{sec:cdd}. }
\label{fig:Dynamics}
\end{figure}

The corresponding current $J''$ of the Burgers vector in this limit is thus written 
\begin{eqnarray}
J''_{ij}=\frac{D}{|\vrho|}\biggl(\sigma_{mn}-\frac{\sigma_{pq}\vrho_{spq}\vrho_{skk}}{\vrho_{uaa}\vrho_{ubb}}\delta_{mn}\biggr)\vrho_{umn}\vrho_{uij},
\label{eq:nGOdynamics1}
\end{eqnarray}
reproducing the glide-only dynamics proposed by Ref.~\onlinecite{AcharyaJMPS06I}.

Substituting the current (Eq.~\ref{eq:nGOdynamics1}) into the free energy dissipation rate (Eq.~\ref{eq:dfdt}) gives
\begin{equation}
\frac{\partial\mathcal{F}}{\partial t}=-\int\!d^3\bx\frac{D}{|\vrho|}\biggl[f^{PK}_if^{PK}_i-\biggl(\frac{d_if_i^{PK}}{|\vect{d}|}\biggr)^2\biggr]\leq0,
\end{equation}
where $f_i^{PK}=\sigma_{mn}\vrho_{imn}$ and $d_i = \vrho_{ikk}$. The equality emerges when the force ${\bf f}^{PK}$ is along the same direction as ${\bf d}$.\\

Unlike the traditional linear dissipative models, our CDD model, coarse grained from microscopic interactions, drives 
the random plastic distortion to non-trivial stress-free states with dislocation wall
singularities, as schematically illustrated in Fig.~\ref{fig:Dynamics}.

Our minimal CDD model, consisting of GNDs evolving under the long-range interaction, provides a framework for
understanding dislocation morphologies at the mesoscale.
Eventually, it can be extended to include vacancies by coupling them to the dislocation current (as discussed in \ref{sec:vacancy}, 
or extended to include disorder, dislocation pinning, and entanglement by adding appropriate interactions to the free
energy functional and refining the effective stress field (as discussed in \ref{sec:disorder}). It has already been extended to 
include SSDs incorporating traditional crystal plasticity theories~\cite{BeaudoinPS2005,AcharyaJMPS06I,AcharyaJMPS06II}. 

\section{\label{sec:CoarseGraining}Coarse Graining}

The discussion in Sec.~\ref{sec:theory} uses the language and conceptual
framework of the condensed matter physics of systems close to equilibrium
-- the generalized ``hydrodynamics'' used to derive equations of motion
for liquids and gases, liquid crystals, superfluids and superconductors,
magnetic materials, etc. In these subjects, one takes the 
broken symmetries and conserved quantities, and systematically writes
the most general evolution laws allowed by symmetry, {\em presuming that
these quantities determine the state of the material}. 
In that framework, the Burgers vector flux $J$ of Eqs.~(\ref{eq:vrhoPDE})
and~(\ref{eq:rhoPDE}) 
would normally be written as a general function of $\rho$ and its gradients,
constrained by symmetries and the necessity that the net energy decreases
with time. Indeed, this was the approach Limkumnerd originally 
took~\cite{LimkumnerdPhD}, but the complexity of the
resulting theory and the multiplicity of terms allowed by symmetry led them to
specialize~\cite{JSPRL06}
to a particular choice motivated by the Peach-Koehler force --- leading
to the equation of motion previously developed by Acharya \etal~\cite{AcharyaJMPS01,AcharyaJMPS05}. 

The assumption that the net continuum dislocation density
determines the evolution, however, is an uncontrolled
  \footnote{There are two uncontrolled approximations we make.
  Here we assume
  that the continuum, coarse-grained dislocation density $\rho = \rho^\Sigma$ determines
  the evolution: we ignore SSDs as unimportant
  on the sub-cellular length-scales of interest to us. Later, we
  shall further assume that the nine independent components of $\rho_{ij}$ all
  are dragged by the stress with the same velocity.}
and probably invalid assumption.
(Ref.~\onlinecite{LangerPRE98} have argued that the chaotic motion
of dislocations may lead to a statistical ensemble that could allow
a systematic theory of this type to be justified, but consensus has
not been reached on whether this will indeed be possible.) The situation
is less analogous to deriving the Navier-Stokes equation (where local
equilibrium at the viscous length is sensible) than to deriving theories
of eddy viscosity in fully developed turbulence (where unavoidable
uncontrolled approximations are needed to subsume swirls on smaller scales
into an effective viscosity of the coarse-grained system). Important features
of how dislocations are arranged in a local region will not
be determined by the net Burgers vector density, and extra state variables
embodying their effects are needed. In the context of dislocation dynamics,
these state variables are usually added as SSDs
and yield surfaces -- although far more complex
memory effects could in principle be envisioned.

Let us write $\rho^0$ as the microscopic dislocation density (the 
sum of line-$\delta$ functions along individual dislocations, as in
Eq.~(\ref{eq:rho_Nye0}) and following equations). For the microscopic
density, allowing both glide and climb, the dislocation current $J^0$ is directly given by the
velocity $\vv^0(\bx)$
of the individual dislocation passing through $\bx$
(see Eq.~\ref{eq:GCdynamics}):
\begin{equation}
J^0_{ij} = v^0_u \vrho^0_{uij}.
\label{eq:current0}
\end{equation}

Let $F^\sigma$ be the microscopic quantity $F^0$ coarse-grained density
over a length-scale $\Sigma$,
\begin{equation}
F^\Sigma_{ij}(\bx) = \int d^3\by F^0_{ij}(\bx+\by) w^\Sigma(\by),
\label{eq:CoarseGraining}
\end{equation}
where $w^\Sigma$ is a smoothing or blurring function. Typically, we use
a normal or Gaussian distribution $\rS$
\begin{equation}
\label{eq:rSN}
w^\Sigma(\by) = N^\Sigma(\by) = (2\pi\Sigma^2)^{-3/2}e^{-y^2/(2\Sigma^2)}.
\end{equation}
For our purposes, we can {\em define} the SSD 
density as the difference between the coarse-grained 
density and the microscopic density~\cite{Hochrainer2010}:
\begin{equation}
\rho^{SSD}(\bx)=\rho^0(\bx)-\rho^{\Sigma}(\bx)=\rho^0(\bx)-\int d^3\by \rho^0(\bx+\by) w^\Sigma(\by).
\end{equation} 
(Ref.~\onlinecite{AcharyaJE11} calls this quantity the dislocation fluctuation tensor
field.)

First, we address the question of SSDs, which we do not include in our simulations. In the
past~\cite{YSCPRL10},
we have argued that they do not contribute to the long-range stresses
that drive the formation of the cell walls, and that the successful
generation of cellular structures in our simplified model suggests that
they are not crucial. Here we go further, and suggest that their density
may be small on the relevant length-scales for cell-wall formation, 
and also that in a theory (like ours) with scale-invariant structures
it would not be consistent to add them separately.

What is the dependence of the SSD density on
the coarse-graining scale? Clearly $\rho^0$ contains all dislocations;
clearly for a bent single crystal of size $L$, $\rho^L$ contains only
those dislocations necessary to mediate the rotation across the crystal
(usually a tiny fraction of the total density of dislocations).
As $\Sigma$ increases past the distance
between dislocations, cancelling pairs of Burgers vectors through the
same grid face will leave the GNDs and join the SSDs. If the dislocation
densities were smoothly varying, as is often envisioned on long length
scales, the SSD density would be roughly independent of $\Sigma$ except
on microscopic scales. But, for a cellular structure with gross inhomogeneities
in dislocation density, the SSD density on the mesoscale may be much lower
than that on the macroscale. Very tangibly, if alternating cell walls
separated by $\ell$ have opposite misorientations (as is quite commonly
observed~\cite{KuhlmannSMM91,HughesMMTA93}), then the SSD 
density for $\Sigma>\ell$ will include most of the dislocations incorporated
into these cell walls, while for $\Sigma < \ell$ the cell walls will
be viewed as geometrically necessary.

How does the GND density within the cell walls compare with the
total dislocation density for a typical material? 
Is it possible that the GNDs dominate
over SSDs in the regime where these cell wall patterns form? Recent
simulations clearly suggest (see Ref.~\onlinecite{BenzergaAM11}~[Figure~5]) that the distinction
between GNDs and SSDs is not clear at the length scale of a micron, and with
reasonable definitions GNDs dominate by at least an order of magnitude over
the residual average SSD density. But what about the experiments? While
more experiments are necessary to clarify this issue, the existing evidence
supports that at mesoscales, SSDs at least are not necessarily dominant. In
particular, Ref.~\onlinecite{HughesAM97} observes that cell boundary
structures exhibit $D_{av}\theta_{av}/b=C$ where $D_{av}$ is the average wall spacing
and $\theta_{av}$ is the average misorientation angle with $C\sim650$ for `geometrically necessary' boundaries
(GNBs) and
$C\sim80$ for `incidental dislocation' boundaries (IDBs). The resulting
dislocation density should scale as
\begin{equation}
\rho_{GND}=\frac{1}{D_{av}h}=\frac{\theta_{av}}{D_{av}b}\sim\frac{C}{D_{av}^2}=
\frac{\theta^2_{av}}{b^2C},
\end{equation}
where $h$ is the average spacing between GNDs in the 
wall~\footnote{Only misorientation mediating dislocations are counted.}. 
There are some 
estimates available from
the literature. Reference~\cite{HughesAM97} tells us for pure aluminum that
$D_{av}$ is often observed to be $D_{av}=1-5\mu m$
which leads to roughly $\rho_{GND}^{GNB}\sim 10^{13}\times(2.6-65)/m^2$ and 
$\rho^{IDB}_{GND}$ one order of magnitude smaller.
Similar estimates in Ref.~\onlinecite{GodfreyAM00}
give $\rho_{GND}^{GNB} = 10^{14}-6\times10^{15}/m^2$ for aluminum at 
von Mises strains of $\epsilon= 0.2$ and $0.6$ respectively. 
The larger von Mises strains, the higher dislocation density.
Typically, in highly deformed aluminum ($\epsilon\sim2.7$), the total dislocation density
is roughly $10^{16}/m^2$ (see Ref.~\onlinecite{HughesPRL98}).
While SSDs within a cell boundary may exist, it is 
clearly far from true that SSDs dominate the dynamics in
these experiments.

These TEM analyses of cell boundary sizes and misorientations have a
misorientation cutoff $\theta_0\sim 2^\circ$~\cite{LiuJAC94}; they analyze the
cell boundaries using a single typical length scale $D_{av}$. Our
model behavior is formally much closer to the fractal
scaling analysis that Ref.~\onlinecite{ZaiserPRL98} used.
How does one identify a cutoff in a theory exhibiting scale invariance
(i.e., with no natural length scale)? Clearly our simulations are 
cut off at the numerical grid spacing, and the scale invariant theory
applies after a few grid spacings. Similarly, if the real materials
are described by a scale-invariant morphology (still an open question), 
the cutoff to the scale invariant regime will be where the granularity
of the dislocations becomes important -- the dislocation spacing, or
perhaps the annihilation length. This is precisely the length scale
at which the dislocations are individually resolved -- at which there
is no separate populations of SSDs and GNDs. Thus ignoring SSDs in our 
theory is at least self-consistent.

So, not only are the SSDs unimportant for the long-range stresses and
appear unnecessary for our (presumably successful) modeling of the formation
of cell walls, but they also may be rare on the sub-cellular
coarse-graining scale we use in our modeling, and it makes sense in
our mesoscale theory for us to omit their effects.


The likelihood that we do not need to incorporate explicit SSDs in our
equations of motion does not mean that our equations are correct. 
The microscopic equation of motion, Eq.~(\ref{eq:current0})
naively looks the same as our `single-velocity' equation of 
motion we use (\eg, Eq.~\ref{eq:GCdynamics}). But, as derived
in~Ref.~\onlinecite{AcharyaJMPS06I}, the coarse-graining procedure 
(Eq.~\ref{eq:CoarseGraining})
leads to a
correction term $L^p$ to the single-velocity equations:
\begin{eqnarray}
\label{eq:JCoarseGrained}
J^\Sigma_{ij} &=& (v^0_s \vrho^0_{sij})^\Sigma \nonumber\\
 &=&v^\Sigma_s \vrho^\Sigma_{sij} 
  +\bigl[(v^0_s \vrho^0_{sij})^\Sigma-v^\Sigma_s \vrho^\Sigma_{sij}\bigr]\nonumber\\
 &=&v^{\Sigma}_s \vrho^{\Sigma}_{sij}+L^p_{ij}.
\end{eqnarray}
Acharya interprets
  \footnote{More precisely, equation~(4) of~Ref.~\onlinecite{AcharyaJMPS06I} contains
  two different definitions for $L^p$; the one in Eq.~(\ref{eq:JCoarseGrained})
  and ${L^p_{ij}}' = \bigl[(v^0_s (\vrho^0_{sij}-\vrho^\Sigma_{sij})^\Sigma\bigr]
   = \bigl[(v^0_s \vrho^{SSD}_{sij})^\Sigma\bigr]$.
  ${L^p}'$ is of course a strain rate due to SSDs,
  but since $\vrho^\Sigma$ varies in space ${L^p}'$ is not equal to $L^p$.
  Ref.~\onlinecite{Acharya12} suggests using a two-variable version of the SSD
  density, $\check{\vrho}^{SSD}(\bx,\bx')=\vrho^0(\bx')-\vrho^{\Sigma}(\bx)$, making
  the two definitions equivalent.}
this correction term $L^p$ as the strain rate due to
SSDs~\cite{AcharyaJMPS06I,AcharyaJMPS06II}, 
and later Beaudoin~\cite{BeaudoinPS2005} 
and others~\cite{AcharyaJMPS10} then
use traditional crystal plasticity SSD evolution laws for it.
Their GNDs thus move according to the same single-velocity laws as ours do,
supplemented by SSDs that evolve by crystal plasticity (and thereby contribute
changes to the GND density). This is entirely appropriate
for scales large compared to the cellular structures, where most of the
dislocations are indeed SSDs. 

Although we argue that SSDs are largely absent at the
nanoscale where we are using our continuum theory, this does not
mean the single-velocity form of our equations of motion can be trusted.
Unlike fluid mixtures, where momentum conservation and
Galilean invariance lead to a shared mean velocity after a few collision
times, the microscopic dislocations are subject to different resolved
shear stresses and are mobile along different glide planes, so
neighboring dislocations may well move in a variety of directions~\cite{Hochrainer2010}.
If so, the microscopic
velocity $\vv^0$ will fluctuate in concert with the microscopic 
Burgers vector density $\vrho^0$ on microscopic scales, and the 
correction $L^p$ will be large. Hence Acharya's correction
term $L^p$ also incorporates multiple velocities for the GND density.
Our single-velocity approximation
(\eg, Eq.~\ref{eq:GCdynamics}) must be viewed as a physically allowed
equation of motion, but a second uncontrolled approximation -- the general
evolution law for the coarse-grained system will be more complex.

Let us be perfectly clear that our arguments, compelling on scales small
compared to the mesoscale cellular structures, should not be viewed as a
critique of the use of SSDs on larger scales. Much of our understanding
of yield stress and work hardening revolves around the macroscopic
dislocation density, which perforce are due to SSDs (since they dominate
on macroscopic scales). We also admire the work 
of Beaudoin, Acharya, and others
which supplements the GND equations we both study with crystal plasticity
rules for the SSDs motivated by Eq.~(\ref{eq:JCoarseGrained}). Surely
on macroscales the SSDs dominate the deformation, and using a single-velocity
law for the GNDs is better than ignoring them altogether, and
we have no particular reason to believe that the contribution of multiple
GND velocities in the evolution laws through $L^p$ will be significant
or dominant.

\section{\label{sec:results}Results}
\subsection{\label{simulation}Two and three dimensional simulations}
We perform simulations in 2D and 3D for the dislocation dynamics of Eq.~(\ref{eq:vrhoPDE}) and Eq.~(\ref{eq:Jeq}),
with dynamical currents defined by CGD~(Eq.~\ref{eq:GCdynamics}), GOD-MDP~(Eq.~\ref{eq:GOdynamics1}),
and GOD-LVP~(Eq.~\ref{eq:nGOdynamics1}). We numerically observe that simulations of Eqs.~(\ref{eq:vrhoPDE}),~(\ref{eq:Jeq})
lead to the same results statistically (i.e., the numerical time step approximations leave the physics invariant). We therefore focus our presentation
on the results of Eq.~(\ref{eq:Jeq}), where the evolving field variable $\BetaP$ is unconstrained.
Our CGD and GOD-MDP models have been quite extensively simulated in one and two dimensions 
and relevant results can be found in Refs.~\onlinecite{YSCPRL10},~\onlinecite{JSPRL06},~and~\onlinecite{JSJMPS07}.
In this paper, we concentrate on periodic grids of spatial extent $L$ in both two~\cite{YSCPRL10} and three dimensions.
The numerical approach we use is a second-order central upwind scheme designed for Hamilton-Jacobi 
equations~\cite{Kurganov01} using finite differences. This method is quite efficient in 
capturing $\delta-$shock singular structures~\cite{ChoiCiSE11}, even though it is 
flexible enough to allow for the use of approximate solvers near the singularities.

\begin{figure*}[htbp]
\centering
\includegraphics[width=0.73\columnwidth]{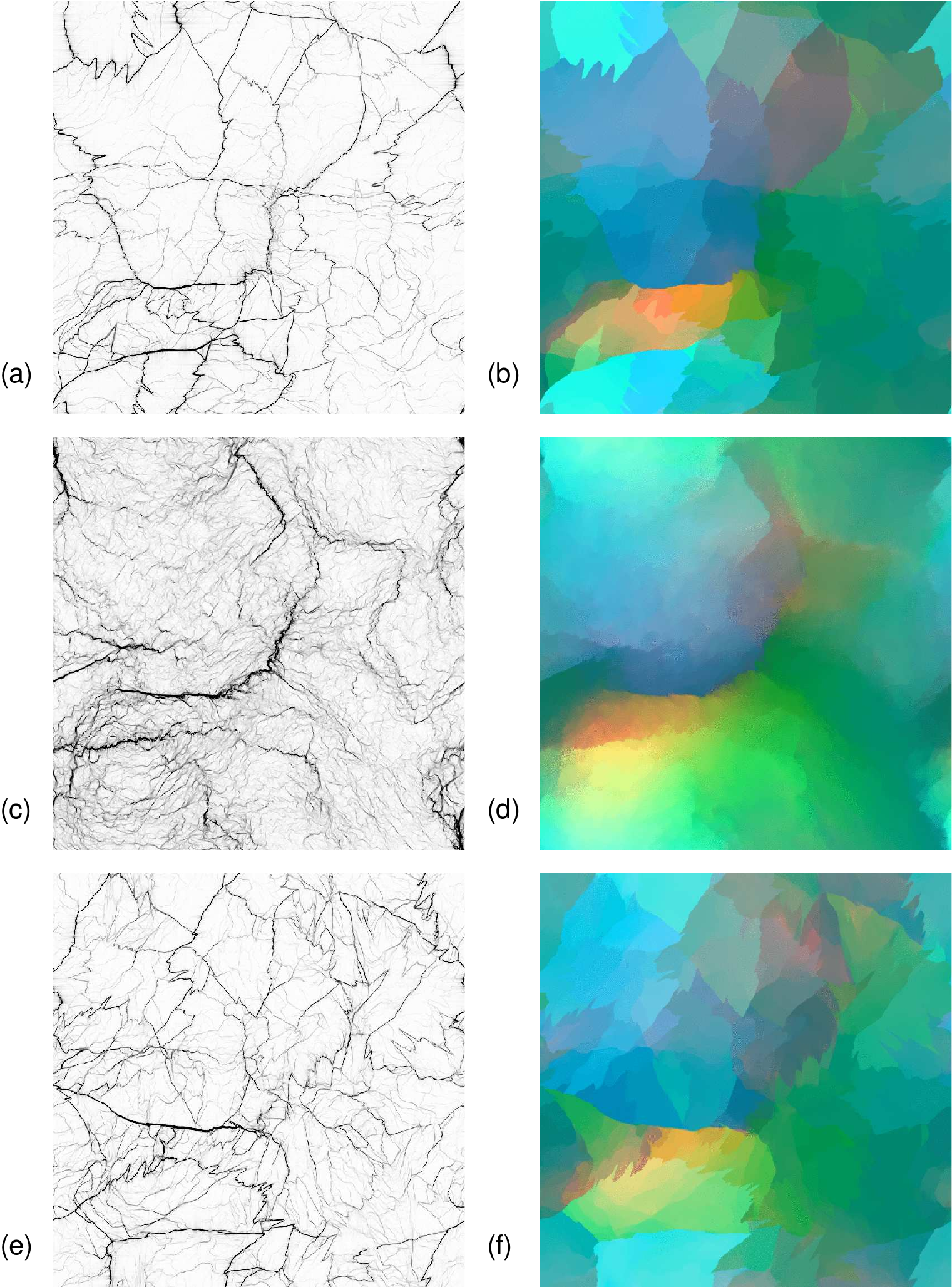}
\caption{(Color online)~{\bf Complex dislocation structures in two dimensions ($1024^2$)} for 
the relaxed states of an initially random distortion. {\it Top:} Dislocation climb is allowed; 
{\it Middle:} Glide only using a mobile dislocation population; {\it Bottom:} Glide only using
a local vacancy pressure. {\it Left:} Net GND density $|\vrho|$ 
plotted linearly in density with dark regions a factor $\sim10^4$ more dense than the lightest visible regions. 
(a) When climb is allowed, the resulting morphologies are sharp, regular, and close to the system scale. 
(c) When climb is forbidden using a mobile dislocation population, 
there is a hierarchy of walls on a variety of length scales, getting
weaker on finer length scales. (e) When climb is removed using a local vacancy pressure,
the resulting morphologies are as sharp as those (a) allowing climb. {\it Right:} Corresponding local crystalline orientation 
maps, with the three
components of the orientation vector $\vect{\Lambda}$ linearly mapped onto a vector of RGB values. Notice the fuzzier cell walls (c)
and (d) suggests a larger fractal dimension.}
\label{fig:relax2Dsim}
\end{figure*}

\begin{figure*}[htbp]
\centering
\includegraphics[width=0.8\columnwidth]{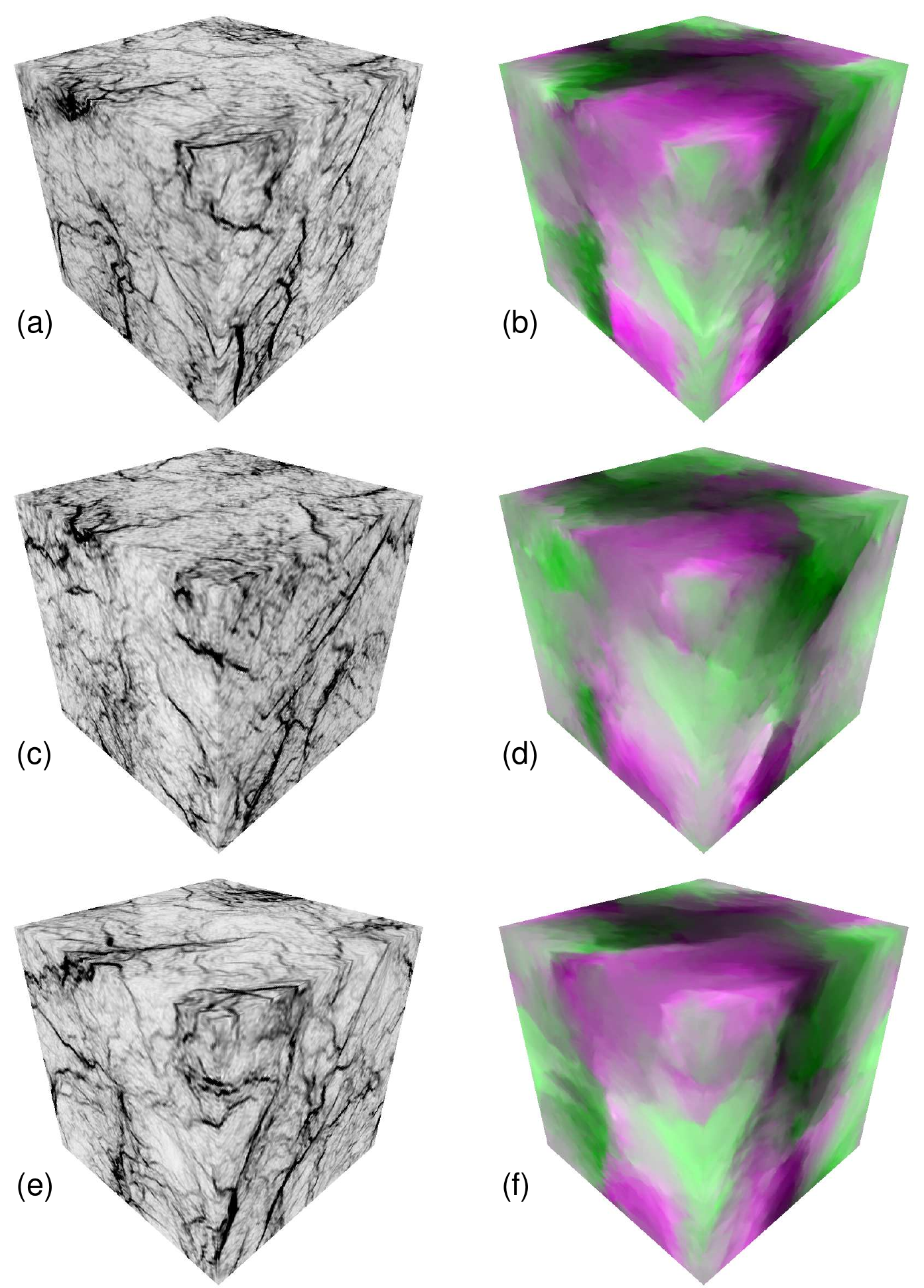}
\caption{(Color online)~{\bf Complex dislocation structures in three dimensions ($128^3$)} for 
the relaxed states of an initially random distortion. Notice these textured views on
the surface of simulation cubes.
{\it Top:} Dislocation climb is allowed; 
{\it Middle:} Glide only using a mobile dislocation population; {\it Bottom:} Glide only using 
a local vacancy pressure. {\it Left:} Net GND density $|\vrho|$ plotted linearly in density with dark regions a 
factor $\sim10^3$ more dense
than the lightest visible regions. 
The cellular structures in (a), (c), and (e) seem similarly fuzzy; our theory in three dimensions
generates fractal cell walls.
{\it Right:} Corresponding local crystalline maps, with the three
components of the orientation vector $\vect{\Lambda}$ linearly mapped onto a vector of RGB values.}
\label{fig:relax3Dsim}
\end{figure*}

\begin{figure*}[htbp]
\centering
\includegraphics[width=1.\columnwidth]{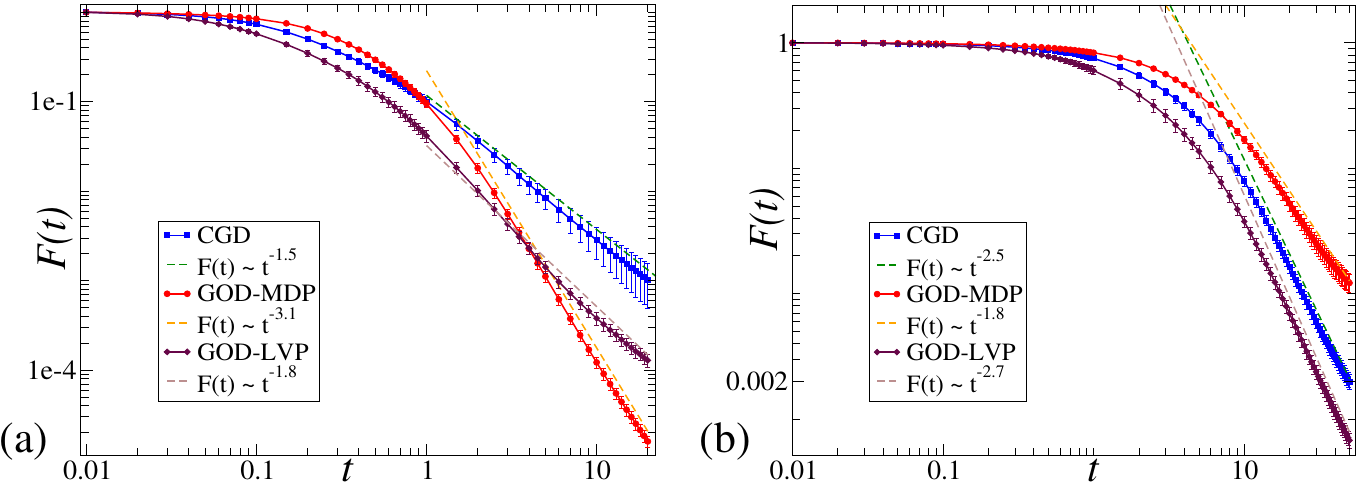}
\caption{(Color online)~{\bf The elastic free energy decreases to zero as a power law in time in both two and three dimensions.}
In both (a) and (b), we show that the free energy $\mathcal{F}$ decays monotonically in time, and goes to zero as a power law 
for CGD, GOD-MDP, and GOD-LVP simulations, as the system relaxes in the absence of external strain.}
\label{fig:Energy3D}
\end{figure*}

Our numerical simulations show a close analogy to those of turbulent flows~\cite{ChoiCiSE11}. 
As in three-dimensional turbulence, defect structures lead to
intermittent transfer of morphology to short length scales. As
conjectured~\cite{Siggia92PFA,Siggia92PRL} for the Euler
equations or the inviscid limit of Navier-Stokes equations, 
our simulations develop singularities in finite time~\cite{JSPRL06,YSCPRL10}. Here
these singularities are $\delta$-shocks representing grain-boundary-like structures emerging from
the mutual interactions among mobile dislocations~\cite{tobepublished}.
 In analogy with turbulence, where the viscosity serves to
smooth out the
vortex-stretching singularities of the Euler equations, we have 
explored the effects of adding an artificial viscosity term to our 
equations of motion~\cite{ChoiCiSE11}.
In the presence of artificial viscosity, our simulations exhibit nice numerical convergence
in all dimensions~\cite{tobepublished}. However, in the limit of vanishing viscosity, the solutions of our dynamics
continue to depend on the lattice cutoff in higher 
dimensions, (our simulations only exhibit numerical
convergence in one dimension). Actually, the fact that the physical system
is cut off by the atomic scale leads to the conjecture that our equations
are in some sense non-renormalizable in the ultraviolet. These issues are discussed in detail in Refs.~\onlinecite{ChoiCiSE11} 
and~\onlinecite{tobepublished}. 
See also Ref.~\onlinecite{AcharyaBIMU11} for global existence
and uniqueness results from an alternative regularization for this type of
equations; it is not known whether these alternative regularizations will
continue to exhibit the fractal scaling we observe.

In the vanishing viscosity limit, our simulations exhibit fractal structure down to the smallest scales. When varying the system size
continuously, the solutions of our dynamics exhibit a convergent set
of correlation functions of the various order parameter fields, which are used to characterize the emergent self-similarity.
This statistical convergence is numerically tested in \ref{sec:sc}.

In both two and three dimensional simulations, we relax the deformed system with 
and without dislocation climb in the absence of external loading. 
Here, the initial plastic distortion
field $\BetaP$ is still a Gaussian random field with correlation length scale $\sqrt{2}L/5\sim0.28L$ and
initial amplitude $\beta_0 = 1$. (In our earlier work~\cite{YSCPRL10}, we described this length as $L/5$, using
a non-standard definition of correlation length scale; see \ref{sec:initialcond}.)
These random initial conditions are explained in \ref{sec:initialcond}.
In 2D, Figure~\ref{fig:relax2Dsim} 
shows that CGD and GOD-LVP simulations (top and bottom) exhibit much sharper, flatter boundaries than
GOD-MDP (middle). This difference is quantitatively described by the large shift
in the static critical exponent $\eta$ in 2D for both CGD and GOD-LVP. 
In our earlier
work~\cite{YSCPRL10}, we announced this difference as providing a sharp distinction between high-temperature, 
non-fractal grain boundaries (for CGD), and low-temperature, fractal cell wall structures (for GOD-MDP).
This appealing message did not survive the 
transition to 3D; Figure~\ref{fig:relax3Dsim} shows
basically indistinguishable complex cellular structures, 
for all three types of dynamics. Indeed, Table~\ref{tab:staticscaling}
shows only a small change in critical exponents, among CGD, GOD-MDP, and GOD-LVP. 
During both two and three dimensional relaxations, their appropriate free energies
monotonically decay to zero as shown in Fig.~\ref{fig:Energy3D}.

\begin{figure*}[htbp]
\centering
\includegraphics[width=1.0\columnwidth]{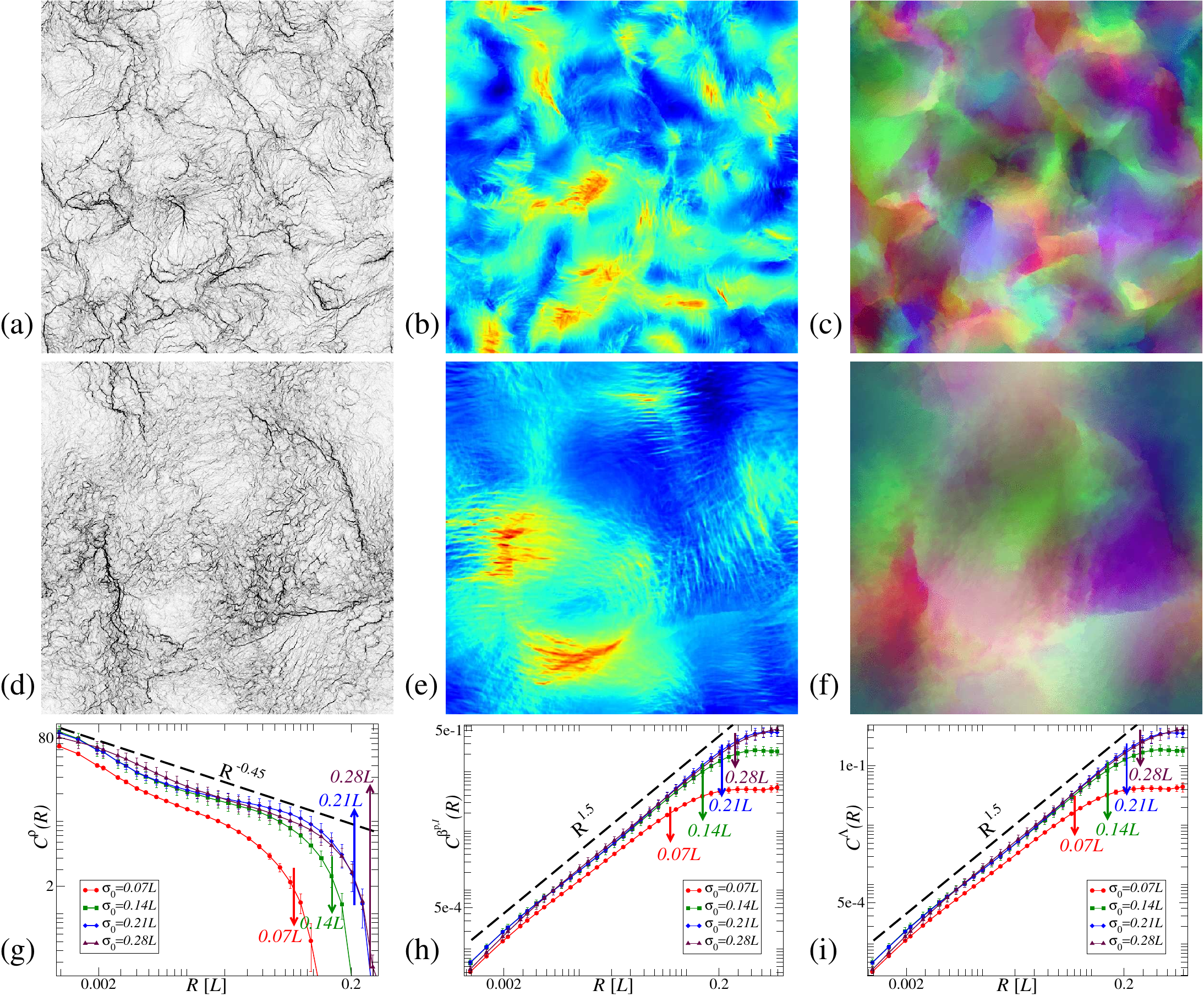}
\caption{(Color online)~{\bf Relaxation with various initial length scales in two dimensions.} 
GNDs are not allowed to climb due to the constraint of a mobile dislocation population in these simulations.
(a), (b), and (c) are the net GND density map $|\vrho|$, the net plastic distortion $|\BetaP|$ (the warmer color indicating the larger distortion), 
and the crystalline orientation map in a fully-relaxed
state evolved from an initial random plastic distortion with correlated length scale $0.07L$. They are compared to the same sequence of plots,
(d), (e), and (f), which are in the relaxed state with the initial length scale $0.21L$ three times as long. Notice the features with the longest wave length reflecting
the initial distortion length scales. 
(g), (h), and (i) are the scalar forms (discussed in Sec.~\ref{sec:corrfunc}) of correlation functions of the GND density $\rho$, the intrinsic
plastic distortion $\BetaPI$, and the crystalline orientation $\Lam$ for well-relaxed states 
with initial length scales varying from $0.07L$ to $0.28L$. They exhibit
power laws independent of the initial length scales, with cutoffs set by the initial lengths. 
(The scaling relation among their critical exponents will be discussed in Sec.~\ref{sec:staticscaling}.)
}
\label{fig:LSCorrFunc}
\end{figure*}

\subsection{\label{sec:corrfuncVI}Self-similarity and initial conditions}
Self-similar structures, as emergent collective phenomena, have been studied in mesoscale crystals~\cite{YSCPRL10}, 
human-scale social network~\cite{SongNature05}, and the astronomical-scale universe~\cite{VergassolaAA94}.
In some models~\cite{VergassolaAA94}, the self-similarity comes from 
{\it scale-free} initial conditions with a power-law spectrum~\cite{PeeblesBook93,ColesBook95}. In our CDD model,
our simulations start from a random plastic distortion with a Gaussian distribution characterized by a single length scale.
The scale-free dislocation structure spontaneously emerges as a result of the deterministic dynamics. 

Our Gaussian random initial condition is analogous to hitting a bulk material randomly with a hammer. 
The hammer head (the dent size scale) corresponds
to the correlated length. 
We need to generate inhomogeneous deformations like random dents, because our theory is deterministic and hence uniform initial
conditions under uniform loading will not develop patterns. 

We have considered alternatives to our imposition of Gaussian random
deformation fields as initial conditions. (a)~As an alternative to
random initial deformations, we could have imposed a more regular
(albeit nonuniform) deformation -- starting with our material bent into
a sinusoidal arc, and then letting it relax. Such simulations produce
more symmetric versions of the fractal patterns we see; indeed, our
Gaussian random initial deformations have correlation lengths `hammer
size' comparable to the system size, so our starting deformations are
almost sinusoidal (although different components have different
phases) -- see \ref{sec:initialcond}. (b)~To explore the
effects of multiple uncorrelated random domains (multiple small dents),
we reduce the Gaussian correlation length as shown in
Fig.~\ref{fig:LSCorrFunc}. We find that the initial-scale deformation
determines the maximal cutoff for the fractal correlations in our model.
In other systems (such as two-dimensional turbulence) one can observe an
`inverse cascade' with fractal structures propagating to long length
scales; we observe no evidence of these here.
(c)~As an alternative to imposing an initial plastic deformation field
and then relaxing, we have explored deforming the material slowly  and
continuously in time. Our preliminary `slow hammering' explorations
turn the Gaussian initial conditions ${\BetaPo}$ into a source term,
modifying Eq.~(\ref{eq:Jeq}) with an additional term to give
$\partial_t \BetaP_{ij} = J_{ij} + \BetaPo_{ij}/\tau$. Our early 
explorations suggest that slow hammering simulations will be 
qualitatively compatible with the relaxation of an initial rapid hammering.
In this paper, to avoid the introduction of the hammering time scale $\tau$,
we focus on the (admittedly less physically motivated) relaxation behavior.

In real materials, initial grain boundaries, impurities, or sample sizes, can be viewed as analogies to our initial dents --- explaining the observation of
dislocation cellular structures both in single crystals and polycrystalline materials.

Figure~\ref{fig:LSCorrFunc} shows relaxation without dislocation climb (due to the constraint of a mobile dislocation population) 
at various initial length scales in 2D. 
From Fig.~\ref{fig:LSCorrFunc}(a) to~(f), the net GND density, the net plastic distortion, and the crystalline orientation map,
measured at two well-relaxed states evolved from different random distortions, all show fuzzy fractal structures, distinguished
only by their longest-length-scale features that originate from the initial conditions. In Fig.~\ref{fig:LSCorrFunc}(g),~(h), and~(i), the
correlation functions of the GND density $\rho$, the intrinsic plastic distortion $\BetaPI$, and the crystalline orientation $\Lam$
are applied to characterize the emergent self-similarity, as discussed in the following section~\ref{sec:corrfunc}. They all
exhibit the same power law, albeit with different cutoffs due to the initial conditions.

\subsection{\label{sec:corrfunc}Correlation functions}
Hierarchical dislocation structures have been observed both experimentally~\cite{YawasakiSM80,MughrabiPMA86,UngarAM86,SchwinkSM92}
and in our simulations~\cite{YSCPRL10}. Early work analyzed experimental cellular structures using the fractal box counting method \cite{ZaiserPRL98}
or by separating the systems into cells and analyzing their sizes and misorientations~\cite{HughesAM97,HughesPRL98,DawsonAM99,HughesPRL01}.
In our previous publication, we analyzed our simulated dislocation patterns using these two methods, and showed 
broad agreement with these experimental analyses~\cite{YSCPRL10}. In fact, lack of the measurements of physical order parameters
leads to incomplete characterization of the emergent self-similarity~\footnote{In these analyses of TEM micrographs, the authors must use an artificial cut-off to facilitate the analysis. This arbitrary scale obscures the scale-free nature behind the emergent dislocation patterns.}. We will not pursue these methods here.

In our view, the emergent self-similarity should best be exhibited by the correlation functions 
of the order parameter fields, such as the GND density $\rho$, the plastic distortion $\BetaP$, and the crystalline
orientation vector $\Lam$. Here we focus on scalar invariants of the various tensor correlation functions.

For the vector correlation function $\MC^{\Lam}_{ij}(\bx)$ (Eq.~\ref{eq:cfrod}), only the sum $\MC^{\Lam}_{ii}(\bx)$ is a 
scalar invariant under three dimensional rotations. For the tensor fields $\rho$ and $\BetaP$, 
their two-point correlation functions are measured in terms of a complete set of three independent scalar invariants, 
which are indicated by `tot' (total), `per' (permutation), and `tr' (trace). 
In searching for the explanation of the lack of scaling~\cite{YSCPRL10} for $\BetaP$ (see Sec.~\ref{sec:corrfuncBetaP} and \ref{sec:betaPH}),
we checked whether these independent invariants might scale independently.
In fact, most of them share a single underlying critical exponent, except for the trace-type scalar invariant of
the correlation function of $\BetaPI$, which go to a constant in well-relaxed states, as discussed in Sec.~\ref{sec:betaPlam}. 
 
\begin{figure*}[htbp]
\centering
\includegraphics[width=1.\columnwidth]{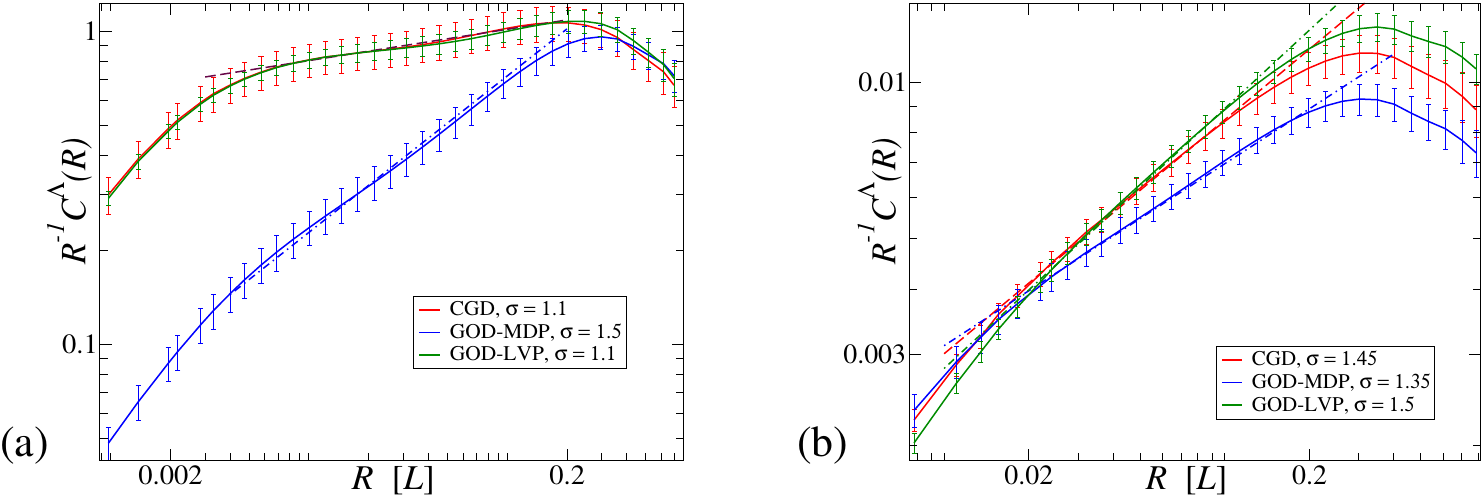}
\caption{(Color online)~{\bf Correlation functions of $\Lam$ in both two and three dimensions.}
In (a) and (b), red, blue, and green lines indicate CGD, GOD-MDP, and GOD-LVP simulations, respectively. 
{\it Left:} Correlation functions of $\Lam$ are measured in relaxed, unstrained $1024^2$ systems; {\it Right:} These correlation functions are
measured in relaxed, unstrained $128^3$ systems.
All dashed lines show estimated power laws quoted in Table~\ref{tab:staticscaling}.}
\label{fig:CorrFunc_Lam}
\end{figure*}

\begin{figure*}[htbp]
\centering
\includegraphics[width=1.\columnwidth]{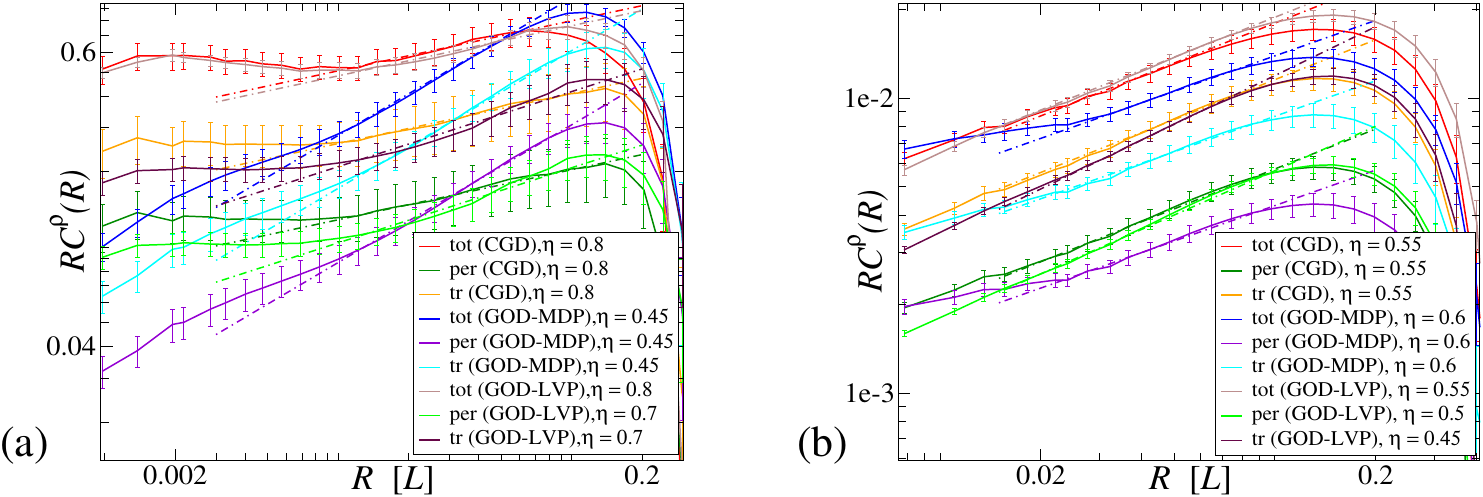}
\caption{(Color online)~{\bf Correlation functions of $\vrho$ in both two and three dimensions.}
{\it Left:} (a) is measured in relaxed, unstrained $1024^2$ systems; {\it Right:} (b) is 
measured in relaxed, unstrained $128^3$ systems.
All dashed lines show estimated power laws quoted in Table~\ref{tab:staticscaling}.
Notice all three scalar forms of the correlation functions of GND density share
the same power law.}
\label{fig:CorrFunc_Rho}
\end{figure*}

\subsubsection{Correlation function of crystalline orientation field} 
As dislocations self-organize themselves into complex structures, the relative differences of the crystalline orientations are
correlated over a long length scale. 

For a vector field, like the crystalline orientation $\Lam$, the natural two-point correlation function is
\begin{eqnarray}
\MC^{\Lam}_{ij}(\bx) &=& \langle(\Lambda_{i}(\bx)-\Lambda_{i}(0))(\Lambda_{j}(\bx)-\Lambda_{j}(0))\rangle\nonumber\\
&=&2\langle\Lambda_i\Lambda_j\rangle - 2\langle\Lambda_{i}(\bx)\Lambda_j(0)\rangle. 
\label{eq:cfrod}
\end{eqnarray}
Note that we correlate changes in $\Lam$ between two points. Just as for the height-height correlation function in surface
growth~\cite{Chaikin1995}, adding a constant to $\Lam(\bx)$ (rotating the sample) leads to an equivalent configuration, so only
differences in rotations can be meaningfully correlated.

It can be also described in Fourier space
\begin{equation}
\widetilde{\MC}^{\Lam}_{ij}(\bk)= 
2\langle\Lambda_i\Lambda_j\rangle(2\pi)^3\delta(\bk)-\frac{2}{V}\widetilde{\Lambda}_{i}(\bk)\widetilde{\Lambda}_{j}(-\bk).
\label{eq:kcfrod}
\end{equation}
In an isotropic medium, we study the scalar invariant formed from $\MC^{\Lambda}_{ij}$
\begin{equation}
\MC^{\Lam}(\bx) = \MC^{\Lam}_{ii}(\bx)=2\langle\Lambda^2\rangle-2\langle\Lambda_i(\bx)\Lambda_i(0)\rangle.
\label{eq:scfrod}
\end{equation}

Figure~\ref{fig:CorrFunc_Lam} shows the correlation functions of crystalline orientations
in both $1024^2$ and $128^3$ simulations. The large shift in critical exponents seen in 2D 
(Fig.~\ref{fig:CorrFunc_Lam}(a)) for both CGD and GOD-LVP is not observed in the fully three
dimensional simulations (Fig.~\ref{fig:CorrFunc_Lam}(b)). 
\begin{figure}[!b]
\centering
\psfrag{betaP}{$R^{-1}\MC^{\BetaP}_{tot}(R)$}
\psfrag{R}{$R\;\;{\rm[}L{\rm]}$}
\includegraphics[width=0.55\columnwidth]{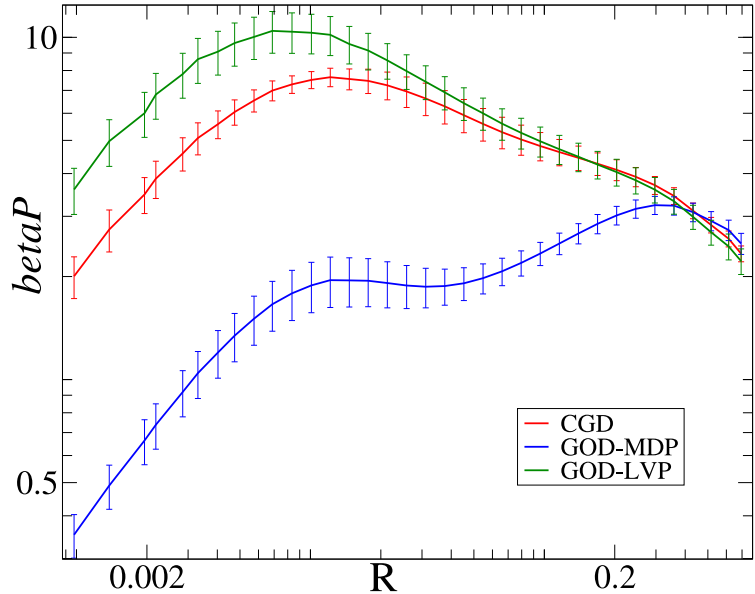}
\caption{(Color online)~{\bf Correlation functions of $\BetaP$ in two dimensions.}
Red, blue, and green lines indicate CGD, GOD-MDP, and GOD-LVP simulations, respectively. 
None of these curves shows a convincing power law.}
\label{fig:CorrFuncBetaP1024}
\end{figure}

\begin{figure*}[htbp]
\centering
\includegraphics[width=1.\columnwidth]{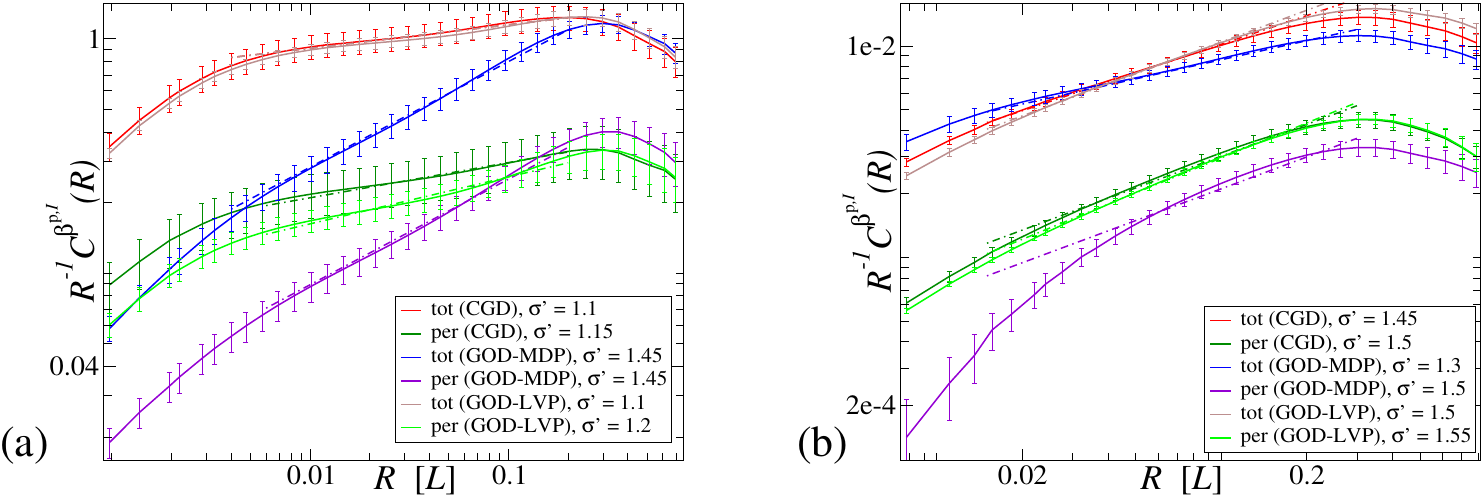}
\caption{(Color online)~{\bf Correlation functions of $\BetaPI$ in both two and three dimensions.}
In (a) and (b), the correlation functions of the intrinsic part of plastic distortion field are shown.
{\it Left:} (a) is measured in relaxed, unstrained $1024^2$ systems; {\it Right:} (b) is measured in
in relaxed, unstrained $128^3$ systems.
All dashed lines show estimated power laws quoted in Table~\ref{tab:staticscaling}.
Notice that we omit the correlation functions of $\MC^{\BetaPI}_{tr}$, which are independent of distance, and
unrelated to the emergent self-similarity, as shown 
in Sec.~\ref{sec:betaPlam}.}
\label{fig:CorrFunc_BetaPI}
\end{figure*}

\subsubsection{Correlation function of GND density field} 
As GNDs evolve into $\delta$-shock singularities, the critical fluctuations of the GND density can be measured
by the two-point correlation function $\MC^{\rho}(\bx)$ of the GND density, which decays as the separating distance between two sites increases.
The complete set of rotational invariants of the correlation function of $\rho$ includes three scalar forms
\begin{eqnarray}
\MC^{\rho}_{tot}(\bx) &=& \langle\rho_{ij}(\bx)\rho_{ij}(0)\rangle,\label{eq:cfrhoA}\\
\MC^{\rho}_{per}(\bx) &=& \langle\rho_{ij}(\bx)\rho_{ji}(0)\rangle,\label{eq:cfrhoB}\\
\MC^{\rho}_{tr}(\bx) &=& \langle\rho_{ii}(\bx)\rho_{jj}(0)\rangle.\label{eq:cfrhoC}
\end{eqnarray}

Figure~\ref{fig:CorrFunc_Rho} shows all the correlation functions of GND density
in both $1024^2$ and $128^3$ simulations. These three scalar forms of the correlation functions
of $\rho$ exhibit the same critical exponent $\eta$, as listed in Table~\ref{tab:staticscaling}.
Similar to the measurements of $\MC^{\Lam}$, the large shift in critical exponents seen in 2D 
(Fig.~\ref{fig:CorrFunc_Rho}(a)) for both CGD and GOD-LVP is not observed in the fully three
dimensional simulations (Fig.~\ref{fig:CorrFunc_Rho}(b)). 

\subsubsection{\label{sec:corrfuncBetaP}Correlation function of plastic distortion field}
The plastic distortion $\BetaP$ is a mixture of both the divergence-free $\BetaPI$ and the curl-free
$\BetaPH$. Figure~\ref{fig:CorrFuncBetaP1024} shows that $\BetaP$ does not appear to be scale invariant, as
observed in our earlier work~\cite{YSCPRL10}. It is 
crucial to study the correlations of the two physical fields, $\BetaPI$
and $\BetaPH$, separately.

Similarly to the crystalline orientation $\Lam$, we correlate the differences between $\BetaPI$ at neighboring
points. The complete set of scalar invariants of correlation functions of $\BetaPI$ thus includes the three scalar forms
\begin{eqnarray}
\MC^{\BetaPI}_{tot}(\bx)&=&\langle(\BetaPI_{ij}(\bx)-\BetaPI_{ij}(0))(\BetaPI_{ij}(\bx)-\BetaPI_{ij}(0))\rangle\nonumber\\
&=&2\langle\BetaPI_{ij}\BetaPI_{ij}\rangle-2\langle\BetaPI_{ij}(\bx)\BetaPI_{ij}(0)\rangle;\label{eq:cfBetaPIA}\\
\MC^{\BetaPI}_{per}(\bx)&=&-\langle(\BetaPI_{ij}(\bx)-\BetaPI_{ij}(0))(\BetaPI_{ji}(\bx)-\BetaPI_{ji}(0))\rangle\nonumber\\
&=&-2\langle\BetaPI_{ij}\BetaPI_{ji}\rangle+2\langle\BetaPI_{ij}(\bx)\BetaPI_{ji}(0)\rangle;\label{eq:cfBetaPIB}\\
\MC^{\BetaPI}_{tr}(\bx)&=&\langle(\BetaPI_{ii}(\bx)-\BetaPI_{ii}(0))(\BetaPI_{jj}(\bx)-\BetaPI_{jj}(0))\rangle\nonumber\\
&=&2\langle\BetaPI_{ii}\BetaPI_{jj}\rangle-2\langle\BetaPI_{ii}(\bx)\BetaPI_{jj}(0)\rangle;\label{eq:cfBetaPIC}
\end{eqnarray}
where an overall minus sign is added to $\MC^{\BetaPI}_{per}$ so as to yield a positive measure.

In Fig.~\ref{fig:CorrFunc_BetaPI}, the correlation functions of the intrinsic plastic distortion $\BetaPI$
in both $1024^2$ and $128^3$ simulations exhibit a critical exponent $\sigma'$. 
These measured critical exponents are shown
in Table~\ref{tab:staticscaling}. We discuss the less physically relevant case of $\BetaPH$ in \ref{sec:betaPH},
Fig.~\ref{fig:CorrFunc_BetaPH}.

\section{\label{sec:staticscaling}Scaling theory}
The emergent self-similar dislocation morphologies are characterized by the rotational invariants of 
correlation functions of physical observables, such as the GND density $\rho$, the crystalline orientation $\Lam$, and
the intrinsic plastic distortion $\BetaPI$. Here we derive the relations expected between these correlation functions, and
show that their critical exponents collapse into a single underlying one through a generic scaling theory. 

In our model, the initial elastic stresses are relaxed via dislocation motion, leading to the formation of
cellular structures. 
In the limit of slow imposed deformations, the elastic stress goes to zero in our model. 
We will use the absence of external stress to simplify our correlation function relations. (Some relations can be valid
regardless of the existence of residual stress.)
Those relations that hold only in stress-free states will be labeled `sf'; 
they will be applicable in analyzing experiments only insofar as residual stresses are small.

\subsection{\label{sec:RelationCF}Relations between correlation functions} 
\subsubsection{$\MC^{\rho}$ and $\MC^\Lam$}
For a stress-free state, we thus ignore the elastic strain term in Eq.~(\ref{eq:rholambda1}) and write 
in Fourier space 
\begin{equation}
\widetilde{\rho}_{ij}(\bk) \sfE -ik_j\widetilde{\Lambda}_i(\bk)+i\delta_{ij}k_k\widetilde{\Lambda}_k(\bk).
\label{eq:sfrho}
\end{equation}

First, we can substitute Eq.~(\ref{eq:sfrho}) into the Fourier-transformed form of the correlation function Eq.~(\ref{eq:cfrhoA}) 
\begin{eqnarray}
\widetilde{\MC}^{\rho}_{tot}(\bk)&\sfE& \frac{1}{V}\biggl(-ik_j\widetilde{\Lambda}_i(\bk)+i\delta_{ij}k_k\widetilde{\Lambda}_k(\bk)\biggr)
\biggl(ik_j\widetilde{\Lambda}_i(-\bk)-i\delta_{ij}k_m\widetilde{\Lambda}_m(-\bk)\biggr)\nonumber\\
&\sfE&\frac{1}{V}(\delta_{ij}k^2+k_ik_j)\widetilde{\Lambda}_{i}(\bk)\widetilde{\Lambda}_{j}(-\bk).
\label{eq:rhocfinKA}
\end{eqnarray}
Multiplying both sides of Eq.~(\ref{eq:kcfrod}) by $(\delta_{ij}k^2+k_ik_j)$ gives
\begin{equation}
(\delta_{ij}k^2+k_ik_j)\widetilde{\MC}^{\Lam}_{ij}(\bk)\sfE-\frac{2}{V}(\delta_{ij}k^2+k_ik_j)\widetilde{\Lambda}_{i}(\bk)\widetilde{\Lambda}_{j}(-\bk).
\label{eq:rhoinlambda}
\end{equation}
Comparing Eq.~(\ref{eq:rhoinlambda}) and Eq.~(\ref{eq:rhocfinKA}), we may 
write $\widetilde{\MC}^{\rho}_{tot}$ in terms of $\widetilde{\MC}^{\Lam}_{ij}$ as
\begin{equation}
\widetilde{\MC}^{\rho}_{tot}(\bk) \sfE -\frac{1}{2}(\delta_{ij}k^2+k_ik_j)\widetilde{\MC}^{\Lam}_{ij}(\bk).
\label{eq:KrhocfA}
\end{equation}

Second, we can substitute Eq.~(\ref{eq:sfrho}) into the Fourier-transformed form of the correlation function Eq.~(\ref{eq:cfrhoB}) 
\begin{equation}
\widetilde{\MC}^{\rho}_{per}(\bk)\sfE\frac{2}{V}k_ik_j\widetilde{\Lambda}_{i}(\bk)\widetilde{\Lambda}_{j}(-\bk).
\label{eq:rhocfinKB}
\end{equation}
Multiplying both sides of Eq.~(\ref{eq:kcfrod}) by $k_ik_j$ and comparing with Eq.~(\ref{eq:rhocfinKB}) gives
\begin{equation}
\widetilde{\MC}^{\rho}_{per}(\bk) \sfE -k_ik_j\widetilde{\MC}^{\Lam}_{ij}(\bk).
\label{eq:KrhocfB}
\end{equation}

Finally, we substitute Eq.~(\ref{eq:sfrho}) into the Fourier-transformed form of the correlation function Eq.~(\ref{eq:cfrhoC}) 
\begin{equation}
\widetilde{\MC}^{\rho}_{tr}(\bk)\sfE\frac{4}{V}k_ik_j\widetilde{\Lambda}_{i}(\bk)\widetilde{\Lambda}_{j}(-\bk).
\label{eq:rhocfinKC}
\end{equation}
Repeating the same procedure of deriving $\widetilde{\MC}^{\rho}_{per}$, we 
write $\widetilde{\MC}^{\rho}_{tr}$ in terms of $\widetilde{\MC}^{\Lam}_{ij}$ as
\begin{equation}
\widetilde{\MC}^{\rho}_{tr}(\bk) \sfE -2k_ik_j\widetilde{\MC}^{\Lam}_{ij}(\bk).
\label{eq:KrhocfC}
\end{equation}

Through an inverse Fourier transform, we convert Eq.~(\ref{eq:KrhocfA}), Eq.~(\ref{eq:KrhocfB}),
and Eq.~(\ref{eq:KrhocfC}) back to real space to find 
\begin{eqnarray}
\MC^{\rho}_{tot}(\bx)&\sfE& \frac{1}{2}\partial^2\MC^{\Lam}(\bx)+\frac{1}{2}\partial_i\partial_j\MC^{\Lam}_{ij}(\bx),\label{eq:cfrholambdaA}\\
\MC^{\rho}_{per}(\bx)&\sfE& \partial_i\partial_j\MC^{\Lam}_{ij}(\bx),\label{eq:cfrholambdaB}\\
\MC^{\rho}_{tr}(\bx) &\sfE& 2\partial_i\partial_j\MC^{\Lam}_{ij}(\bx).
\label{eq:cfrholambdaC}
\end{eqnarray}

\subsubsection{\label{sec:betaPlam}$\MC^{\BetaPI}$ and $\MC^\Lam$}
The intrinsic part of the plastic distortion field is directly related to the GND density field. In stress-free states, 
the crystalline orientation vector can fully describe the GND density. We thus can connect $\MC^{\BetaPI}$ to $\MC^{\Lam}$.

First, substituting $\KBetaPI_{ij}=-i\varepsilon_{ilm}k_l\widetilde{\rho}_{mj}/k^2$ into the Fourier-transformed form 
of Eq.~(\ref{eq:cfBetaPIA}) gives
\begin{eqnarray}
\widetilde{\MC}^{\BetaPI}_{tot}\!\!(\bk)\!\!&=&2\langle\BetaPI_{ij}\BetaPI_{ij}\rangle(2\pi)^3\delta(\bk)-\frac{2}{V}
\biggl(-i\epsilon_{ilm}\frac{k_l}{k^2}\widetilde{\rho}_{mj}(\bk)\biggr)
\biggl(i\epsilon_{ist}\frac{k_s}{k^2}\widetilde{\rho}_{tj}(-\bk)\biggr)\nonumber\\
&=&\!\!2\langle\BetaPI_{ij}\BetaPI_{ij}\rangle(2\pi)^3\delta(\bk)\!-\!\frac{2}{k^2}\biggl(\!\frac{1}{V}
\widetilde{\rho}_{mj}(\bk)\widetilde{\rho}_{mj}(\!-\bk)\!\biggr).
\label{eq:betaPIinrhoA}
\end{eqnarray} 
During this derivation, some terms vanish due to the geometrical constraint on $\rho$, Eq.~(\ref{eq:cons_rho}).
Multiplying $-k^2/2$ on both sides of Eq.~(\ref{eq:betaPIinrhoA}) and applying the Fourier-transformed form of Eq.~(\ref{eq:cfrhoA}) gives
\begin{equation}
-\frac{k^2}{2}\widetilde{\MC}^{\BetaPI}_{tot}(\bk)=\widetilde{\MC}^{\rho}_{tot}(\bk).
\label{eq:betaPIinrhoA1}
\end{equation}
In stress-free states, we can substitute Eq.~(\ref{eq:KrhocfA}) into Eq.~(\ref{eq:betaPIinrhoA1}) 
\begin{equation}
-\frac{k^2}{2}\widetilde{\MC}^{\BetaPI}_{tot}(\bk)\sfE\widetilde{\MC}^{\rho,sf}_{tot}(\bk)=-\frac{1}{2}\biggl(\delta_{ij}k^2+k_ik_j\biggr)
\widetilde{\MC}^{\Lam}_{ij}(\bk),
\end{equation}
which is rewritten after multiplying $-2/k^2$ on both sides
\begin{equation}
\widetilde{\MC}^{\BetaPI}_{tot}(\bk)\sfE\widetilde{\MC}^{\Lam}(\bk)+\frac{k_ik_j}{k^2}
\widetilde{\MC}^{\Lam}_{ij}(\bk).
\label{eq:betaPIcfA}
\end{equation}

Second, substituting $\KBetaPI_{ij}=-i\varepsilon_{ilm}k_l\widetilde{\rho}_{mj}/k^2$ into the Fourier-transformed form of 
Eq.~(\ref{eq:cfBetaPIB}) gives
\begin{eqnarray}
\widetilde{\MC}^{\BetaPI}_{per}\!\!(\bk)\!\!&=&\!\!-2\langle\BetaPI_{ij}\BetaPI_{ji}\rangle(2\pi)^3\delta(\bk)+\frac{2}{V}
\biggl(-i\epsilon_{ilm}\frac{k_l}{k^2}\widetilde{\rho}_{mj}(\bk)\biggr)
\biggl(i\epsilon_{jst}\frac{k_s}{k^2}\widetilde{\rho}_{ti}(-\bk)\biggr)\nonumber\\
&=&\!\!-2\langle\BetaPI_{ij}\BetaPI_{ji}\rangle(2\pi)^3\delta(\bk)-\frac{2}{Vk^4}k_ik_j\widetilde{\rho}_{mj}(\bk)\widetilde{\rho}_{mi}(\!-\bk)\nonumber\\
&&+\frac{2}{k^2}
\widetilde{\MC}^{\rho}_{tot}(\bk)
-\frac{2}{k^2}\widetilde{\MC}^{\rho}_{tr}(\bk),
\label{eq:betaPIinrhoB}
\end{eqnarray} 
where we skip straightforward but tedious expansions 
and the geometrical constraint on $\rho$, Eq.~(\ref{eq:cons_rho}).
Notice that this relation is correct even in the presence of stress. 

In stress-free states, we substitute Eqs.~(\ref{eq:sfrho}),~(\ref{eq:KrhocfA}),~(\ref{eq:KrhocfC}) 
into Eq.~(\ref{eq:betaPIinrhoB}), and ignore the constant zero wavelength term
\begin{eqnarray}
\widetilde{\MC}^{\BetaPI}_{per}\!\!(\bk)&\sfE&-\frac{2k_ik_j}{Vk^4}\biggl(-ik_j\widetilde{\Lambda}_m(\bk)+i\delta_{mj}k_k\widetilde{\Lambda}_k(\bk)\biggr)
\biggl(ik_i\widetilde{\Lambda}_m(-\bk)-i\delta_{mi}k_n\widetilde{\Lambda}_n(-\bk)\biggr)\nonumber\\
&&-\frac{1}{k^2}(k^2\delta_{ij}+k_ik_j)
\widetilde{\MC}^{\Lam}_{ij}(\bk)+\frac{4}{k^2}k_ik_j
\widetilde{\MC}^{\Lam}_{ij}(\bk)\nonumber\\
&\sfE&2\frac{k_ik_j}{k^2}\widetilde{\MC}^{\Lam}_{ij}(\bk).
\label{eq:betaPIinrhoB1}
\end{eqnarray} 

Finally, substituting $\KBetaPI_{ij}=-i\varepsilon_{ilm}k_l\widetilde{\rho}_{mj}/k^2$ into the Fourier-transformed form 
of Eq.~(\ref{eq:cfBetaPIC}) gives
\begin{eqnarray}
\widetilde{\MC}^{\BetaPI}_{tr}\!\!(\bk)\!\!&=&\!\!2\langle\BetaPI_{ii}\BetaPI_{jj}\rangle(2\pi)^3\delta(\bk)-\frac{2}{V}
\biggl(-i\epsilon_{ilm}\frac{k_l}{k^2}\widetilde{\rho}_{mi}(\bk)\biggr)
\biggl(i\epsilon_{jst}\frac{k_s}{k^2}\widetilde{\rho}_{tj}(-\bk)\biggr)\nonumber\\
&=&\!\!2\langle\BetaPI_{ii}\BetaPI_{jj}\rangle(2\pi)^3\delta(\bk)+\frac{2}{Vk^4}k_ik_j\widetilde{\rho}_{mi}(\bk)\widetilde{\rho}_{mj}(\!-\bk)\nonumber\\
&&-\frac{2}{k^2}
\widetilde{\MC}^{\rho}_{tot}(\bk)+\frac{2}{k^2}\widetilde{\MC}^{\rho}_{per}(\bk),
\label{eq:betaPIinrhoC}
\end{eqnarray} 
valid in the presence of stress.
Here we repeat a similar procedure as was used to derive in Eq.~(\ref{eq:betaPIinrhoB}). 

In stress-free states, we substitute Eqs.~(\ref{eq:sfrho}),~(\ref{eq:KrhocfA}),~(\ref{eq:KrhocfB}) 
into Eq.~(\ref{eq:betaPIinrhoC})
\begin{eqnarray}
\widetilde{\MC}^{\BetaPI}_{tr}\!\!(\bk)&\sfE&2\langle\BetaPI_{ii}\BetaPI_{jj}\rangle(2\pi)^3\delta(\bk)
+\frac{1}{k^2}(k^2\delta_{ij}+k_ik_j)
\widetilde{\MC}^{\Lam}_{ij}(\bk)-\frac{2}{k^2}k_ik_j
\widetilde{\MC}^{\Lam}_{ij}(\bk)\nonumber\\
&&+\frac{2k_ik_j}{Vk^4}\biggl(-ik_i\widetilde{\Lambda}_m(\bk)+i\delta_{mi}k_k\widetilde{\Lambda}_k(\bk)\biggr)
\biggl(ik_j\widetilde{\Lambda}_m(-\bk)-i\delta_{mj}k_n\widetilde{\Lambda}_n(-\bk)\biggr)\nonumber\\
&\sfE&2\langle\BetaPI_{ii}\BetaPI_{jj}\rangle(2\pi)^3\delta(\bk),
\label{eq:betaPIinrhoC1}
\end{eqnarray} 
which is a trivial constant in space.

Through an inverse Fourier transform, Eqs.~(\ref{eq:betaPIcfA}),~(\ref{eq:betaPIinrhoB1}), and~(\ref{eq:betaPIinrhoC1}) 
can be converted back to real space, giving
\begin{eqnarray}
\MC^{\BetaPI}_{tot}\!(\bx)\!\!\!&\sfE&\!\!\MC^{\Lam}\!(\bx)\!\!+\!\!\frac{1}{4\pi}\!\!\!\int\!\!d^3\bx'\!\biggl(\!\frac{\delta_{ij}}{R^3}\!-\!3\frac{R_iR_j}{R^5}\!\biggr)\!\MC^{\Lam}_{ij}(\bx'),
\label{eq:betaPIcfRA}\\
\MC^{\BetaPI}_{per}\!(\bx)\!\!\!&\sfE&\!\!\frac{1}{2\pi}\int\!\!d^3\bx'\biggl(\frac{\delta_{ij}}{R^3}-3\frac{R_iR_j}{R^5}\biggr)\MC^{\Lam}_{ij}(\bx'),
\label{eq:betaPIcfRB}\\
\MC^{\BetaPI}_{tr}\!(\bx)\!\!\!&\sfE&\!\!2\!\!\int\!\!d^3\bx'\BetaPI_{ii}(\bx')\BetaPI_{jj}(\bx')\!=\!2\langle\BetaPI_{ii}\BetaPI_{jj}\rangle,
\label{eq:betaPIcfRC}
\end{eqnarray}
where $\vect{R}=\bx'-\bx$.
According to Eqs.~(\ref{eq:betaPIcfA}) and~(\ref{eq:betaPIinrhoB1}), we can extract a relation
\begin{equation}
\MC^{\BetaPI}_{per}(\bx)-2\MC^{\BetaPI}_{tot}(\bx)+2\MC^{\Lam}(\bx)\sfE const.
\label{eq:betaPIrho_R1}
\end{equation}

\begin{table*}[htbp]
\caption{{\bf Critical exponents for correlation functions at stress-free states.}(C.F. and S.T. represent `Correlation Functions' and
`Scaling Theory', respectively.)}
\scalebox{1}{
\begin{ruledtabular}
\begin{tabular}{cccccccc}
\multirow{3}{*}{C.F.}&\multirow{3}{*}{S.T.}&\multicolumn{6}{c}{Simulations}\\
&&\multicolumn{2}{c}{Climb\&Glide}&\multicolumn{2}{c}{Glide Only~(MDP)}&\multicolumn{2}{c}{LVP Glide Only~(LVP)}\\
&&2D($1024^2$)&3D($128^3$)&2D($1024^2$)&3D($128^3$)&2D($1024^2$)&3D($128^3$)\\
\hline
$\MC^{\rho}_{tot}$&$\eta$&$0.80\pm0.30$&$0.55\pm0.05$&$0.45\pm0.25$&$0.60\pm0.20$&$0.80\pm0.30$&$0.55\pm0.05$\\
$\MC^{\rho}_{per}$&$\eta$&$0.80\pm0.20$&$0.55\pm0.05$&$0.45\pm0.20$&$0.60\pm0.20$&$0.70\pm0.30$&$0.50\pm0.05$\\
$\MC^{\rho}_{tr}$&$\eta$&$0.80\pm0.20$&$0.55\pm0.05$&$0.45\pm0.20$&$0.60\pm0.10$&$0.70\pm0.30$&$0.45\pm0.05$\\
$\MC^{\Lam}$&$2-\eta$&$1.10\pm0.65$&$1.45\pm0.25$&$1.50\pm0.30$&$1.35\pm0.25$&$1.10\pm0.65$&$1.50\pm0.25$\\
$\MC^{\BetaPI}_{tot}$&$2-\eta$&$1.10\pm0.60$&$1.45\pm0.15$&$1.45\pm0.25$&$1.30\pm0.20$&$1.10\pm0.60$&$1.50\pm0.20$\\
$\MC^{\BetaPI}_{per}$&$2-\eta$&$1.15\pm0.45$&$1.50\pm0.25$&$1.45\pm0.25$&$1.50\pm0.50$&$1.20\pm0.45$&$1.55\pm0.25$\\
\end{tabular}
\end{ruledtabular}}
\label{tab:staticscaling}
\end{table*}

We can convert Eq.~(\ref{eq:betaPIinrhoA1}) through an inverse Fourier transform
\begin{equation}
\MC^{\rho}_{tot}(\bx)=\frac{1}{2}\partial^2\MC^{\BetaPI}_{tot}(\bx),
\label{eq:betaPIrho_R}
\end{equation}
or 
\begin{equation}
\MC^{\BetaPI}_{tot}(\bx)=-\frac{1}{2\pi}\int d^3\bx'\frac{\MC^{\rho}_{tot}(\bx')}{R},
\label{eq:betaPIrho_R0}
\end{equation}
valid in the presence of residual stress.

\subsection{Critical exponent relations}
When the self-similar dislocation structures emerge, the correlation functions of all 
physical quantities are expected to exhibit scale-free power laws. We consider the simplest
possible scenario, where single variable scaling is present to reveal the minimal number of underlying critical exponents.

First, we define the critical exponent $\eta$ as the power law describing the asymptotic decay of
$\MC^{\rho}_{tot}(\bx)\sim |\bx|^{-\eta}$, one of the correlation functions for the GND 
density tensor (summed over components).
If we rescale the spatial variable $\bx$ by a factor $b$, the correlation function $\MC^{\rho}$
is rescaled by the power law as
\begin{equation}
\MC^{\rho}_{tot}(b\bx)=b^{-\eta}\MC^{\rho}_{tot}(\bx).
\label{eq:exp_Rho}
\end{equation}

Similarly, the correlation function of the crystalline orientation field $\Lam$ is described by
a power law, $\MC^{\Lam}(\bx)\sim |\bx|^{\sigma}$, where $\sigma$ is its critical exponent. We
repeat the rescaling by the same factor $b$ 
\begin{equation}
\MC^{\Lam}(b\bx)=b^{\sigma}\MC^{\Lam}(\bx).
\label{eq:exp_Lam}
\end{equation}

Since $\MC^{\rho}_{tot}$ can be written in terms of $\MC^{\Lam}$, Eq.~(\ref{eq:cfrholambdaA}),
we rescale this relation by the same factor $b$
\begin{equation}
\MC^{\rho}_{tot}(b\bx) \sfE \frac{1}{2}\biggl[\frac{\partial}{b}\biggr]^2\MC^{\Lam}(b\bx)+\frac{1}{2}
\biggl[\frac{\partial_i}{b}\biggr]\biggl[\frac{\partial_j}{b}\biggr]\MC^{\Lam}_{ij}(b\bx).
\label{eq:vrho_lam}
\end{equation}
Substituting Eq.~(\ref{eq:exp_Lam}) into Eq.~(\ref{eq:vrho_lam}) gives
\begin{eqnarray}
\MC^{\rho}_{tot}(b\bx)& \sfE &b^{\sigma-2}\biggl[\frac{1}{2}\partial^2\MC^{\Lam}(\bx)+\frac{1}{2}
\partial_i\partial_j\MC^{\Lam}_{ij}(\bx)\biggr]\nonumber\\
&\sfE&b^{\sigma-2}\MC^{\rho}_{tot}(\bx).
\label{eq:vrho_lam_new}
\end{eqnarray}
Comparing with Eq.~(\ref{eq:exp_Rho}) gives a relation between $\sigma$ and $\eta$
\begin{equation}
\sigma = 2-\eta.
\end{equation}

We can repeat the same renormalization group procedure to analyze the critical exponents of
the other two scalar forms of the correlation functions of the GND density field. Clearly,
$\MC^{\rho}_{per}$ and $\MC^{\rho}_{tr}$ share the same critical exponent $\eta$ with $\MC^{\rho}_{tot}$.

Also, we can define the critical exponent $\sigma'$ as the power law describing the asymptotic growth of
$\MC^{\BetaPI}_{tot}(\bx)\sim|\bx|^{\sigma'}$, one of the correlation functions for the intrinsic part of
the plastic distortion field. We can rescale 
the correlation function $\MC^{\BetaPI}$
\begin{equation}
\MC^{\BetaPI}_{tot}(b\bx)=b^{\sigma'}\MC^{\BetaPI}_{tot}(\bx).
\label{eq:exp_BetaPI}
\end{equation}
We rescale the relation Eq.~(\ref{eq:betaPIrho_R}) by the same factor $b$, and substitute Eq.~(\ref{eq:exp_BetaPI}) into it
\begin{eqnarray}
\MC^{\rho}_{tot}(b\bx)&=&\frac{1}{2}\biggl[\frac{\partial}{b}\biggr]^2\MC^{\BetaPI}_{tot}(b\bx)=b^{\sigma'-2}
\biggl[\frac{1}{2}\partial^2\MC^{\BetaPI}_{tot}(\bx)\biggr]\nonumber\\
&=&b^{\sigma'-2}\MC^{\rho}_{tot}(\bx).
\end{eqnarray} 
Comparing with Eq.~(\ref{eq:exp_Rho}) also gives a relation between $\sigma'$ and $\eta$ 
\begin{equation}
\sigma' = 2-\eta.
\end{equation}

Since both $\MC^{\BetaPI}_{tot}$ and $\MC^{\Lam}$ share the same critical exponent $2-\eta$, it is clear that $\MC^{\BetaPI}_{per}$, the other
scalar form of the correlation functions of the intrinsic plastic distortion field, also shares this critical 
exponent, according to Eq.~(\ref{eq:betaPIrho_R1}).

Thus the correlation functions of three physical
quantities (the GND density $\rho$, the crystalline orientation $\Lam$, and 
the intrinsic plastic distortion $\BetaPI$) all share the same underlying universal critical exponent $\eta$
for self-similar morphologies, in the case of zero residual stress, and still hold in the limit of slow 
imposed deformation. Table~\ref{tab:staticscaling} verifies 
the existence of single underlying critical exponent in both two and three dimensional simulations for each type of dynamics.
Imposed strain, studied in Ref.~\onlinecite{YSCPRL10}, could in principle change $\eta$, but the scaling relations derived here should
still apply. The strain, of course, breaks the isotropic symmetry, allowing even more allowed correlation functions to be measured.

\subsection{\label{sec:scalescaling} Coarse graining, correlation functions, and cutoffs}

Our dislocation density $\rho$, as discussed in
Sec.~\ref{sec:CoarseGraining}, is a coarse-grained average over some
distance $\Sigma$ -- taking the discrete microscopic dislocations and 
yielding a continuum field expressing their flux in different directions.
Our power laws and scaling will be cut off in some way at this 
coarse-graining scale. For our simulations, the correlation functions
extend down to a few times the numerical grid spacing (depending on the
numerical diffusion in the algorithm we use). For experiments, the 
correlation functions will be cut off in ways that are determined by
the instrumental resolution. Since the process of coarse-graining is
at the heart of the renormalization-group methods we rely upon to 
explain the emergent scale invariance in our model, we make an initial
exploration here of how coarse-graining by the Gaussian blur of
Eq.~(\ref{eq:CoarseGraining}) and Eq.~(\ref{eq:rSN}) affects the 
$\rho^\Sigma-\rho^\Sigma$ correlation function.

Following Eq.~(\ref{eq:CoarseGraining}), 
\begin{eqnarray}
\MC^{\rho^\Sigma}_{tot}(\bx)&=&\langle\rho_{ij}^\Sigma(\bx)\rho_{ij}^\Sigma(0)\rangle\nonumber\\
&=&\frac{1}{V}\frac{1}{(2\pi\Sigma^2)^3}\int d^3\by\int d^3\bz\rho_{ij}^0(\by+\bz)e^{-z^2/(2\Sigma^2)}
\nonumber\\
&&\times\int d^3\bz'\rho_{ij}^0(\by+\bx+\bz')e^{-z'^2/(2\Sigma^2)}.
\label{eq:crhoCoarseGrain}
\end{eqnarray}
By changing variables $\bs=\by+\bz$ and $\bD=\bz'-\bz$, we integrate out the variable $\bz$ of Eq.~(\ref{eq:crhoCoarseGrain})
\begin{eqnarray}
\MC^{\rho^\Sigma}_{tot}(\bx)&=&\frac{1}{8\pi^{3/2}\Sigma^3}\frac{1}{V}\int d^3\bD\int d^3\bs\rho_{ij}^0(\bs)\rho_{ij}^0(\bs
+\bD+\bx)e^{-\Delta^2/(4\Sigma^2)}\nonumber\\
&=&\frac{1}{8\pi^{3/2}\Sigma^3}\int d^3\bD \MC^{\rho^0}_{tot}(\bx+\bD)e^{-\Delta^2/(4\Sigma^2)}.
\label{eq:crhoCoarseGrain1}
\end{eqnarray}

In our simulating system, the correlation functions of GND density can be described
by a power-law (Eq.~\ref{eq:exp_Rho}), $\MC^{\rho}_{tot}(\bx)=g|\bx|^{-\eta}$, where
$g$ is a constant. Thus, Eq.~(\ref{eq:crhoCoarseGrain1}) is  
\begin{equation}
\MC^{\rho^\Sigma}_{tot}(\bx)=\frac{g}{8\pi^{3/2}\Sigma^3}\int d^3\bD |\bx+\bD|^{-\eta}e^{-\Delta^2/(4\Sigma^2)}.
\label{eq:crhoCoarseGrain2}
\end{equation}
This correlation function of the coarse-grained GND density at the given scale $\Sigma$ is
a power-law smeared by a Gaussian distribution.

\begin{figure*}[thbp]
\centering
\includegraphics[width=.6\columnwidth]{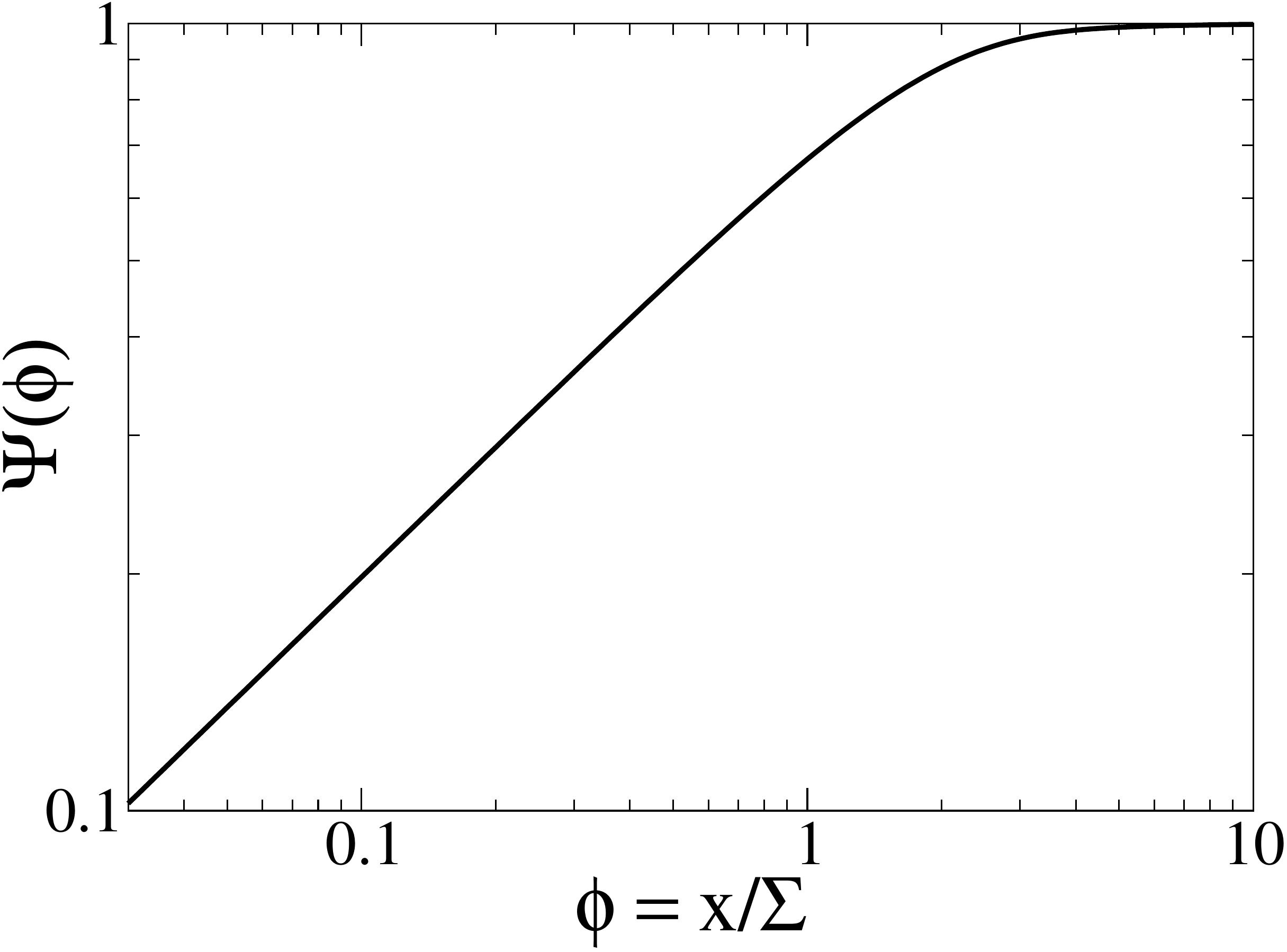}
\caption{{\bf Scaling function of the correlation function of coarse-grained GND density 
$\rho^\Sigma.$} 
We calculate the correlation function of the coarse-grained GND density at the given scale $\Sigma$.
Theoretically, its scaling function remains a power-law at the small coarse-graining length scale, and
flattens out to be 1 as the correlation length of the system is far larger than the coarse-graining
scale.}
\label{fig:CoarseGrainCorrFunc}
\end{figure*}

Since the scalar field of the coarse-grained correlation function is rotational invariant, we assume 
that $\bx$ is aligned along the $x$ axis, $\bx=(x,0,0)$. Then we could evaluate the integral of 
Eq.~(\ref{eq:crhoCoarseGrain2}) in cylindrical coordinates $\bD=(X,r,\theta)$
\begin{eqnarray}
\MC^{\rho^\Sigma}_{tot}(x,\Sigma)&=&\frac{g}{8\pi^{3/2}\Sigma^3}\int_0^{\infty}\!\!2\pi rdr\int_{-\infty}^{\infty}\!\!dX
|(x+X)^2+r^2|^{-\eta/2}e^{-(X^2+r^2)/(4\Sigma^2)}\nonumber\\
&=& \frac{g}{2^{1+\eta}\pi^{1/2}}\Sigma^{-\eta-1}\!\!\int_{-\infty}^{\infty}\!\!dX e^{(X^2+2xX)/(4\Sigma^2)}
\Gamma\bigl(1-\eta/2,(x+X)^2/(4\Sigma^2)\bigr).\nonumber\\
\label{eq:crhoCoarseGrain3}
\end{eqnarray}
We can rewrite this coarse-grained correlation functions Eq.~(\ref{eq:crhoCoarseGrain3}) as a power-law multiplied by a scaling function
\begin{equation}
\MC^{\rho^\Sigma}_{tot}(x,\Sigma)=g|x|^{-\eta}\Psi(x/\Sigma),
\end{equation}
where the scaling function $\Psi(\cdot)$
(Fig.~\ref{fig:CoarseGrainCorrFunc}) equals
\begin{equation}
\Psi(\phi)=\frac{1}{2^{1+\eta}\pi^{1/2}}|\phi|^\eta\int_{-\infty}^{\infty}
\!\!ds e^{s(s+2\phi)/4}\Gamma\bigl(1-\eta/2,(s+\phi)^2/4\bigr).
\end{equation}

\section{\label{sec:conclusion}Conclusion}
In our earlier works~\cite{JSPRL06,YSCPRL10, ChoiCiSE11}, we have 
proposed a flexible framework of CDD to study complex mesoscale phenomena of 
collective dislocation motion. Traditionally, deterministic CDDs have missed the 
experimentally ubiquitous feature of cellular
pattern formation. Our CDD models have made progress in 
that respect. In the beginning, we focused our efforts on describing coarse-grained 
dislocations that naturally develop dislocation cellular structures in ways that 
are consistent with experimental observations of scale invariance and fractality, 
a target achieved in Ref.~\onlinecite{YSCPRL10}. However, that paper studied only 2D, instead of the more realistic 3D.

In this manuscript, we go further in many aspects of the theory
extending the results of our previous work:

We provide a derivation of our theory that explains the differences with traditional
theories of plasticity. In addition to our previously studied climb-glide~(CGD) and glide-only~(GOD-MDP) models, 
we extend our construction in order to incorporate
vacancies, and re-derive~\cite{AcharyaJMPS06I} a different glide-only dynamics~(GOD-LVP) which we show
exhibits very similar behavior in 2D to our CGD model. It is worth mentioning that in this way, 
the GOD-LVP and the CGD dynamics become statistically similar in 2D, while the previously 
studied, less physical, GOD-MDP model provides rather different behavior in 2D~\cite{YSCPRL10}.

We present 3D simulation results here for the first time, showing
qualitatively different behavior from that of 2D. In 3D, all three types of 
dynamics -- CGD, GOD-MDP and GOD-LVP -- show similar non-trivial fractal patterns 
and scaling dimensions. 
Thus our 3D analysis shows that the flatter `grain boundaries' we observe in the 2D simulations
are not intrinsic to our dynamics, but are an artifact of the artificial $z$-independent initial
conditions. Experimentally, grain boundaries are indeed flatter and cleaner than cell walls, and
our theory no longer provides a new explanation for this distinction. We expect that the dislocation
core energies left out of our model would flatten the walls, and that adding disorder or entanglement
would prevent the low-temperature glide-only dynamics from flattening as much.

We also fully describe, in a statistical sense, multiple correlation functions -- the 
local orientation, the plastic distortion, the GND density -- their symmetries
and their mutual scaling relations. Correlation functions of important
physical quantities are categorized and analytically shown to share
one stress-free exponent. The anomaly in the correlation functions of $\BetaP$, which
was left as a question in our previous publication~\cite{YSCPRL10}, has been discussed and
explained. All of these correlation functions and properties are 
verified with the numerical results of the dynamics that we extensively discussed. 

As discussed in Sec.~\ref{sec:intro}, our model is an immensely simplified
caricature of the deformation of real materials. How does it connect to 
reality?

First, we show that a model for which the dynamics is driven only
by elastic strain
produces realistic cell wall structures even while ignoring 
slip systems, crystalline anisotropy~\cite{HughesPRL98}, pinning,
junction formation, and
SSDs.  The fact that low-energy dislocation
structures (LEDS) provides natural explanations for many properties of these
structures has long been emphasized by
Ref.~\onlinecite{KuhlmannMSE87}.  Intermittent flow,
forest interactions, and pinning will in general impede access
to low energy states. These real-world features, our model suggests, can be
important for the morphology of the cell wall structures but are not the
root cause of their formation nor of their evolution under stress (discussed
in previous work~\cite{YSCPRL10}). 

One must note, however, that strain
energy minimization does not provide the explanation for wall structures
in our model material.  Indeed, there is an immense space of dislocation 
densities which make the strain energy zero~\cite{JSPRB07}, including
many continuous densities. Our dynamics relaxes into a small subset of
these allowed structures -- it is the dynamics that leads to cell
structure formation here, not purely the energy. In discrete dislocation
simulations and real materials, the quantization of the Burgers vector leads
to a weak logarithmic energetic preference for sharp walls. This
$-\mu {\mathbf b}/(4\pi (1-\nu)) \theta \log \theta$ energy of low-angle grain
boundaries yields a $\log 2$ preference for one wall of angle $\theta$
rather than two walls of angle $\theta/2$. This leads to a `zipping' together of
low angle grain boundaries.  Since $\mathbf{b}\to 0$ in a
continuum theory, this preference is missing from our model. Yet, we still
find cell wall formation suggesting that such mechanisms are
not central to cell wall formation.  

Second, how should we connect our fractal cell wall structures with those
(fractal or non-fractal) seen in experiments? Many qualitatively different
kinds of cellular structures are seen in experiments -- variously termed
cell block structures, mosaic structures, ordinary cellular structures, \dots.
Ref.~\onlinecite{HansenMMTA2011} recently categorized these structures
into three types, and argue that the orientation of the stress with respect
to the crystalline axes largely determines which morphology is exhibited.
The cellular structures in our model, which ignores crystalline anisotropy,
likely are the theoretical progenitors of all of these morphologies. In
particular, Hansen's type~1 and type~3 structures incorporate both 
`geometrically necessary' and `incidental dislocation' boundaries
(GNBs and IDBs),
while type~2 structures incorporate only the latter. Our simulations cannot 
distinguish between these two types, and indeed qualitatively look similar
to Hansen's type~2 structures. One should note that the names of these
boundaries are misleading -- the `incidental' boundaries do 
mediate geometrical rotations, with the type~2 boundaries at a given
strain having similar average misorientations to the geometrically 
necessary boundaries of type~1 structures~\citep[Figure~8]{HansenMMTA2011}.
It is commonly asserted that the IDBs are formed by statistical trapping
of stored dislocations; our model suggests that stochasticity
is not necessary for their formation. 

Third, how is our model compatible with traditional plasticity, which
focuses on the total density of dislocation lines? 
Our model evolves the net dislocation density, ignoring the {\em geometrically
unnecessary} or {\em statistically stored} dislocations with cancelling
Burgers vectors. These latter dislocations are important for yield stress and
work hardening on macroscales, but are invisible to our theory (since they
do not generate stress). Insofar as the cancellation of Burgers vectors
on the macroscale is due to cell walls of opposing misorientations on 
the mesoscale, there needs to be no conflict here. Also our model remains
agnostic about whether cell boundaries include significant components of
geometrically unnecessary dislocations. However, our model does assume
that the driving force for cell boundary formation is the motion of
GNDs, as opposed to (for example) 
inhomogeneous flows of SSDs.

There still remain many fascinating mesoscale experiments, such as dislocation 
avalanches~\cite{MiguelNature01, Dimiduk06Science}, 
size-dependent hardness (smaller is stronger)~\cite{Uchic04Science},
and complex anisotropic loading~\cite{GracioMSE91,GracioIJP03}, that we
hope to emulate. We intend in the future to include several relevant additional
ingredients to our dynamics, such as vacancies~(\ref{sec:vacancy}),
impurities~(\ref{sec:disorder}),
immobile dislocations/SSDs and slip systems, to reflect real materials.

\section*{Acknowledgement}
We would like to thank A. Alemi, P. Dawson, R. LeVeque, M. Miller, E. Siggia, A. Vladimirsky, D. Warner, M. Zaiser and S. Zapperi 
for helpful and inspiring discussions on
plasticity and numerical methods over the last three years. We were supported by the Basic Energy Sciences (BES) program of DOE through
DE-FG02-07ER46393. Our work was partially supported by the National Center for
Supercomputing Applications under MSS090037 and utilized the Lincoln and Abe clusters.

\appendix
\section{\label{sec:physParameters}Physical quantities in terms of the 
plastic distortion tensor $\BetaP$}
In an isotropic infinitely large medium, the local deformation $\vu$, the elastic distortion $\BetaE$ and the
internal long-range stress $\sigma^{\rm int}$ can be expressed~\cite{TMB91,JSPRL06} in terms of the plastic distortion field
$\BetaP$ in Fourier space:
\begin{eqnarray}
\widetilde{\vu}_i(\bk)&=&N_{ikl}(\bk)\widetilde{\beta}^{\rm p}_{kl}(\bk),\nonumber\\
N_{ikl}(\bk)&=&-\frac{i}{k^2}(k_{k}\delta_{il}
+k_{l}\delta_{ik})-i\frac{\nu k_i\delta_{kl}}{(1-\nu)k^2}
+i\frac{k_ik_kk_l}{(1-\nu)k^4};
\label{eq:ku}\\
\widetilde{\beta}_{ij}^{\rm e}(\bk)&=&T_{ijkl}(\bk)\widetilde{\beta}_{kl}^{\rm p}(\bk),\nonumber\\
T_{ijkl}(\bk)&=&\frac{1}{k^2}(k_{i}k_k\delta_{jl}
+k_{i}k_l\delta_{jk}-k^2\delta_{ik}\delta_{jl})\nonumber\\
&&+\frac{k_{i}k_{j}}{(1-\nu)k^4}(\nu k^2\delta_{kl}-k_{k}k_{l});
\label{eq:kbetaE}\\
\widetilde{\sigma}_{ij}^{\rm int}(\bk)
&=&M_{ijmn}(\bk)\widetilde{\beta}_{mn}^{\rm p}(\bk),\nonumber\\
M_{ijmn}(\bk)&=&\frac{2u\nu}{1-\nu}\Bigl(\frac{k_{m}k_{n}\delta_{ij}+k_{i}k_{j}\delta_{mn}}{k^2}
-\delta_{ij}\delta_{mn}\Bigr)\nonumber\\
&&+u\Bigl(\frac{k_{i}k_{m}}{k^2}\delta_{jn}+\frac{k_{j}k_{n}}{k^2}\delta_{im}-\delta_{im}\delta_{jn}\Bigr)
\nonumber\\
&&+u\Bigl(\frac{k_{i}k_{n}}{k^2}\delta_{jm}+\frac{k_{j}k_{m}}{k^2}\delta_{in}-\delta_{in}\delta_{jm}\Bigr)
\nonumber\\
&&-\frac{2u}{1-\nu}\frac{k_{i}k_{j}k_{m}k_{n}}{k^4}.
\label{eq:ksigma}
\end{eqnarray}
All these expressions are valid for systems with periodic boundary conditions.
 
According to the definition Eq.~(\ref{eq:LambdaDef}) of the crystalline orientation $\Lam$, we can replace
$\omega^{\rm e}$ with $\BetaE$ and $\epsilon^{\rm e}$ by using the elastic distortion tensor decomposition Eq.~(\ref{eq:betaEDecom}) 
\begin{equation}
\Lambda_i=\frac{1}{2}\varepsilon_{ijk}(\BetaE_{jk}-\epsilon_{jk}^{\rm e}).
\end{equation} 
Here the permutation factor acting on the symmetric elastic strain tensor gives zero.
Hence we can express the crystalline orientation vector $\Lam$ in terms of $\BetaP$ by using Eq.~(\ref{eq:kbetaE})
\begin{eqnarray}
\widetilde{\Lambda}_{i}(\mathbf {k})
&=&\frac{1}{2}\varepsilon_{ijk}\biggl\{\frac{1}{k^2}(k_{j}k_s\delta_{kt}
+k_{j}k_t\delta_{ks}-k^2\delta_{js}\delta_{kt})\nonumber\\
&&+\frac{k_{j}k_{k}}{(1-\nu)k^4}(\nu k^2\delta_{st}-k_{s}k_{t})
\biggr\}\widetilde{\beta}^{\rm p}_{st}(\mathbf{k})\nonumber\\
&=&\frac{1}{2k^2}(\varepsilon_{ijt}k_{j}k_s
+\varepsilon_{ijs}k_{j}k_t-k^2\varepsilon_{ist})\widetilde{\beta}^{\rm p}_{st}(\mathbf {k}).
\end{eqnarray}

\section{\label{sec:varmethod}Energy dissipation rate}
\subsection{\label{sec:FourierF}Free energy in Fourier space}
In the absence of external stress, the free energy $\F$ is the elastic energy caused by the 
internal long-range stress 
\begin{equation}
\F = \int\!d^3\bx\,\,\frac{1}{2}\sigma_{ij}^{\rm int}\epsilon_{ij}^{\rm e}=\int\!d^3\bx\,\,\frac{1}{2}C_{ijmn}\epsilon_{ij}^{\rm e}\epsilon_{mn}^{\rm e},
\label{eq:HFE0}
\end{equation}
where the stress is $\sigma_{ij}^{\rm int}=C_{ijmn}\epsilon_{mn}^{\rm e}$, with $C_{ijmn}$ the stiffness tensor.

Using the symmetry of $C_{ijmn}$ and ignoring large rotations, $\epsilon_{ij}^{\rm e}=(\BetaE_{ij}+\BetaE_{ji})/2$, we can rewrite the elastic energy $\F$ in terms of $\BetaE$ 
\begin{equation}
\F = \int\!d^3\bx\,\,\frac{1}{2}C_{ijmn}\BetaE_{ij}\BetaE_{mn}.
\label{eq:HFE1}
\end{equation}

Performing a Fourier transform on both $\BetaP_{ij}$ and $\BetaP_{mn}$ simultaneously gives
\begin{equation}
\F = \int\!d^3\bx\int\frac{d^3\bk}{(2\pi)^3}\int\frac{d^3\bk'}{(2\pi)^3}\,\,e^{i(\bk+\bk')\bx}
\biggl(\frac{1}{2}C_{ijmn}\KBetaE_{ij}(\bk)\KBetaE_{mn}(\bk')\biggr).
\label{eq:HFE2}
\end{equation}
Integrating out the spatial variable $\bx$ leaves a $\delta-$function $\delta(\bk+\bk')$ in Eq.~(\ref{eq:HFE2}). We hence integrate out
the k-space variable $\bk'$ 
\begin{equation}
\F = \int\frac{d^3\bk}{(2\pi)^3}\,\,\frac{1}{2}C_{ijmn}\KBetaE_{ij}(\bk)\KBetaE_{mn}(-\bk).
\label{eq:HFE3}
\end{equation}

Substituting Eq.~(\ref{eq:kbetaE}) into Eq.~(\ref{eq:HFE3}) gives 
\begin{eqnarray}
\F \!\!&=&\!\! \int\frac{d^3\bk}{(2\pi)^3}\,\,\frac{1}{2}\bigl(C_{ijmn}T_{ijpq}(\bk)T_{mnst}(-\bk)\bigr)\KBetaP_{pq}(\bk)\KBetaP_{st}(-\bk)\nonumber\\
&=&\!\!-\int\frac{d^3\bk}{(2\pi)^3}\,\,\frac{1}{2}M_{pqst}(\bk)\KBetaP_{pq}(\bk)\KBetaP_{st}(-\bk),
\label{eq:HFE4}
\end{eqnarray}
where we skip straightforward but tedious simplifications.

When turning on the external stress, we repeat the same procedure used in Eq.~(\ref{eq:HFE2}), yielding
\begin{equation}
\F^{\rm ext}=-\int d^3\bx\,\,\sigma_{ij}^{\rm ext}\BetaP_{ij}=-\int\frac{d^3\bk}{(2\pi)^3}\,\,\Fourier{\sigma}_{ij}^{\rm ext}(\bk)\KBetaP_{ij}(-\bk).
\label{eq:HFE5}
\end{equation}

\subsection{\label{sec:calDFDrho}Calculation of energy functional derivative with respect to the GND density $\vrho$}
According to Eq.~(\ref{eq:vrho_betaP}), the infinitesimal change of the variable $\delta\vrho$ is given
in terms of $\delta\BetaP$
\begin{equation}
\delta\vrho_{ijk} = -g_{ijls}\partial_l\bigl(\delta\BetaP_{sk}\bigr).
\label{eq:EF5}
\end{equation}

Substituting Eq.~(\ref{eq:EF5}) into Eq.~(\ref{eq:EF0}) and applying integration by parts, the infinitesimal change of $\F$ 
is hence rewritten in terms of $\BetaP$
\begin{equation}
\delta\F[\BetaP] = \int\!d^3\bx\,\,g_{ijls}\partial_l\Biggl(\frac{\delta\F}{\delta\vrho_{ijk}}\Biggr)\delta\BetaP_{sk}.
\label{eq:EF6}
\end{equation}
According to Eq.~(\ref{eq:EFBetaP}), it suggests 
\begin{equation}
\delta\F[\BetaP] = \int\!d^3\bx\,\,\frac{\delta\F}{\delta\BetaP_{sk}}\delta\BetaP_{sk}
=\int\!d^3\bx\,\,(-\sigma_{sk})\delta\BetaP_{sk}.
\label{eq:EF7}
\end{equation}
Comparing Eq.~(\ref{eq:EF6}) and Eq.~(\ref{eq:EF7}) implies 
\begin{equation}
g_{ijls}\partial_l\Biggl(\frac{\delta\F}{\delta\vrho_{ijk}}\Biggr)=-\sigma_{sk},
\label{eq:sigmarho}
\end{equation}
up to a total derivative which we ignore due to the use of periodic boundary conditions.

\subsection{\label{sec:dfdt}Derivation of energy dissipation rate}
We can apply variational methods to calculate the dissipation rate of the free energy. 
As is well known, the general elastic energy $\E$ in a crystal can
be expressed as $\E = \frac{1}{2}\int\!d^3\bx\,\,\sigma_{ij}\epsilon_{ij}^{\rm e}$, with $\epsilon_{ij}^{\rm e}$ the elastic
strain. An infinitesimal change of $\E$ is: 
\begin{equation}
\delta\E=\frac{1}{2}\int\!d^3\bx\,\,\sigma_{ij}\delta\epsilon_{ij}^{\rm e}+\frac{1}{2}\int\!d^3\bx\,\,\delta\sigma_{ij}\epsilon_{ij}^{\rm e}
=\int\!d^3\bx\,\,\sigma_{ij}\delta\epsilon_{ij}^{\rm e},
\end{equation}
where we use $\sigma_{ij}\delta\epsilon_{ij}^{\rm e}=C_{ijkl}\epsilon_{kl}^{\rm e}\delta\epsilon_{ij}^{\rm e}=\delta\sigma_{ij}\epsilon_{ij}^{\rm e}$.

So the infinitesimal change of the free energy Eq.~(\ref{eq:HFR}) is 
\begin{equation}
\delta \F =
\int\!d^3\bx\,\,\biggl(\sigma_{ij}^{\rm int}\delta\epsilon_{ij}^{\rm e}-\sigma_{ij}^{\rm ext}\delta\epsilon_{ij}^{\rm p}\biggr).
\end{equation}
We apply the relation $\epsilon^{\rm e}=\epsilon-\epsilon^{\rm p}$, where $\epsilon^{\rm p}$ is the plastic strain and $\epsilon$ is
the total strain:
\begin{equation}
\delta \F =
\int\!d^3\bx\,\,\biggl(\sigma_{ij}^{\rm int}\delta\epsilon_{ij}-\sigma_{ij}^{\rm int}\delta\epsilon^{\rm p}_{ij}-\sigma_{ij}^{\rm ext}\delta\epsilon^{\rm p}\biggr).
\label{eq:dF1}
\end{equation}
Using the symmetry of $\sigma_{ij}$ and ignoring large rotations, $\epsilon_{ij}=\frac{1}{2}(\partial_iu_j+\partial_ju_i)$, we
can rewrite the first term of Eq.~(\ref{eq:dF1}) as $\int\!d^3\bx\,\,\sigma_{ij}^{\rm int}\delta(\partial_{i}u_j)$.
Integrating by parts yields 
$\int\!d^3\bx\,\,\bigl(\partial_{i}(\delta u_j\sigma_{ij}^{\rm int})-\delta u_j \partial_i\sigma_{ij}^{\rm int}\bigr)$.
We can convert the first volume integral to a surface integral, which vanishes for an infinitely large system. Hence
\begin{equation}
\delta \F = 
\int\!d^3\bx\,\,\biggl(\partial_i\sigma_{ij}^{\rm int}\delta u_j-(\sigma_{ij}^{\rm int}+\sigma_{ij}^{\rm ext})\delta\epsilon^{\rm p}_{ij}\biggr).
\label{eq:dF2}
\end{equation}
The first term of Eq.~(\ref{eq:dF2}) is zero assuming instantaneous elastic relaxation due to the local force equilibrium condition,
\begin{equation}
\delta \F=-\int\!d^3\bx\,\,(\sigma_{ij}^{\rm int}+\sigma_{ij}^{\rm ext})\delta\beta_{ij}^{\rm p},
\label{eq:dFdbetaP}
\end{equation}
using the symmetry of $\sigma_{ij}$ and $\epsilon^{\rm p}_{ij}=\frac{1}{2}(\BetaP_{ij}+\BetaP_{ji})$.

The free energy dissipation rate is thus $\delta\F/\delta t$ for $\delta\BetaP_{ij}=\frac{\partial\BetaP}{\partial t}\delta t$,
hence 
\begin{eqnarray}
\frac{\partial\mathcal{F}}{\partial t}
&=&-\int\!d^3\bx\,\,(\sigma_{ij}^{\rm int}+\sigma_{ij}^{\rm ext})\frac{\partial\BetaP_{ij}}{\partial t}\nonumber\\
&=&-\int\!d^3\bx\,\,(\sigma_{ij}^{\rm int}+\sigma_{ij}^{\rm ext})J_{ij}.
\label{eq:dfdt}
\end{eqnarray}

When dislocations are allowed to climb, substituting the CGD current Eq.~(\ref{eq:GCdynamics}) into Eq.~(\ref{eq:dfdt}) implies 
that the free energy dissipation rate is strictly negative
\begin{eqnarray}
\frac{\partial\mathcal{F}}{\partial t}&=&-\int\!d^3\bx\,\,(\sigma_{ij}^{\rm int}+\sigma_{ij}^{\rm ext})\bigl[v_l\vrho_{lij}
\bigr]\nonumber\\
&=&-\int\!d^3\bx\,\,\frac{|\vrho|}{D}v^2\leq0. 
\end{eqnarray}
 
When removing dislocation climb by considering the mobile dislocation population,
we substitute Eq.~(\ref{eq:GOdynamics1}) into Eq.~(\ref{eq:dfdt}) to guarantee that 
the rate of the change of the free energy density is also the negative of a perfect
square 
\begin{eqnarray}
\frac{\partial\mathcal{F}}{\partial t}&=&-\int d^3\bx(\sigma_{ij}^{\rm int}+\sigma_{ij}^{\rm ext})\Biggl[v'_l\bigl(\vrho_{lij}-\frac{1}{3}\delta_{ij}\vrho_{lkk}\bigr)\Biggr]\nonumber\\
&=&-\int\!d^3\bx\,\,\frac{|\vrho|}{D}v'^2\leq0. 
\label{eq:dfdt2}
\end{eqnarray}

\section{\label{sec:modelex}Model Extensions: Adding vacancies and disorder to CDD}

\subsection{\label{sec:vacancy}Coupling vacancy diffusion to CDD}
In plastically deformed crystals at low temperature, dislocations usually move only in the glide plane because vacancy diffusion is almost frozen out. 
When temperature increases, vacancy diffusion leads to dislocation climb out of the glide plane. At intermediate temperatures, slow
vacancy diffusion can enable local creep. The resulting dynamics should couple the vacancy and dislocation fields in non-trivial ways. Here
we couple the vacancy diffusion to the dislocation motion in our CDD model.

We introduce an order parameter field $c(\bx)$, indicating the vacancy concentration density at the point $\bx$. The free energy
$\F$ is thus expressed
\begin{equation}
\F=\F^{Dis}+\F^{Vac}=\int d^3\bx\biggl(\frac{1}{2}\sigma_{ij}\epsilon_{ij}^{\rm e}+\frac{1}{2}\alpha (c-c_0)^2\biggr),
\label{eq:vacF}
\end{equation}
where $\alpha$ is a positive material parameter related to the vacancy creation energy, and $c_0$ is the overall equilibrium vacancy
concentration density.

Assuming that GNDs share the velocity $\vv$ in an infinitesimal volume, we write the current $J$ for GNDs
\begin{equation}
J_{ij}=v_u\vrho_{uij}.
\label{eq:vacDJ}
\end{equation}
The current trace $J_{ii}$ describes the rate of volume change, which acts as a source and sink
of vacancies. The coupling dynamics for vacancies is thus given as
\begin{equation}
\partial_t c=\gamma\nabla^2c+J_{ii},
\label{eq:vacVJ}
\end{equation}
where $\gamma$ is a positive vacancy diffusion constant.

The infinitesimal change of the free energy $\F$ (Eq.~\ref{eq:vacF}) is
\begin{equation}
\delta\F=\int d^3\bx\biggl(\frac{\delta\F^{Dis}}{\delta\BetaP_{ij}}\delta\BetaP_{ij}+\frac{\delta\F^{Vac}}{\delta c}\delta c\biggr).
\end{equation}
We apply Eq.~(\ref{eq:dFdbetaP}) and $\delta\F^{Vac}/\delta c = \alpha (c-c_0)$ 
\begin{equation}
\delta\F=\int d^3\bx\biggl(-\sigma_{ij}\delta\BetaP_{ij}+\alpha (c-c_0)\delta c\biggr).
\end{equation}

The free energy dissipation rate is thus $\delta\F/\delta t$ for $\delta\BetaP_{ij}=\frac{\partial \BetaP}{\partial t}\delta t$
and $\delta c=\frac{\partial c}{\partial t}\delta c$, hence
\begin{equation}
\frac{\partial\F}{\partial t}=-\int d^3\bx\biggl(\sigma_{ij}\frac{\partial\BetaP_{ij}}{\partial t}-\alpha (c-c_0)\frac{\partial c}{\partial t}\biggr).
\label{eq:vacJ0}
\end{equation}

Substituting the current $J$ (Eq.~\ref{eq:vacDJ}) and Eq.~(\ref{eq:vacVJ}) into Eq.~(\ref{eq:vacJ0}) gives
\begin{eqnarray}
\frac{\partial\F}{\partial t}&=&-\int d^3\bx\bigl(\sigma_{ij}(v_u\vrho_{uij})-\alpha (c-c_0)(\gamma\nabla^2c+v_u\vrho_{uii})\bigr)\nonumber\\
&=&-\int d^3\bx\bigl((\sigma_{ij}-\alpha (c-c_0)\delta_{ij})\vrho_{uij}\bigr)v_u-\int d^3\bx\alpha\gamma(\nabla c)^2,
\end{eqnarray}
where we integrate by parts by assuming an infinitely large system.

If we choose the velocity $v_u=\frac{D}{|\vrho|}\bigl(\sigma_{ij}-\alpha (c-c_0)\delta_{ij}\bigr)\vrho_{uij}$, ($D$ is a positive material
dependent constant and $1/|\vrho|$ is added for the same reasons as discussed in Sec.~\ref{sec:GCD}), 
the free energy is guaranteed to decrease monotonically.
The coupling dynamics for both GNDs and vacancies is thus
\begin{equation}
\left\{
\begin{array}{l l}
\partial_t\BetaP_{ij}=\frac{D}{|\vrho|}\bigl(\sigma_{mn}-\alpha (c-c_0)\delta_{mn}\bigr)\vrho_{umn}\vrho_{uij},\\
\partial_t c = \gamma\nabla^2 c + \frac{D}{|\vrho|}\bigl(\sigma_{mn}-\alpha (c-c_0)\delta_{mn}\bigr)\vrho_{umn}\vrho_{ukk}.
\end{array}
\right.
\label{eq:vd_dynamics}
\end{equation}
This dynamics gives us a clear 
picture of the underlying physical mechanism: the vacancies contribute an extra hydrostatic pressure $p=-\alpha(c-c_0)$.

\subsection{\label{sec:disorder}Coupling disorder to CDD}

In real crystals, the presence of precipitates or impurities results in a force pinning 
nearby dislocations. We can mimic this effect by incorporating a spatially varying random potential field $V(\bx)$. 

In our CDD model, we can add the interaction energy between GNDs and random disorder into the free energy $\F$ (Eq.~\ref{eq:HFR})
\begin{equation}
\F=\F_{E}+\F_{I}
=\int d^3\bx\biggl(\frac{1}{2}\sigma_{ij}^{\rm int}\epsilon_{ij}^{\rm e}-\sigma_{ij}^{\rm ext}\epsilon_{ij}^{\rm p}
+V(\bx)|\vrho|\biggr),
\label{eq:newF}
\end{equation} 
where $\F_E$ indicates the elastic free energy corresponding to the integral of the first two terms, and $\F_I$ indicates
the interaction energy, the integral of the last term. 

An infinitesimal change of the free energy is written
\begin{equation}
\delta\F=\delta\F_E+\delta\F_I=\int d^3\bx\biggl(\frac{\delta\F_E}
{\delta\BetaP_{ij}}\delta\BetaP_{ij}+\frac{\delta\F_I}{\delta\BetaP_{sk}}\delta\BetaP_{sk}\biggr).
\label{eq:FEFI}
\end{equation}

In an infinitely large system, Eq.~(\ref{eq:dFdbetaP}) gives 
\begin{equation}
\frac{\delta\F_E}{\delta\BetaP_{ij}}=-(\sigma_{ij}^{\rm int}+\sigma_{ij}^{\rm ext}),
\label{eq:FEBetaP}
\end{equation}
and Eq.~(\ref{eq:EF6}) implies
\begin{eqnarray}
\delta\F_I 
&=&\int d^3\bx g_{ijls}\partial_l\Bigl(\frac{\delta\F_I}{\delta\vrho_{ijk}}\Bigr)\delta\BetaP_{sk}\nonumber\\
&=&\int d^3\bx g_{ijls}\partial_l\Bigl(V(\bx)\frac{\vrho_{ijk}}{|\vrho|}\Bigr)\delta\BetaP_{sk}.
\label{eq:FIRho}
\end{eqnarray}

Substituting Eq.~(\ref{eq:FEBetaP}) and Eq.~(\ref{eq:FIRho}) into Eq.~(\ref{eq:FEFI}) gives
\begin{eqnarray}
\delta\F&=&-\int d^3\bx\biggl(\sigma_{ij}^{\rm int}+\sigma_{ij}^{\rm ext}-g_{mnli}
\partial_l\Bigl(V(\bx)\frac{\vrho_{mnj}}{|\vrho|}\Bigr)\biggr)\delta\BetaP_{ij}\nonumber\\
&=&-\int d^3\bx\sigma^{\rm eff}_{ij}\delta\BetaP_{ij}.
\end{eqnarray}
where the effective stress field is $\sigma^{\rm eff}_{ij}=\sigma_{ij}^{\rm int}+\sigma_{ij}^{\rm ext}-g_{mnli}
\partial_l\Bigl(V(\bx)\frac{\vrho_{mnj}}{|\vrho|}\Bigr)$.

By replacing $\sigma_{ij}$ with $\sigma^{\rm eff}_{ij}$ in the equation of motion of either allowing climb
(Eq.~\ref{eq:GCdynamics}) or removing climb (Eqs.~\ref{eq:GOdynamics1} and~\ref{eq:nGOdynamics1}),
we achieve the new CDD model that models GNDs interacting with disorder. 

\begin{figure*}[htbp]
\centering
\includegraphics[width=1.\columnwidth]{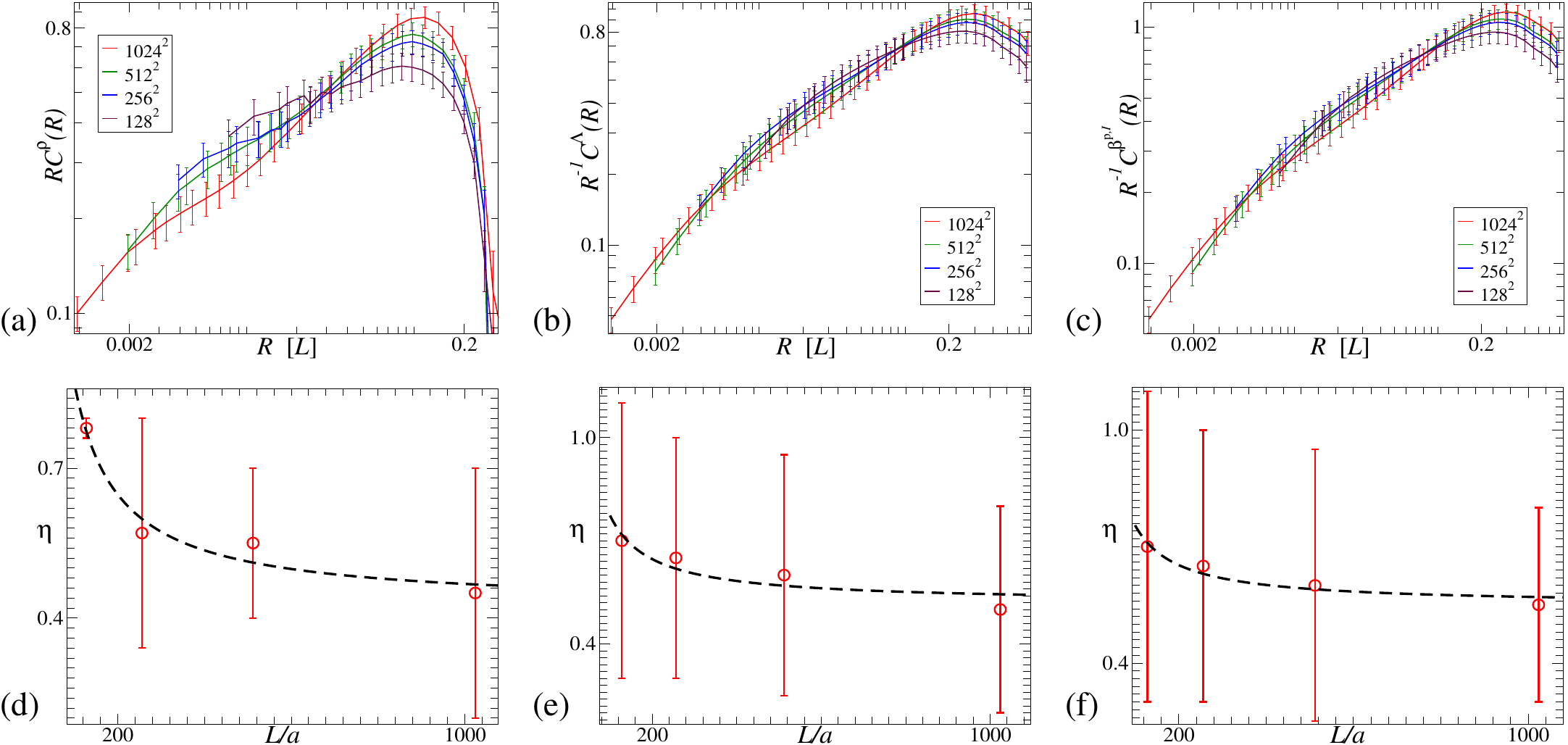}
\caption{(Color online)~{\bf Statistical convergence of correlation functions of $\Lambda$, $\rho$ and $\BetaPI$ by varying lattice sizes
in two dimensions.} 
We compare correlation functions of relaxed glide-only states (GOD-MDP) at resolutions from $128^2$ to $1024^2$ systems. 
{\it Top:} We see that the correlation functions in all cases exhibit similar power laws in (a), (b), and (c); {\it Bottom:} (d), (e), and (f) show a
single underlying critical exponent which appears to converge with increasing resolution, where $a$ is the grid spacing. 
The black dashed lines are guides to the eye.}
\label{fig:Convergence}
\end{figure*}

\begin{figure*}[htbp]
\centering
\includegraphics[width=1.\columnwidth]{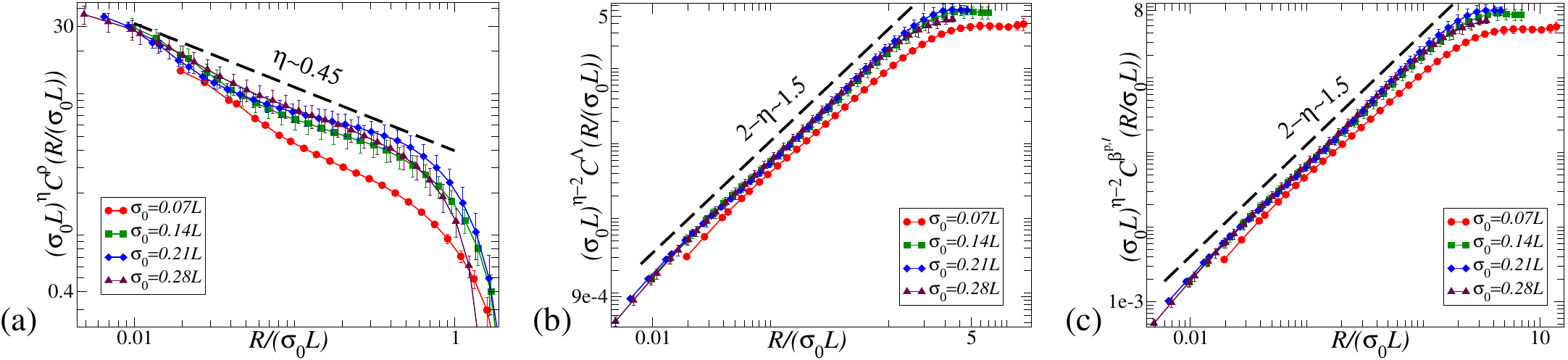}
\caption{(Color online)~{\bf Statistical convergence of correlation functions of $\Lambda$, $\rho$ and $\BetaPI$ by varying the initial
length scales in $1024^2$ simulations.} 
We measure correlation functions of relaxed glide-only states (GOD-MDP) at initial correlated lengths from $0.07L$ to $0.28L$.
In (a), (b), and (c), the radial-length variable $R$ is rescaled by their initial correlation lengths, and the corresponding correlation
functions are divided by the same lengths to the exhibiting powers. They roughly collapse into the scaling laws. Notice that the power laws
measured in the state with the initial correlated length $0.07L$ get distorted due to the small outer cutoff.}
\label{fig:LSFSS}
\end{figure*}

\section{\label{sec:detail}Details of the Simulations}
\subsection{\label{sec:sc}Finite size effects}
Although we suspect that our simulations don't have weak solutions~\cite{tobepublished}, we can show that these solutions
converge statistically. We use two ways to exhibit the statistical convergence. 

When we continue to decrease the grid spacing to zero (the
continuum limit), we show the statistical convergence of correlation functions of $\rho$, $\Lam$, and $\BetaPI$,
with a slow expected drift of apparent exponents with system size, see Fig.~\ref{fig:Convergence}. 

We can also decrease the initial correlated length scales in a large two dimensional simulation. Since the emergent
self-similar structures are always developed below the initial correlated lengths, as discussed in Sec.~\ref{sec:corrfuncVI}, this is similar 
to decreasing the system size by reducing the initial correlated lengths. In Fig.~\ref{fig:LSFSS}, the correlation functions of 
$\rho$, $\Lam$, and $\BetaPI$ collapse into a single scaling curve, using finite size scaling.

\subsection{\label{sec:initialcond}Gaussian random initial conditions}
Gaussian random fields are extensively used in physical modelings to mimic stochastic fluctuations with a 
correlated length scale. In our simulations, we construct an initially random plastic distortion, a nine-component tensor
field, where every component is an independent Gaussian random field sharing a underlying length scale. 

We define a Gaussian random field $f$ with correlation length $\sigma_0$ by convolving white noise
$\langle\xi(\bx)\xi(\bx')\rangle=\delta(\bx-\bx')$ with a Gaussian of width $\sigma_0$:
\begin{eqnarray}
f(\bx)=\int d^3\bx'\xi(\bx')e^{-(\bx-\bx')^2/\sigma_0^2}.
\label{eq:newf1}
\end{eqnarray}
In Fourier space, this can be done as a multiplication:
\begin{equation}
\widetilde{f}(\bk)=e^{-\sigma_0^2k^2/4}\widetilde{\xi}(\bk).
\label{eq:newf}
\end{equation}

The square $\widetilde{f}(\bk)\widetilde{f}(-\bk)=e^{-\sigma_0^2k^2/2}$ implies that
the correlation function $\langle f(\bx)f(\bx')\rangle=(2\pi\sigma^2_0)^{-3/2}e^{-(\bx-\bx')^2/(2\sigma_0^2)}$.

\begin{figure*}[htbp]
\centering
\includegraphics[width=1.\columnwidth]{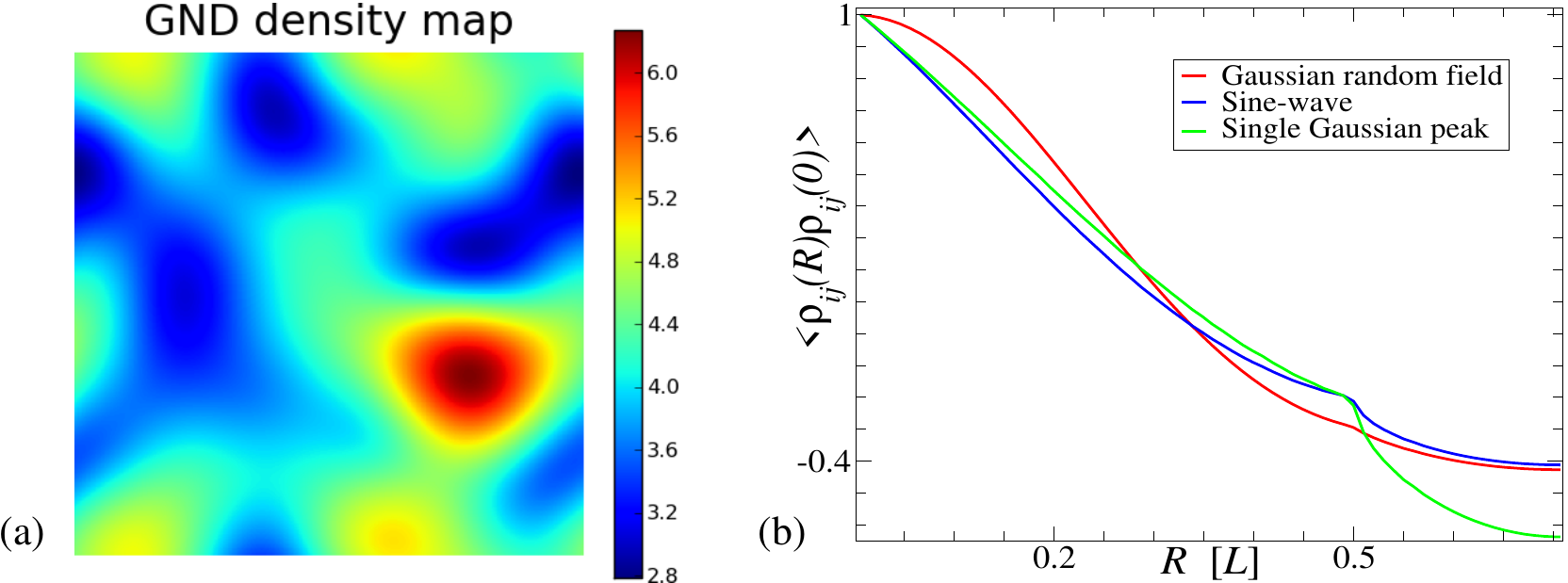}
\caption{(Color online)~{\bf Gaussian random initial conditions with the correlated length scale $0.28L$ in two dimensions.} 
(a) shows the initial net GND density map; (b) exhibits the correlation functions of $\rho$ under various initial conditions, 
where we compare the Gaussian random field to both a sinusoidal wave and a single periodic superposition of Gaussian peaks. 
The kink arises due to the edges and corners of the square unit cell. }
\label{fig:Initial}
\end{figure*}

In our simulations, the initial plastic distortion tensor field $\BetaP$ is constructed in Fourier space
\begin{equation}
\KBetaP_{ij}(\bk)=e^{-\sigma_0^2k^2/4}\widetilde{\zeta}_{ij}(\bk),
\label{eq:initBetaP}
\end{equation}
where the white noise signal $\zeta$ is characterized as $\langle\zeta_{(i,j)}(\bx)\zeta_{(i,j)}(\bx')\rangle=A_{(i,j)}\delta(\bx-\bx')$,
and in Fourier space $\frac{1}{V}\widetilde{\zeta}_{(i,j)}(\bk)\widetilde{\zeta}_{(i,j)}(-\bk)=A_{(i,j)}$. 
(We use $(i,j)$ to indicate a component of the tensor field, to avoid the Einstein summation rule.)
The correlation function of each component of $\BetaPI$ is thus expressed in Fourier space
\begin{eqnarray}
\widetilde{\MC}^{\BetaPI}_{(i,j)}&=&2\langle\BetaPI_{(i,j)}\BetaPI_{(i,j)}\rangle(2\pi)^3\delta(\bk)
-\frac{2}{V}\KBetaP_{(i,j)}(\bk)\KBetaP_{(i,j)}(-\bk)\nonumber\\
&=&2\langle\BetaPI_{(i,j)}\BetaPI_{(i,j)}\rangle(2\pi)^3\delta(\bk)
-2A_{(i,j)}e^{-\sigma_0^2k^2/2},
\end{eqnarray} 
where the Gaussian kernel width $\sigma_0$, as a standard length scale, defines the correlation length of our simulation. (In our earlier
work, we use a non-standard definition for the correlation length, so our $\sigma_0$ equals the old length scale times $\sqrt{2}$.)

According to Eq.~(\ref{eq:betaP_rho}) and Eq.~(\ref{eq:initBetaP}), we can express the initial GND density field $\rho$ in Fourier space
\begin{equation}
\Krho_{ij}(\bk)=-i\varepsilon_{ilm}e^{-\sigma_0^2k^2/4}k_l\widetilde{\zeta}_{mj}(\bk).
\end{equation}
The scalar invariant $\MC^{\rho}_{tot}$ of the correlation function of $\rho$ is thus expressed in Fourier space
\begin{eqnarray}
\MC^{\rho}_{tot}(\bk)&=&\frac{1}{V}\Krho_{ij}(\bk)\Krho_{ij}(-\bk)\nonumber\\&=&
\frac{1}{V}e^{-\sigma_0^2k^2/2}\bigl(k^2\delta_{mn}-k_mk_n)\widetilde{\zeta}_{mj}(\bk)\widetilde{\zeta}_{nj}(-\bk).
\label{eq:initBetaPCF}
\end{eqnarray}

The resulting initial GND density is not Gaussian correlated, unlike the initial plastic distortion. Figure~\ref{fig:Initial}
exhibits the initial GND density map due to the Gaussian random plastic distortions with the correlation length $0.28L$, and its correlation function.
We compare the latter to the correlation functions of both a sinusoidal wave and a single periodic superposition of Gaussian peaks. The
similarity of the three curves shows that 
our Gaussian random initial condition at $\sigma_0\sim0.28L$ approaches the largest effective correlation length possible for periodic boundary
conditions.

\begin{figure*}[htbp]
\centering
\includegraphics[width=0.85\columnwidth]{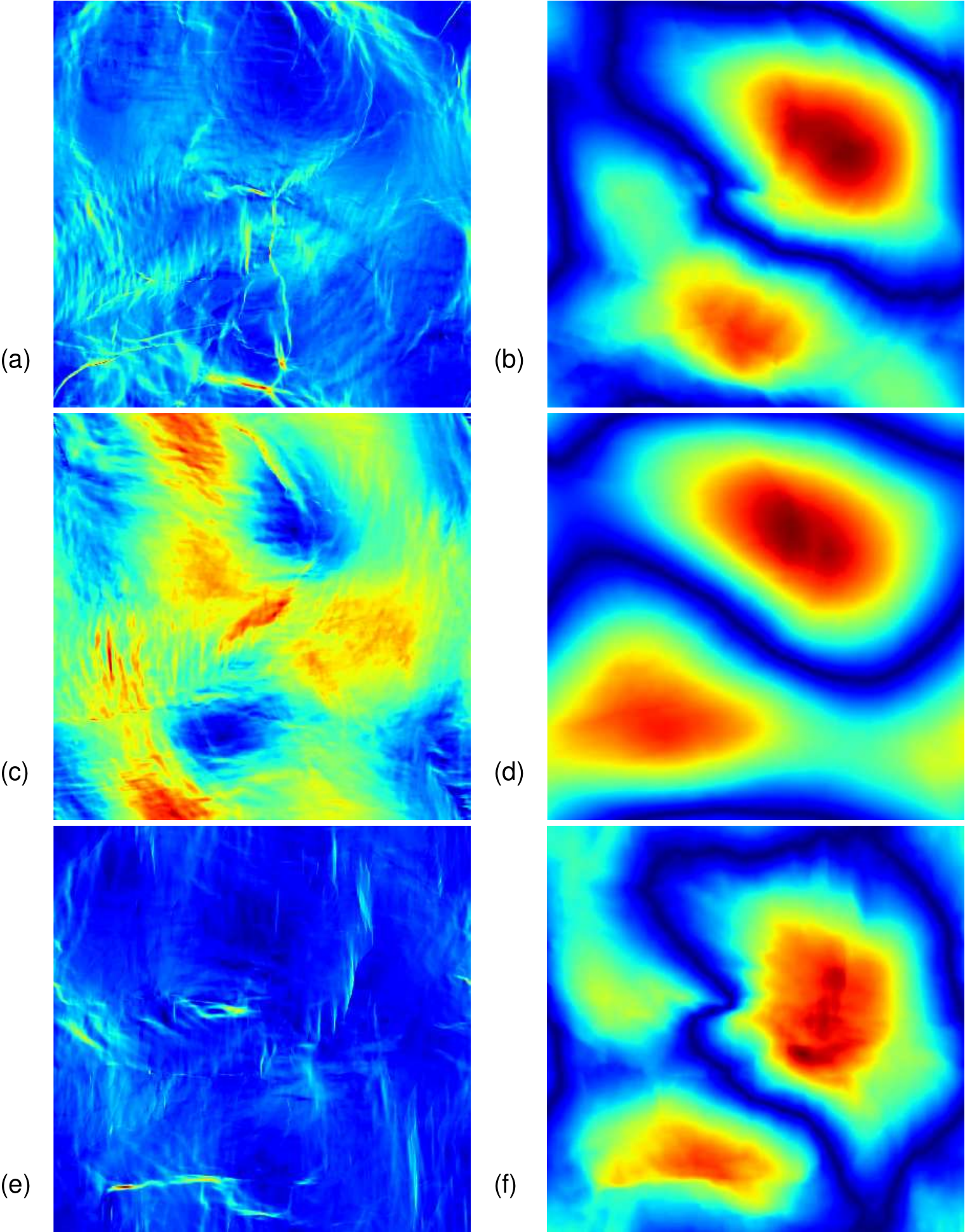}
\caption{(Color online)~{\bf Strain-history-dependent fields $\BetaPH$ and $\bm\psi$ in two dimensions}
for the relaxed states. {\it Top:} Dislocation climb is allowed; {\it Middle:} Glide-only using a mobile dislocation population; 
{\it Bottom:} Glide-only using a local vacancy pressure. {\it Left:} The strain-history-dependent plastic
distortion $|\BetaPH|$. (a), (c), and (e) exhibit patterns reminiscent of self-similar dislocation structures. 
{\it Right:} The strain-history-dependent plastic deformation $|\bm\psi|$. (b), (d), and (f) exhibit smooth patterns with a little distortion, which
are not fractal.}
\label{fig:HistoryDep}
\end{figure*}
\begin{figure*}[htbp]
\centering
\includegraphics[width=1.\columnwidth]{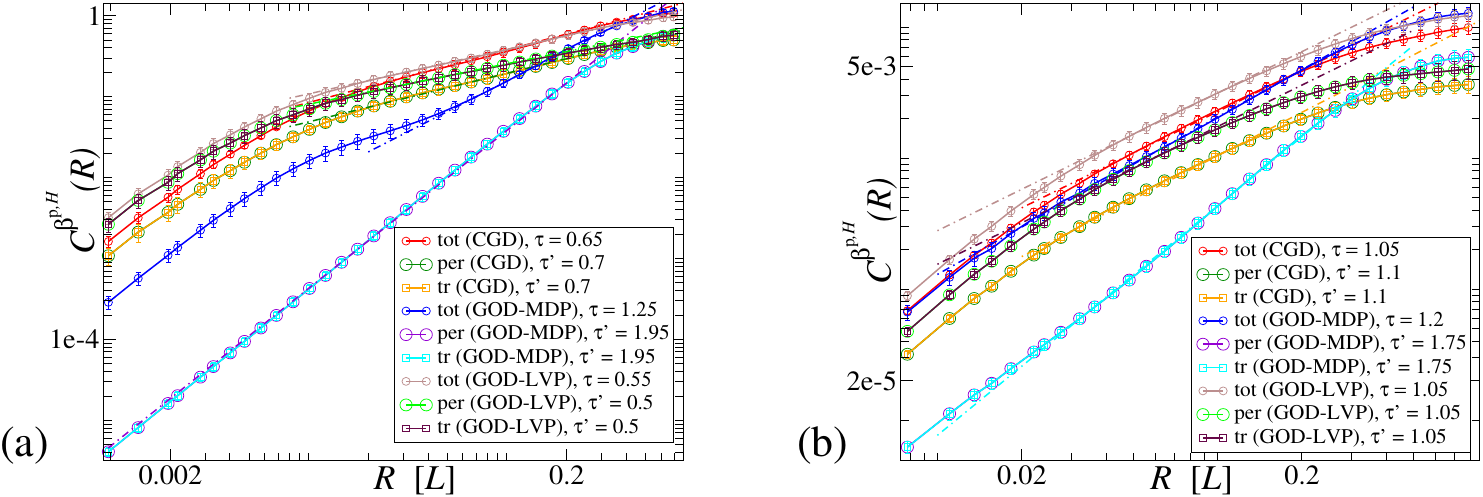}
\caption{(Color online)~{\bf Correlation functions of $\BetaPH$ in both two and three dimensions.}
In both (a) and (b), the correlation functions of the strain-history-dependent part of the plastic distortion $\BetaPH$ are shown.
{\it Left:} (a) is measured in relaxed, unstrained $1024^2$ systems; {\it Right:} (b) is measured in
in relaxed, unstrained $128^3$ systems.
All dashed lines show estimated power laws quoted in Table~\ref{tab:histscaling}.
}
\label{fig:CorrFunc_BetaPH}
\end{figure*}
\begin{figure*}[htbp]
\centering
\includegraphics[width=1.\columnwidth]{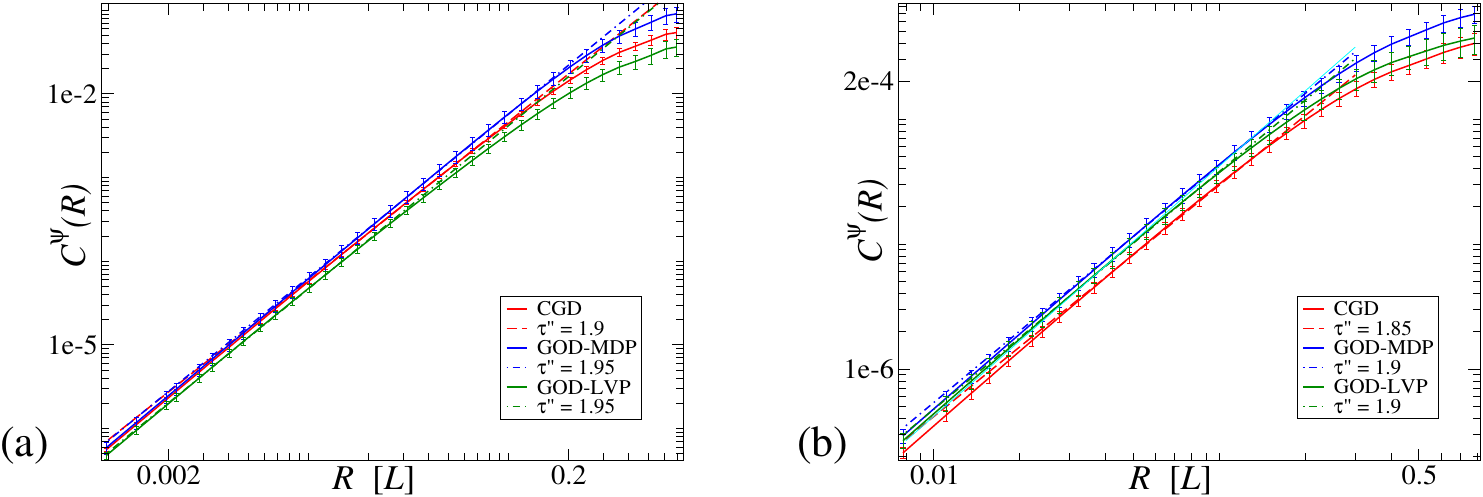}
\caption{(Color online)~{\bf Correlation functions of $\bm\psi$ in both two and three dimensions.}
In (a) and (b), the correlation functions of the strain-history-dependent deformation $\bm\psi$ are shown.
Red, blue, green lines indicate CGD, GOD-MDP, and GOD-LVP, respectively. 
{\it Left:} (a) is measured in relaxed, unstrained $1024^2$ systems; {\it Right:} (b) is
measured in relaxed, unstrained $128^3$ systems.
All dashed lines show estimated power laws quoted in Table~\ref{tab:histscaling}.}
\label{fig:CorrFunc_Psi}
\end{figure*}
\begin{table*}[!htbp]
\caption{{\bf Critical exponents for correlation functions of strain-history-dependent fields at stress-free states.}(C.F. and Exp. represent
`Correlation Functions' and `Exponents', respectively.)}
\scalebox{1.}{
\begin{ruledtabular}
\begin{tabular}{cccccccc}
\multirow{3}{*}{C.F.}&\multirow{3}{*}{Exp.}&\multicolumn{6}{c}{Simulations}\\
&&\multicolumn{2}{c}{Climb\&Glide}&\multicolumn{2}{c}{Glide~Only~(MDP)}&\multicolumn{2}{c}{Glide~Only~(LVP)}\\
&&2D($1024^2$)&3D($128^3$)&2D($1024^2$)&3D($128^3$)&2D($1024^2$)&3D($128^3$)\\
\hline
$\MC^{\BetaPH}_{tot}$&$\tau$&$0.65\pm1.00$&$1.05\pm0.65$&$1.25\pm0.60$&$1.20\pm0.50$&$0.55\pm1.10$&$1.05\pm0.65$\\
$\MC^{\BetaPH}_{per}$&$\tau'$&$0.70\pm0.95$&$1.10\pm0.60$&$1.95\pm0.05$&$1.75\pm0.15$&$0.50\pm1.15$&$1.05\pm0.70$\\
$\MC^{\BetaPH}_{tr}$&$\tau'$&$0.70\pm0.95$&$1.10\pm0.60$&$1.95\pm0.05$&$1.75\pm0.15$&$0.50\pm1.15$&$1.05\pm0.70$\\
$\MC^{\bm\psi}$&$\tau''$&$1.90\pm0.10$&$1.85\pm0.15$&$1.95\pm0.05$&$1.90\pm0.10$&$1.95\pm0.05$&$1.90\pm0.10$\\
\end{tabular}
\end{ruledtabular}}
\label{tab:histscaling}
\end{table*}

\section{\label{sec:othercorrfunc}Other correlation functions unrelated to static scaling theory}
\subsection{\label{sec:betaPH}Correlation functions of the strain-history-dependent plastic deformation and distortion fields}
The curl-free strain-history-dependent part of the plastic distortion field, as shown in 
Fig.~\ref{fig:HistoryDep}(a), (c), and~(e), exhibits structures reminiscent of self-similar morphology. We correlate their differences at neighboring points
\begin{eqnarray}
\MC^{\BetaPH}_{tot}(\bx)&=&\langle(\BetaPH_{ij}(\bx)-\BetaPH_{ij}(0))(\BetaPH_{ij}(\bx)-\BetaPH_{ij}(0))\rangle,\label{eq:cfBetaPHA}\\
\MC^{\BetaPH}_{per}(\bx)&=&\langle(\BetaPH_{ij}(\bx)-\BetaPH_{ij}(0))(\BetaPH_{ji}(\bx)-\BetaPH_{ji}(0))\rangle,\label{eq:cfBetaPHB}\\
\MC^{\BetaPH}_{tr}(\bx)&=&\langle(\BetaPH_{ii}(\bx)-\BetaPH_{ii}(0))(\BetaPH_{jj}(\bx)-\BetaPH_{jj}(0))\rangle.\label{eq:cfBetaPHC}
\end{eqnarray}

Consider also the deformation field $\bm\psi$ (shown in Fig.~\ref{fig:HistoryDep}(b), (d), and~(f)) of Eq.~(\ref{eq:betaPinrho}) whose gradient gives the 
strain-history-dependent plastic deformation $\BetaPH$. 
Similarly to the crystalline orientation $\Lam$, we correlate differences of $\bm\psi$. 
The unique rotational invariant of its two-point correlation functions is written
\begin{equation}
\MC^{\bm\psi}(\bx) = 2\langle\psi^2\rangle-2\langle\psi_i(\bx)\psi_i(0)\rangle.
\label{eq:rcfpsi}
\end{equation}

In Fig.~\ref{fig:CorrFunc_BetaPH}, the correlation functions of the strain-history-dependent
plastic distortion $\BetaPH$ in both $1024^2$ and $128^3$ simulations show critical exponents $\tau$ and $\tau'$.
Although apparently unrelated to the previous underlying critical exponent $\eta$, this exponents $\tau$ and $\tau'$ quantify the 
fractality of the strain-history-dependent plastic distortion.
Figure~\ref{fig:CorrFunc_Psi} shows the correlation functions of the strain-history-dependent deformation
$\bm\psi$, with the critical exponent $\tau''$ close to $2$, which implies a smooth non-fractal field, shown in
Fig.~\ref{fig:HistoryDep}(c) and~(d).
All measured critical exponents are 
listed in Table~\ref{tab:histscaling}.

Figure~\ref{fig:CorrFunc_BetaPH} shows the power-law dependence of the rotational invariants $\MC^{\BetaPH}_{per}$ and $\MC^{\BetaPH}_{tr}$ (they overlap). 
According to the definition $\KBetaPH_{ij}=ik_i\widetilde{\psi}_j$, we can write down the Fourier-transformed forms of Eq.~(\ref{eq:cfBetaPHB}) 
and Eq.~(\ref{eq:cfBetaPHC}) respectively 
\begin{eqnarray}
\widetilde{\MC}^{\BetaPH}_{per}(\bk)\!\!&=&\!\!2\langle\BetaPH_{ij}\BetaPH_{ji}\rangle(2\pi)^3\delta(\bk)
-\frac{2}{V}k_ik_j\widetilde{\psi}_j(\bk)\widetilde{\psi}_i(-\bk),
\label{eq:betaPIIcfkB}\\
\widetilde{\MC}^{\BetaPH}_{tr}(\bk)\!\!&=&\!\!2\langle\BetaPH_{ii}\BetaPH_{jj}\rangle(2\pi)^3\delta(\bk)
-\frac{2}{V}k_ik_j\widetilde{\psi}_{i}(\bk)\widetilde{\psi}_j(-\bk).
\label{eq:betaPIIcfkC}
\end{eqnarray}
Except the zero-wavelength terms, the same functional forms shared by these two rotational scalars explain the observed overlapping power laws.

\subsection{\label{sec:corrfunc_stress}Stress-stress correlation functions}
As the system relaxes to its final stress-free state, we can measure the fluctuations of the internal 
elastic stress fields, using a complete set of two rotational invariants of correlation functions
\begin{eqnarray}
\MC^{\sigma}_{tot} (\bx) = \langle\sigma_{ij}^{\rm int}(\bx)\sigma_{ij}^{\rm int}(0)\rangle,\\
\MC^{\sigma}_{tr} (\bx) = \langle\sigma_{ii}^{\rm int}(\bx)\sigma_{jj}^{\rm int}(0)\rangle;
\label{eq:SigmaCorr}
\end{eqnarray} 
and in Fourier space
\begin{eqnarray}
\widetilde{\MC}^{\sigma}_{tot} (\bk) = \frac{1}{V}\widetilde{\sigma}_{ij}^{\rm int}(\bk)\widetilde{\sigma}_{ij}^{\rm int}(-\bk),
\label{eq:KSigmaCorrA}\\
\widetilde{\MC}^{\sigma}_{tr} (\bk) = \frac{1}{V}\widetilde{\sigma}_{ii}^{\rm int}(\bk)\widetilde{\sigma}_{jj}^{\rm int}(-\bk).
\label{eq:KSigmaCorrB}
\end{eqnarray} 
Because $\sigma_{ij}$ is symmetric, these two correlation functions form a complete set of linear invariants under rotational transformations. 

\subsection{\label{sec:spectrum}Energy density spectrum}
The average internal elastic energy $\E$ is written 
\begin{eqnarray}
\E &=& \frac{1}{V}\int d^d\bx\biggl[\frac{1}{2}\sigma_{ij}^{\rm int}\epsilon_{ij}^{\rm e}\biggr]\nonumber\\
&=&\frac{1}{V}
\int d^d\bx\frac{1}{4\mu}\biggl[\sigma_{ij}^{\rm int}\sigma_{ij}^{\rm int}-\frac{\nu}{1+\nu}\sigma_{ii}^{\rm int}\sigma_{jj}^{\rm int}\biggr],
\label{eq:Espec}
\end{eqnarray}
where, in an isotropic bulk medium, the elastic strain $\epsilon^{\rm e}$ is expressed in terms of $\sigma^{\rm int}$,
\begin{equation}
\epsilon^{\rm e}_{ij} = \frac{1}{2\mu}\biggl(\sigma^{\rm int}_{ij}-\frac{\nu}{1+\nu}\delta_{ij}\sigma^{\rm int}_{kk}\biggr).
\label{eq:stiff}
\end{equation}

\begin{figure*}[htbp]
\centering
\psfrag{Cs(k)}{$\widetilde{C}^{\sigma}(k)$}
\psfrag{ks}{$k\;\;\mathbf{[2\pi]}$}
\psfrag{es(k)}{$E(k)$}
\psfrag{ke}{$k\;\;\mathbf{[2\pi]}$}
\psfrag{Cr(k)}{$\widetilde{\MC}^{\rho^E}(k)$}
\psfrag{kr}{$k\;\;\mathbf{[2\pi]}$}
\includegraphics[width=1.\columnwidth]{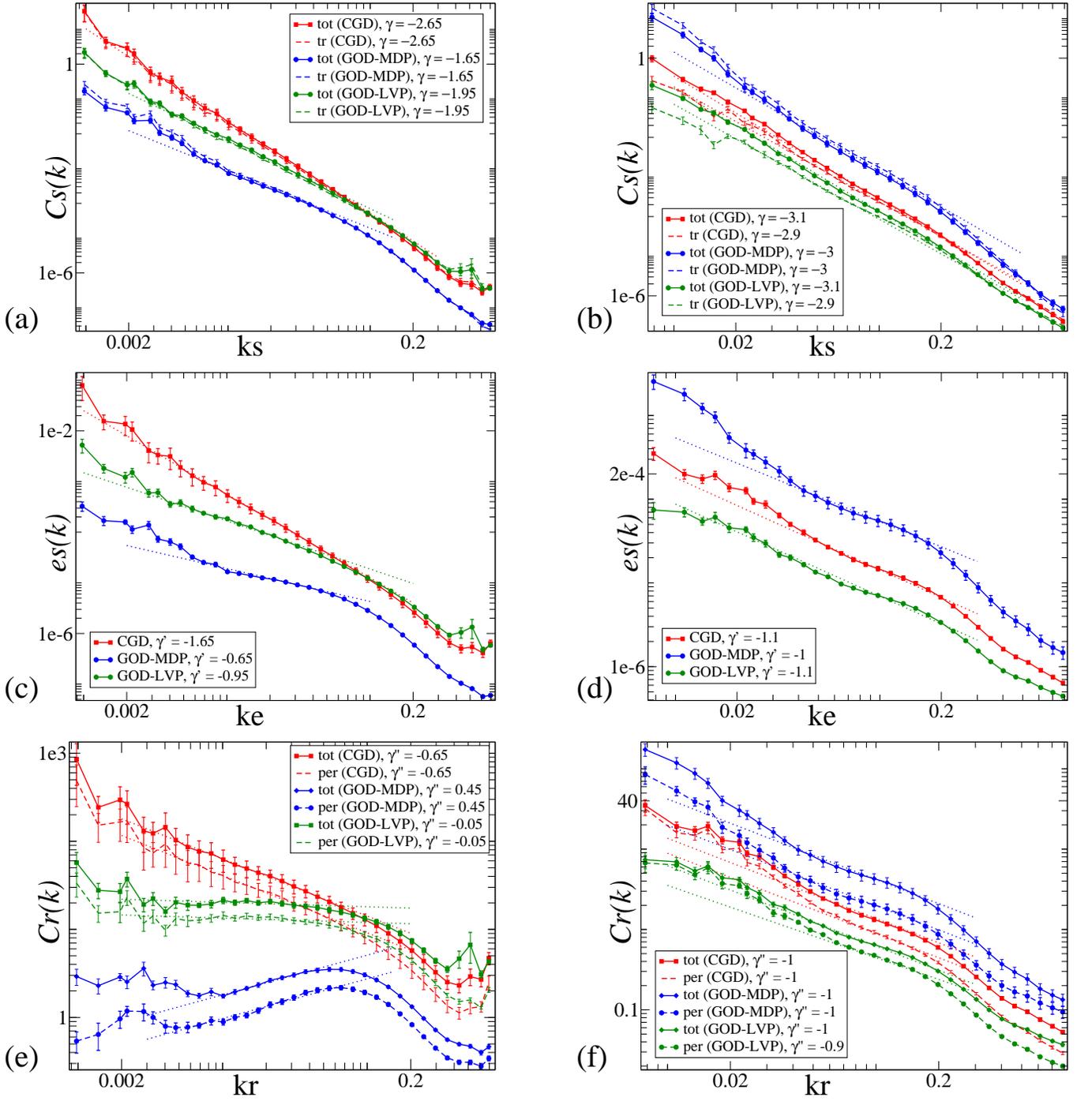}
\caption{(Color online)~{\bf Stress-stress correlation functions $\widetilde{C}^{\sigma}(k)$, elastic energy spectrum $E(k)$, 
correlation functions of the stress-full part of GND density $\widetilde{\MC}^{\rho^E}(k)$.} Red, blue, and green lines indicate
CGD, GOD-MDP, and GOD-LVP, respectively. All dashed lines show estimated power laws quoted
in Table~\ref{tab:otherscaling}. }
\label{fig:OtherCF}
\end{figure*}
\begin{table*}[!htbp]
\caption{{\bf Power-laws relations} among $\widetilde{C}^{\sigma}(k)$, $E(k)$, and $\widetilde{\MC}^{\rho^E}(k)$. (d represents the dimension; P.Q. and
S.T. represent `Physical Quantities' and `Scaling Theory', respectively.)}
\scalebox{1}{
\begin{ruledtabular}
\begin{tabular}{cccccccc}
\multirow{3}{*}{P.Q.}&\multirow{3}{*}{S.T.}&\multicolumn{6}{c}{Simulations}\\
&&\multicolumn{2}{c}{Climb\&Glide}&\multicolumn{2}{c}{Glide Only~(MDP)}&\multicolumn{2}{c}{Glide Only~(LVP)}\\
&&2D($1024^2$)&3D($128^3$)&2D($1024^2$)&3D($128^3$)&2D($1024^2$)&3D($128^3$)\\
\hline
$\widetilde{C}^{\sigma}_{tot}(k)$&$\gamma$&$-2.65$&$-3.1$&$-1.65$&$-3.0$&$-1.95$&$-3.1$\\
$\widetilde{C}^{\sigma}_{tr}(k)$&$\gamma$&$-2.65$&$-2.9$&$-1.65$&$-3.0$&$-1.95$&$-2.9$\\
$E(k)$&$\gamma+d-1$&$-1.65$&$-1.1$&$-0.65$&$-1.0$&$-0.95$&$-1.1$\\
$\widetilde{\MC}^{\rho^E}_{tot}(k)$&$\gamma+2$&$-0.65$&$-1.0$&$0.45$&$-1.0$&$-0.05$&$-1.0$\\
$\widetilde{\MC}^{\rho^E}_{per}(k)$&$\gamma+2$&$-0.65$&$-1.0$&$0.45$&$-1.0$&$-0.05$&$-0.9$\\
\end{tabular}
\end{ruledtabular}}
\label{tab:otherscaling}
\end{table*}

We can rewrite Eq.~(\ref{eq:Espec}) in Fourier space
\begin{eqnarray}
\E &=& \frac{1}{V} 
\int\frac{d^d\bk}{(2\pi)^d}\frac{1}{4\mu}\biggl[\widetilde{\sigma}_{ij}^{\rm int}(\bk)\widetilde{\sigma}_{ij}^{\rm int}(-\bk)-
\frac{\nu}{1+\nu}\widetilde{\sigma}_{ii}^{\rm int}(\bk)\widetilde{\sigma}_{jj}^{\rm int}(-\bk)\biggr].
\label{eq:EspecK}
\end{eqnarray}

Substituting Eq.~(\ref{eq:KSigmaCorrA}) and Eq.~(\ref{eq:KSigmaCorrB}) into Eq.~(\ref{eq:EspecK}) gives 
\begin{equation}
\E = 
\int\frac{d^d\bk}{2^{d+2}\pi^d}\frac{1}{\mu}\biggl[\widetilde{\MC}^{\sigma}_{tot}(\bk)-
\frac{\nu}{1+\nu}\widetilde{\MC}^{\sigma}_{tr}(\bk)\biggr]
\label{eq:EspecK2}
\end{equation}
If the stress-stress correlation functions are isotropic, we can integrate out the angle variable of Eq.~(\ref{eq:EspecK2})
\begin{equation}
\E = 
\int_{0}^{\infty}\!\!dk\frac{f(d)}{\mu}k^{d-1}\biggl[\widetilde{\MC}^{\sigma}_{tot}(k)-
\frac{\nu}{1+\nu}\widetilde{\MC}^{\sigma}_{tr}(k)\biggr],
\label{eq:EspecK3}
\end{equation}
where $f(d)$ is a constant function over the dimension $d$,
\begin{equation}
f(d) = 
\begin{cases}
1/(8\pi) & d=2,\\
1/(8\pi^2) & d=3.
\end{cases}
\end{equation}

Writing the elastic energy density in terms of the energy density spectrum $\E(t)=\int_{0}^{\infty}E(k,t)dk$ implies
\begin{equation}
E(k) = 
\frac{f(d)}{\mu}k^{d-1}\biggl[\widetilde{\MC}^{\sigma}_{tot}(k)-
\frac{\nu}{1+\nu}\widetilde{\MC}^{\sigma}_{tr}(k)\biggr].
\label{eq:EspecK4}
\end{equation}

\subsection{\label{sec:Erho_cf}Correlation function of the stressful part of GND density}
According to Eq.~(\ref{eq:rholambda1}), the stressful part of GND density is defined as
\begin{equation}
\rho^E_{ij}(\bx) = \varepsilon_{isl}\partial_s\epsilon_{lj}^{\rm e}(\bx).
\label{eq:stressfulrho}
\end{equation}
Substituting Eq.~(\ref{eq:stiff}) into Eq.~(\ref{eq:stressfulrho}) gives
\begin{equation}
\rho^{E}_{ij} = \frac{1}{2\mu}\varepsilon_{isl}\partial_s\biggl(\sigma_{lj}^{\rm int}-\frac{\nu}{1+\nu}\delta_{lj}\sigma_{mm}^{\rm int}\biggl).
\label{eq:rhosf}
\end{equation}

The complete set of rotational invariants of the correlation function of $\rho^E$ includes three scalar forms
\begin{eqnarray}
\MC^{\rho^E}_{tot}(\bx) &=& \langle\rho_{ij}^E(\bx)\rho_{ij}^E(0)\rangle,\label{eq:sfrho0}\\
\MC^{\rho^E}_{per}(\bx) &=& \langle\rho_{ij}^E(\bx)\rho_{ji}^E(0)\rangle,\label{eq:sfrho1}\\
\MC^{\rho^E}_{tr}(\bx) &=& \langle\rho_{ii}^E(\bx)\rho_{jj}^E(0)\rangle,\label{eq:sfrho2}
\end{eqnarray}
where $\MC^{\rho^E}_{tr}(\bx)$ is always zero due to $\rho^{E}_{ii}=0$.

Substituting Eq.~(\ref{eq:rhosf}) into both Eqs.~(\ref{eq:sfrho0}) and~(\ref{eq:sfrho1}) and applying the Fourier transform gives
\begin{eqnarray}
\widetilde{\MC}^{\rho^E}_{tot}(\bk) &=&\frac{1}{4\mu^2V}\varepsilon_{isl}(ik_s)\biggl(\widetilde{\sigma}_{lj}^{\rm int}(\bk)-\frac{\nu}{1+\nu}\delta_{lj}\widetilde{\sigma}_{mm}^{\rm int}(\bk)\biggl)
\nonumber\\
&&\times\varepsilon_{ipq}(-ik_p)\biggl(\widetilde{\sigma}_{qj}^{\rm int}(-\bk)-\frac{\nu}{1+\nu}\delta_{qj}\widetilde{\sigma}_{nn}^{\rm int}(-\bk)\biggl)\nonumber\\
&=&\frac{k^2}{4\mu^2}\biggl(\frac{1}{V}\widetilde{\sigma}_{lj}^{\rm int}(\bk)\widetilde{\sigma}_{lj}^{\rm int}(-\bk)\biggr)
-\frac{\nu k^2}{2\mu^2(1+\nu)^2}
\biggl(\frac{1}{V}\widetilde{\sigma}_{mm}^{\rm int}(\bk)\widetilde{\sigma}_{nn}^{\rm int}(-\bk)\biggr),\nonumber\\
\label{eq:kcfErho1}\\
\widetilde{\MC}^{\rho^E}_{per}(\bk) &=&\frac{1}{4\mu^2V}\varepsilon_{isl}(ik_s)
\biggl(\widetilde{\sigma}_{lj}^{\rm int}(\bk)-\frac{\nu}{1+\nu}
\delta_{lj}\widetilde{\sigma}_{mm}^{\rm int}(\bk)\biggl)\nonumber\\&&\times
\varepsilon_{jpq}(-ik_p)\biggl(\widetilde{\sigma}_{qi}^{\rm int}(-\bk)-\frac{\nu}{1+\nu}\delta_{qi}\widetilde{\sigma}_{nn}^{\rm int}(-\bk)\biggl)\nonumber\\
&=&\frac{k^2}{4\mu^2}\biggl(\frac{1}{V}\widetilde{\sigma}_{lj}^{\rm int}(\bk)\widetilde{\sigma}_{lj}^{\rm int}(-\bk)\biggr)
-\frac{(1+\nu^2) k^2}{4\mu^2(1+\nu)^2}
\biggl(\frac{1}{V}\widetilde{\sigma}_{mm}^{\rm int}(\bk)\widetilde{\sigma}_{nn}^{\rm int}(-\bk)\biggr),\nonumber\\
\label{eq:kcfErho2}
\end{eqnarray}

where we make use of the equilibrium condition $\partial_i\sigma_{ij}=0$ and thus $k_i\widetilde{\sigma}_{ij}=0$. 
Substituting Eqs.~(\ref{eq:KSigmaCorrA}) and~(\ref{eq:KSigmaCorrB}) into Eqs.~(\ref{eq:kcfErho1}) and~(\ref{eq:kcfErho2}) 
\begin{equation}
\widetilde{\MC}^{\rho^E}_{tot}(\bk)=\frac{k^2}{4\mu^2}\biggl[\widetilde{\MC}^{\sigma}_{tot}(\bk)
-\frac{2\nu}{(1+\nu)^2}\widetilde{\MC}^{\sigma}_{tr}(\bk)\biggr],
\end{equation}
\begin{equation}
\widetilde{\MC}^{\rho^E}_{per}(\bk)=\frac{k^2}{4\mu^2}\biggl[\widetilde{\MC}^{\sigma}_{tot}(\bk)
-\frac{1+\nu^2}{(1+\nu)^2}\widetilde{\MC}^{\sigma}_{tr}(\bk)\biggr].
\end{equation}
Here we can ignore the angle dependence if the stress-stress correlation functions are isotropic.

\subsection{Scaling relations}
According to Eq.~(\ref{eq:EspecK4}), the term $k^{d-1}$ suggests that the power-law exponent relation between 
$E$ and $\widetilde{\MC}^{\sigma}$ is
\begin{equation}
\gamma' = \gamma+d-1.
\end{equation}
Again, both Eqs.~(\ref{eq:kcfErho1}) and~(\ref{eq:kcfErho2}) imply 
that the power-law exponent relation between $\widetilde{\MC}^{\rho^E}$ and 
$\widetilde{\MC}^{\sigma}$ is 
\begin{equation}
\gamma'' = \gamma+2,
\end{equation}
regardless of the dimension.

Table~\ref{tab:otherscaling} shows a nice agreement between predicted scaling and numerical measurements for power-law exponents
of $\widetilde{C}^{\sigma}$, $E$, and $\widetilde{\MC}^{\rho^E}$. These relations are valid in the presence of residual stress.

During the relaxation processes, the elastic free energy follows a power-law decay in time asymptotically, seen in Fig.~\ref{fig:Energy3D}. All
the above measured correlation functions of elastic quantities share the same power laws in Fourier space, albeit with decaying magnitudes in time.


\end{document}